\documentclass[prl, aps, twocolumn, amssymb, superscriptaddress]{revtex4-2}
\usepackage{amssymb}
\usepackage{amsbsy}
\usepackage{amsmath}
\usepackage{amsthm}
\usepackage{graphicx}
\usepackage{graphics}
\usepackage{setspace}
\usepackage{array}
\usepackage{color}
\usepackage{xcolor}
\usepackage{fontenc}
\usepackage{textcomp}
\usepackage{bm}
\usepackage{pifont}
\usepackage[bookmarks=false,linkcolor=blue,urlcolor=blue,colorlinks,citecolor=blue]{hyperref}
\usepackage{soul}
\usepackage[T1]{fontenc}
\usepackage{mathdots}
\usepackage{makecell}

\usepackage{dsfont}
\usepackage{bbold}

\newcommand{\Nabla}{\bm{\nabla}}
\newcommand{\kk}{{\bs k}}
\newcommand{\pp}{{\bs p}}

\newcommand{\rr}{{\bs r}}

\newcommand{\vv}{\bs v}
\newcommand{\dd}{{\rm d}}
\newcommand{\bs}{\boldsymbol}

\newcommand{\blue}[1]{{\color{blue}{#1}}}

\newcommand{\bsub}{\begin{subequations}}
\newcommand{\esub}{\end{subequations}}

\definecolor{darkred}{rgb}{0.8,0,0}
\definecolor{royalblue}{rgb}{0.0, 0.14, 0.4}
\definecolor{magenta}{cmyk}{0,.9,0,0.2}
\definecolor{amethyst}{rgb}{0.6, 0.4, 0.8}
\definecolor{cadmiumgreen}{rgb}{0.0, 0.42, 0.24}
\definecolor{deepcarmine}{rgb}{0.66, 0.13, 0.24}
\definecolor{forestgreen}{rgb}{0.13, 0.55, 0.13}

\newcommand{\beginsupplement}{
        \setcounter{table}{0}
        \renewcommand{\thetable}{S\arabic{table}}
        \setcounter{figure}{0}
        \renewcommand{\thefigure}{S\arabic{figure}}
        \setcounter{equation}{0}
        \renewcommand{\theequation}{S\arabic{equation}}
        \setcounter{section}{0}
        \renewcommand{\thesection}{\Alph{section}}
        \setcounter{subsection}{0}
        \renewcommand{\thesubsection}{\arabic{subsection}}
}

\begin{document}
\title{Nonlinear thermal and thermoelectric transport from quantum geometry}

\author{Yuan Fang}
\thanks{These authors contributed equally}
\affiliation{Department of Physics and Astronomy,  Extreme Quantum Materials Alliance, Smalley Curl Institute, Rice University, Houston, Texas 77005, USA}
\author{Shouvik Sur}
\thanks{These authors contributed equally}
\affiliation{Department of Physics and Astronomy,  Extreme Quantum Materials Alliance, Smalley Curl Institute, Rice University, Houston, Texas 77005, USA}
\author{Yonglong Xie}
\affiliation{Department of Physics and Astronomy,  Extreme Quantum Materials Alliance, Smalley Curl Institute, Rice University, Houston, Texas 77005, USA}
\author{Qimiao Si}
\affiliation{Department of Physics and Astronomy,  Extreme Quantum Materials Alliance, Smalley Curl Institute, Rice University, Houston, Texas 77005, USA}

\begin{abstract}
Quantum geometry may enable the development of quantum phases ranging from superconductivity to correlated topological states. One powerful probe of quantum geometry is the nonlinear Hall response which detects Berry curvature dipole in systems with time-reversal invariance and broken inversion symmetry. With broken time-reversal symmetry, this response is also associated with quantum metric dipole.
Here we investigate nonlinear thermal and thermoelectric responses, which provide a wealth of new information about quantum geometry. In particular, we uncover a web of connections between these quantities that parallel the standard Wiedemann-Franz and Mott relations. Implications for the studies of a variety of topological systems, including Weyl-Kondo semimetals and Bernal bilayer graphene, are discussed.
\end{abstract}

\maketitle

\blue{\emph{Introduction}}--- 
In crystalline settings, quantum geometry characterizes the adiabatic evolution of electronic states, and may play an important role in enabling novel quantum phases such as correlated topological states and superconductivity~\cite{provost1980riemannian,Mar97.1,Peotta2015,verma2025quantum}.
Berry curvature, the antisymmetric component of the quantum geometry tensor, characterizes the topology of individual bands. 
In the presence of time-reversal symmetry ($\cal T$), and when the inversion symmetry (${\cal P}$) is broken, its dipole is manifested through nonlinear Hall effect~\cite{deyo2009semiclassical,Fu2015BCD,low2015topological,ma2019observation,Kang2019nonlinear,Kumar2021NLHE,kaplan2025}.
For example, in the case of Weyl-Kondo semimetals~\cite{lai2018weyl,grefe2020weyl, dzsaber2017kondo}, this effect and its fully nonequilibrium counterpart~\cite{sur2024fully} have played an important role in the experimental characterization~\cite{Dzsaber2021Giant}.
In systems with broken time-reversal symmetry, the dipoles of the quantum metric,  the symmetric components of the quantum geometry tensor, also contribute to the nonlinear Hall effect~\cite{Gao2014Field,Wang2021Intrinsic,Liu2021Intrinsic,PhysRevLett.127.277202,kaplan2024unification}.
Especially when the inversion symmetry ${\cal P}$ is broken as well, but the product ${\cal PT}$ is preserved, the Berry curvature's contributions to anomalous (charge) Hall effect vanish, both at the linear and nonlinear orders, due to Kramers degeneracy of the bands (though there can still be spin-Hall current~\cite{murakami2003dissipationless}).
This leaves the quantum metric as the sole contributor to intrinsic anomalous nonlinear Hall effect in these systems, which has  been measured experimentally~\cite{QMDnature,QMDscience}.
Since these effects only provide slices of information about the quantum geometry, further probes are highly desirable and can help the overall cause of overcoming the general difficulty of probing the quantum geometry tensor~\cite{Neupert2013Measuring,Kang2025}.

\begin{figure}[t]
\centering
\includegraphics[width=\linewidth]{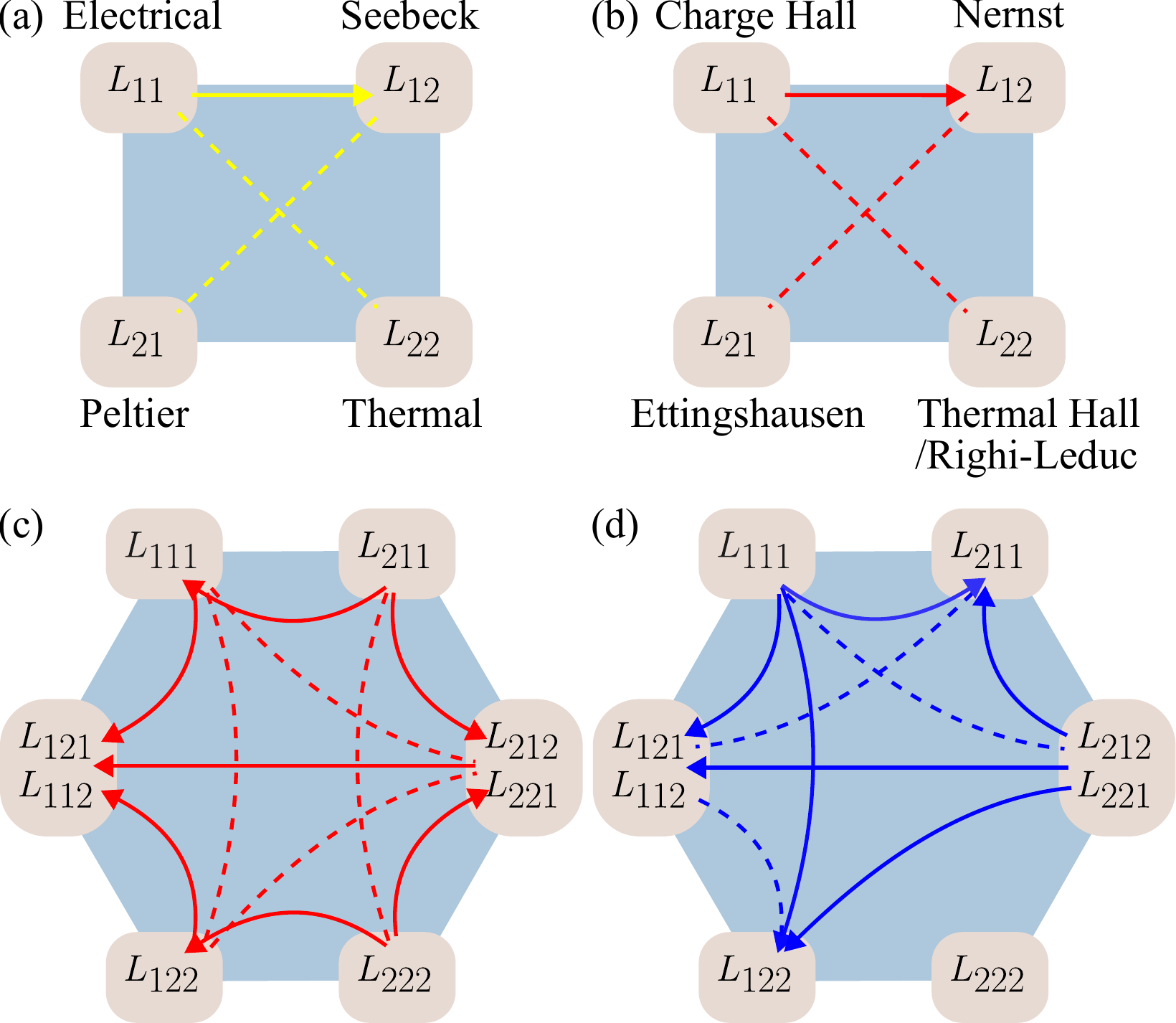}
\caption{(a) Relations for linear order longitudinal responses.  
(b) Relations for linear order transverse responses.
(c) Relations for second order transverse responses with $\cal T$ symmetry.
(d) Relations for second order transverse responses with $\cal PT$ symmetry.
Solid lines indicate Mott type relations and arrows show the direction of derivative.
Dashed lines indicate Wiedemann-Franz type relations.
Here, red and blue colors represent Berry curvature and quantum metric effects, respectively,  while yellow color represents Drude contributions.
}
\label{fig:relations}
\end{figure}

We are thus motivated to consider the role of quantum geometry to intrinsic nonlinear thermal and thermoelectric transport. 
A particular focus of our analysis is on whether, and in what way, the nonlinear electrical, thermal, and thermoelectric transport coefficients are fundamentally related~\cite{Zeng2020Fundamental}.
Such relationships encode the nature of the quantum geometry that underlies the mechanism for nonlinear transport.
Given the crucial role of symmetries in shaping topological properties, a systematic framework is required to understand how symmetries govern the relationships among various nonlinear transport coefficients.

In this paper, we use a framework that treats temperature gradient and electric field on equal footing, and demonstrate a set of distinct generalized Wiedemann-Franz and Mott relations for the nonlinear transport coefficients.
For definiteness, our focus will be on systems with $\mathcal{T}$ and $\mathcal{PT}$ symmetries.
We apply our results to Bernal bilayer graphene and show that the concurrence of enhanced density of states and high concentration of both components of the quantum geometry makes this system as an ideal setting to test the predictions we make in this work.
We also discuss the implications of our results for the non-centrosymmetric and time-reversal-symmetric Weyl-Kondo semimetals, in which an enhanced spontaneous (nonlinear) Hall effect has been demonstrated~\cite{Dzsaber2021Giant} and analyzed~\cite{sur2024fully}.

\blue{\emph{Conventional relations among electrical, thermal, and thermoelectric coefficients}}--- 
We begin by defining the general nonlinear response coefficients. The charge and heat currents are expanded as power series in the applied electric field and thermal driving force,
\begin{equation}
\label{eqn:sig_def}
j^c_l = \sum_{n,m\geq0} L_{l, \underbrace{1 \ldots 1}_{n\ \text{times}},\,\underbrace{2 \ldots 2}_{m\ \text{times}}}^{c, a_1\dots a_n b_1\dots b_m} ~ E_{a_1}\dots E_{a_n} (\nabla_{b_1} T)\dots (\nabla_{b_m}T) 
\end{equation}
where $j^c_l$ refers to the $c$-th spatial component of the charge ($l = 1$) or heat ($l=2$) current $\bs j_l$, and $a_j$ ($b_j$) labels the spatial components of the applied electric fields (temperature gradients).
In linear response, the electric and heat currents take the form 
$j^a_1 = L^{a,b}_{1,1} E_b + L^{a,b}_{1,2} (\nabla_b T)$ and 
$j_2^a = L^{a,b}_{2,1} E_b + L^{a,b}_{2,2} (\nabla_b T)$.
Here the first lower index $1(2)$ denotes charge(heat) current, while the index after the comma labels the driving fields, namely the electric field $\boldsymbol{E}$ and the temperature gradient $\boldsymbol{\nabla} T$.
The corresponding transport coefficients are connected by the Wiedemann–Franz law, the Onsager reciprocity relations, and the Mott relation 
(see Figs.~\ref{fig:relations}\,(a) and (b) for a graphical representation).
The Wiedemann–Franz law~\cite{franz1853ueber, jonson1980mott} and the Mott 
relation~\cite{Mott1969, smrcka1977transport} state that,
\begin{align}
L^{a,b}_{2,2} = \mathfrak{L} T L^{a,b}_{1,1}; \qquad 
L_{1,2}^{a,b} = e \mathfrak{L} T\, \partial_\mu L_{1,1}^{a,b},
\end{align}
where $\mathfrak{L} = \pi^2 k_B^2 / 3e^2$ is the Lorenz number, and $\mu$ is the chemical potential.
Here, $L^{a,b}_{1,1}$, $L^{a,b}_{2,2}$, and $L_{1,2}^{a,b}$ are the charge-Hall, thermal-Hall, and Seebeck (thermopower) coefficients (the transverse component of $L_{1,2}^{a,b}$ is also referred to as the Nernst coefficient), respectively.
Finally, the Onsager reciprocity relation states that $L_{2,1}^{ab} = T L_{1,2}^{ba}$~\cite{Onsager1931, callen1948application}, where $L_{21}^{ab}$ is the Peltier thermoelectric coefficients.
These relationships have been central to characterizing the Drude transport of Fermi liquid systems~\cite{ashcroft1976solid,Girvin_Yang_2019,mahan2013many}, and deviations from them at sufficiently low temperatures have been interpreted as signatures of new physics beyond the Fermi liquid description~\cite{Pfa12.1,ghawri2022breakdown}.
These relations are  expected to hold for the linear-response 
anomalous conductivities of $\cal T$-broken systems~\cite{Xiao2006Berry,onoda2008quantum, pu2008mott}.

\blue{\emph{General formalism for nonlinear responses}}---
We now outline the approach for determining the nonlinear transport coefficients of a partially filled band and their relationships.
To be definite, we address the \emph{weak-field regime} (as opposed to the fully nonequilibrium regime~\cite{sur2024fully}), where $|\bs E|$ and $|\bs \nabla T|$ are small compared to both the Fermi energy and the energy gap that separates the partially-full band from all other bands, and focus on the intrinsic responses controlled by the quantum geometry.
In the weak-field limit, the effect of completely empty and full bands on the intrinsic nonlinear transport is captured through their contributions to the quantum metric (to be discussed below).
Within our formalism, there exist three independent sources of nonlinearity, viz. nonequilibrium distribution function, group velocity, and Berry curvature density. 

The nonequilibrium distribution function is obtained from the Boltzmann equation in the relaxation time approximation,
\begin{align}
\{ \partial_t + \dot \rr \cdot \nabla_{\rr} + \dot \kk \cdot \nabla_{\kk} \} f(\rr,\kk,t) = 
-\frac{f(\rr,\kk,t) -f_0(\kk)}{\tau},
\end{align}
where we have assumed a uniform momentum relaxation time, $\tau$, and $f_0$ is the equilibrium distribution function. 
By utilizing the semiclassical equations of motion, $\dot \rr = \frac{1}{\hbar} \frac{\partial \epsilon_n(\kk)}{\partial \kk} - e \bs E \times \bs \Omega$ and $\dot \kk = -\frac{e}{\hbar} \bs E \,$,
we expand $f$ in terms of $\bs E$ and $\bs \nabla T$, 
\begin{align}
f = f_0 + \sum_{n,m} f^{(n,m)} E^n (\nabla T/T)^m
\end{align}
and consider the direct current (DC) limit. 
The coefficients in the expansion, $f^{(n,m)}$, are derived in the supplementary materials (SM)~\cite{sm}. 

The group velocity and Berry curvature density can be expanded in terms of the electric field~\cite{kaplan2024unification,Fang2024Quantum,Wen2025Thermal, sm}:
\begin{align}
 v_{n\kk}^{a\prime} &= \frac{1}{\hbar} \frac{\partial\epsilon_{n\kk}}{ \partial k_a} + \frac{1}{\hbar}  \partial_{k_a} G_n^{bc} F_bF_c + O(F^3) \,, \label{eqn:v_correction} \\
\Omega^{c\prime}_{n\kk} &= \Omega^{c}_{n\kk} + F_d \epsilon^{abc} \partial_{k_a} G^{bd}_n + O(F^2) \,, \label{eqn:omega_correction}
\end{align}
where the band-normalized quantum metric $G_n^{ab}=\sum_{m\neq n}(A_{nm}^a A_{mn}^b+A_{mn}^a A_{nm}^b)/(2\varepsilon_{nm})$ with $A_{nm}^a$ being the inter-band Berry connection $A^a_{nm} = -i\langle u_{n\bs{k}}|\partial_{k_a} u_{m\bs{k}}\rangle$, $\varepsilon_{nm}$ the energy difference.
Here the effective force $\mathbf{F}=-e\mathbf{E}+(\epsilon-\mu)\nabla T/T$, where the temperature gradient acts as a statistical force and is therefore treated on an equal footing with the electric field~\cite{sm,Luttinger1964}.
As a result, the quantum metric dipole enters both corrections, giving rise to second-order response contributions from both group velocity and Berry curvature.
Although we exclusively focus on the quantum geometric contribution to transverse response functions in the main text, in the SM~\cite{sm} we also discuss the structure of the Drude terms~\footnote{The contributions from the Drude terms to transverse response can be eliminated by a suitable (anti-)symmetrization.}.

Following Ref.~\cite{Matsumoto2011Theoretical}, the anomalous contributions to the currents are given by 
\begin{align}
& \bs j_1 =e \Nabla \times \frac{1}{\hbar} \int_\kk \int^\infty_{\epsilon_\kk} d\epsilon f(\bs k, T, \bs E, \bs \nabla T)  ~\bs{\Omega}_\kk \\
& \bs j_2 = -\Nabla \times \frac{1}{\hbar} \int_\kk \int^\infty_{\epsilon_\kk} d\epsilon (\epsilon-\mu)  f(\bs k, T, \bs E, \bs \nabla T) ~\bs{\Omega}_\kk,
\end{align}
where $\int_\kk\equiv \int_{\kk \in \text{BZ}} d^2k/(2\pi)^2$, the Berry curvature density is defined as $\bs{\Omega}_\kk = -i\langle \Nabla u_\kk |\times| \Nabla u_\kk \rangle$~\footnote{An equivalent approach is to consider the orbital magnetization current up to magnetic quadrupole moment induced by Berry curvature~\cite{Xiao2010RMP,Xiao2005Berry,Xiao2006Berry,Gao2018Orbital,Oji1985,Nakai2019Nonreciprocal}.}, $\epsilon_\kk$ is the dispersion of the partially-full band, $\mu$ is the chemical potential, and we have reinstated the full functional dependencies of the nonequilibrium distribution function for completeness. 
The formalism applies in the weak-field regime, where local equilibrium holds and the Fermi surface is only weakly perturbed.
In order to obtain the $\bs \nabla T$-driven current, we keep the leading order terms in $T$ for the anomalous transport coefficients through the Sommerfeld expansion, since we focus on the weak-field regime~\footnote{These integrals can be expressed in terms of polylogarithm functions~\cite{Zagier2007}.} (see SM~\cite{sm}).
Although general expressions of the transport coefficients can be obtained, their finiteness and inter-relations depend on the symmetries of the system. 

Using this approach, we now turn to the central component of our work, {\it viz.} the relationships between thermal, themoelectric and electrical nonlinear responses that generalize the standard Wiedemann-Franz and Mott relationships. Below we present our results for the two symmetry settings we outlined earlier. The details of the derivation can be found in the SM~\cite{sm}.

\blue{\emph{$\cal T$ symmetric case with Berry curvature dipole} ---} In $\cal T$-symmetric but $\cal{P}$-broken (and, hence, $\cal{P T}$-broken) systems the Berry curvature is odd in $\bs k$, and the intrinsic anomalous response at the linear order vanishes in all channels.
This situation still allows a finite intrinsic anomalous response at higher order in the drive~\cite{deyo2009semiclassical, Fu2015BCD, low2015topological}. 
Due to the presence of the $\cal T$-symmetry, the quantum metric's contributions to the second order terms in $\bs j_l$ vanish.
Thus, all second order transport coefficients are controlled by the Berry curvature dipole.
Given the distinct source of the topological response, we address what new relationships, if any, develop  between the transport coefficients at the second order.
We note that the na\"{i}ve generalization of the Wiedemann-Franz and Mott relation to the second order is generally invalid. 
In particular, there is in general no universal relation in each pair $\{L_{1,11}^{a,b c}, L_{2,22}^{a,b c}\}$ and $\{L_{1,22}^{a,b c}, \partial_\mu L_{1,11}^{a,b c}\}$, as already observed in Refs.~\cite{Zeng2020Fundamental,wang2022quantum}.

At the second order, eight different transport coefficients are possible.
We find $L_{1,21}^{a,b c}=L_{1,12}^{a,c b}$, $L_{2,21}^{a,b c}=L_{2,12}^{a,c b}$, which reduce the total number of apparently independent coefficients to six.
As one of our key results in this paper, we find that these six coefficients are interrelated, leading to a web of relationships among the second-order anomalous transport coefficients which replaces the relationships present at the linear order. 
In particular, \emph{three} Wiedemann-Franz type, 
\begin{align}
L_{2,22}^{a,b c} &= \mathfrak{L}  L_{2,11}^{a,b c}  
\label{WF-BCD1} \\
L_{1, 22}^{a,b c} &= \mathfrak{L} L_{1,11}^{a,b c} \label{WF-BCD2}\\
L_{2, 12}^{a,b c} &= -4 \mathfrak{L} 
 T L_{1, 11}^{a,b c}, \label{WF-BCD3}
\end{align}
and \emph{two} Mott type, 
\begin{align}
L_{1,11}^{a,b c} &= -e (L_{2, 11}^{a,b c})' \label{Mott-BCD1}\\
L_{1, 12}^{a,b c} &= -\frac{e}{2}(L_{2, 12}^{a,b c})', \label{Mott-BCD2}
\end{align}
relationships constrain the six different transport coefficients~\footnote{We note that Eqs.~(\ref{WF-BCD3}) and (\ref{Mott-BCD2}) require $\bs E \times \bs \nabla T = 0$}. 
By using these minimal set of relations, it is possible to relate other unlisted pairs. The entire web is summarized by Fig.~\ref{fig:relations}(c).

\blue{\emph{$\cal PT$ symmetric case with quantum metric dipole}}--- In systems that simultaneously break $\cal T$ and $\cal P$ but preserve $\cal{PT}$ the bands possess Kramers degeneracy at all $\bs k$-points in the Brillouin zone. 
Consequently, the net Berry curvature of each pair of  Kramers degenerate bands mutually cancel, and a simple Berry-curvature-controlled topological response does not exist. 
By contrast, the contribution from the quantum metric remains non-trivial, and it can sustain anomalous charge-Hall effect at nonlinear orders~\cite{QMDnature,QMDscience,kaplan2024unification, das2023intrinsic,oh2024thermoelectric}.

We find that intrinsic nonlinear thermoelectric coefficients due to  contributions from the quantum metric dipole are also non-trivial at the second order~\cite{sm}.
Because the source of the anomalous response in this case is distinct from the last two cases, we expect a different set of relationships among the transport coefficients. 
Indeed, we find \emph{three} Wiedemann-Franz type,
\begin{align}
L^{a,bc}_{2,21} &= \mathfrak{L}T L^{a,bc}_{1,11} \\
L_{2, 11}^{a,b c} &= -T L_{1, 21}^{a,b c}  \\ 
L_{1, 21}^{a,b c} &= -eT L_{1, 22}^{a,b c} 
\end{align}
and \emph{one} Mott type,
\begin{align}
L_{1, 21}^{a,b c} &= e (L_{2, 21}^{a,b c})' 
\end{align}
relation~\footnote{Note that the Drude contributions mix with the contributions from the quantum metric. However, these contributions can be distinguished, as discussed in section B.2 of the SM~\cite{sm}.}.
These relations, originating in the quantum-metric dipole, are highlighted in Fig.~\ref{fig:relations}(d).

\begin{figure}[t]
\centering
\includegraphics[width=\linewidth]{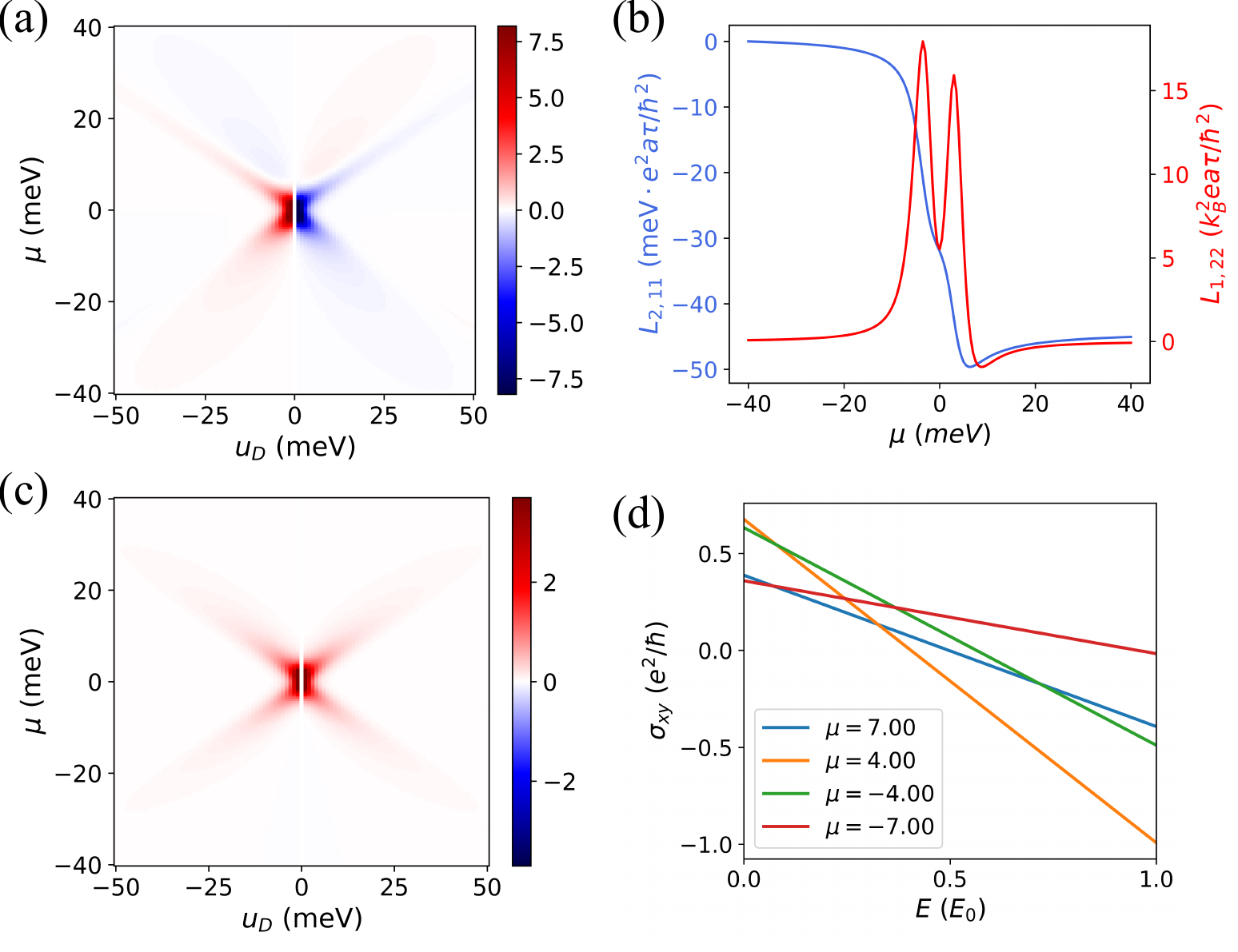}
\caption{
(a) Berry curvature dipole at $\delta=0.5$ in the phase diagram of $(u_D,\mu)$. 
(b) Berry curvature dipole induced nonlinear thermoelectric coefficients, $L_{2,11}^{x,yy}$  and $L_{1,22}^{x,yy}$, at $\delta=0.5$, $u_D=-5\,{\rm meV}$.  
(c) Quantum metric dipole at single valley with $\delta=0$.
(d) Hall conductivity at single valley with $\delta=0$, $u_D=-5\,{\rm meV}$. The intercept (slope) is entirely controlled by the net Berry curvature (quantum metric dipole). 
Here we used unit $E_0=1{\rm mV}/a\approx 4{\rm V/\mu m}$ where $a$ is the lattice constant.
More details on our choice of parameters are summarized in the SM~\cite{sm}.
}
\label{fig:pd}
\end{figure}

\blue{\emph{Application to gate-tunable 2D materials}}--- 
The nonlinear coefficients can be experimentally determined through DC transport measurements.
Here, we highlight the application of our results to gate-tunable 2D materials.
A key advantage of these systems is the ability to control the chemical potential via an external gate voltage, enabling direct tests of Mott-type relations.
In particular, we consider Bernal bilayer graphene, which not only hosts a rich variety of interaction-driven phases~\cite{zhou2022isospin, seiler2022quantum, de2022cascade}, but also have been found to exhibit a large nonlinear Hall conductivity~\cite{chichinadze2024observation}.

The low energy effective Hamiltonian for the electronic states in valley  $\zeta = \pm$ is given by~\cite{sm,Koshino2011Landau},
\begin{align}
& H_\zeta = 
- \frac{v^2}{\gamma_1} \left[ (k_x^2 - k_y^2)\, \sigma_x + 2 \zeta k_x k_y\, \sigma_y \right]
+ v_3 \left( \zeta k_x\, \sigma_x + k_y\, \sigma_y \right) \nonumber \\
&+ \frac{2 v v_4}{\gamma_1} (k_x^2 + k_y^2)\, \sigma_0
+ \frac{u_D}{2} \sigma_z + \delta v_3 \left( \zeta k_x\, \sigma_x - k_y\, \sigma_y \right) 
\end{align}
where $v_i$, $\gamma_i$ are hopping parameters, and $\sigma_j$ are Pauli matrices acting on the sublattice index. 
We note that due to the spin-orbital coupling being negligible, the Hamiltonian is spin-agnostic. 
Here, we have included a displacement field (a strain term), $u_D$ ($\delta$), which breaks the inversion ($C_{3z}$) symmetry.
Over an extended range of $u_D$ and $\delta$ (see Fig.~S6 of SM~\cite{sm}), both the quantum geometric features and the density of states remain peaked near the band edge at $\mu = \pm u_D/2$, as illustrated in Fig.~\ref{fig:pd}(a) and (c) for the $(u_D, \mu)$ plane.
This feature greatly enhances all anomalous nonlinear responses in bilayer graphene, making it ideal to extract quantum metric dipole and to test our generalized relations.

The magnetic layer group of the fully symmetric phase with  valley-degeneracy and  $(u_D, \delta) = (0, 0)$ is $p\bar{3}m11'$~\cite{bradley2010mathematical}. 
For this group, both the Berry curvature dipole, $\int_\kk f \partial_{k_a} \Omega_z$, and the quantum metric dipole, $\int_\kk f \partial_{k_a}G^{b c}$, vanish. 
Moreover, due the presence of the valley and spin degeneracies, the net Berry curvature also vanishes. 

Here, we consider the setting with $\cal T$-symmetry (with the valley degeneracy preserved), but with the $C_{3z}$ (inversion ${\cal P}$) symmetry broken by a nonzero strain $\delta$ (nonzero displacement field $u_D$).
The magnetic layer group is reduced to  $pm11'$.
(Note that this setting is realized through spontaneous symmetry breaking, in the so-called ``Sym$_4$'' phase of the system~\cite{zhou2022isospin, chichinadze2024observation}, without explicitly applying an external strain.)
While the $\cal T$-symmetry prevents the quantum metric dipole from contributing to the thermoelectric response, the Berry curvature dipole sustains a non-trivial Mott relationship, 
\begin{equation}
    L_{1,22}^{a,bc} = -e\mathfrak{L}(L_{2,11}^{a,bc})'
\end{equation}
as shown in Fig.~\ref{fig:pd}(b).
We can utilize the tunability of the gate voltage to portray the Mott-type relationship between the nonlinear Nernst~\cite{Wu2021nernst, Zeng2020Fundamental} and Ettingshausen coefficients, $L_{1, 22}^{x,yy}$ and $L_{2, 11}^{x,yy}$, respectively [cf. Fig.~\ref{fig:relations}(c)]. We reiterate that this Mott relationship is unique to the nonlinear response, and replaces the standard reciprocity relationship $L_{2,1}^{a,b} = T L_{1,2}^{a,b}$ in the case of linear response.

To develop the utility of this model system further, we note that it allows us to illustrate a mechanism to isolate the quantum geometry contribution to intrinsic anomalous transport in the absence of $\cal T$. 
We consider the case with spontaneously broken valley symmetry~\cite{zhou2022isospin}. 
Here, the $C_{3z}$ symmetry is preserved, but the inversion symmetry is broken by the displacement field $u_D\neq0$ (while no strain is applied, $\delta=0$).
This corresponds to the magnetic layer group $p3m1$.
Due to the broken $\cal T$ and $\cal{PT}$ symmetries, an intrinsic anomalous charge-Hall conductivity is present at linear order of the applied fields. 
At second order, however, the $C_{3z}$ symmetry constrains the contributions from the Berry curvature dipole to zero, while allowing the quantum metric dipole to contribute non-trivially.
Thus, the  intrinsic anomalous charge-Hall conductivity due to an electric field applied along $\hat y$, $\sigma_{xy}(E_y) \equiv j^x_1(E_y)/E_y$, takes the form,
\begin{align}
    \sigma_{xy}(E_y) = L_{1,1}^{x, y} +    L_{1, 1 1}^{x, y y} E_y,
\end{align}
where  $L_{1,1}^{x,y}$ ($L_{1, 1 1}^{x, yy}$) is solely controlled by the Berry curvature (quantum metric dipole). 
The intercept and the slope of the low-temperature $\sigma_{xy}$ vs. $E_y$ plot reveal the respective quantum geometric content of the system, as demonstrated in Fig.~\ref{fig:pd}(d).
Thus, the nonlinear charge Hall response in this setting provides a direct probe of the quantum metric dipole $\int_\kk f_0 \partial_{k_x}G^{yy}$.

\blue{\emph{Onsager reciprocity relationship at nonlinear orders}}--- 
By assuming microscopic time-reversal symmetry, Onsager's reciprocity relations play a central role in relating charge and heat transport close to thermodynamic equilibrium~\cite{Onsager1931}. 
Since these relations are intended for  linear responses, they need not continue to apply at nonlinear orders.
Indeed, we find a na\"{i}ve generalization to the second order would imply $L_{1,22}^{a, bc} \propto L_{2, 11}^{a, bc}$, which our explicit calculations show to be invalid in $\mathcal{T}$-symmetric and $\mathcal{P}$-broken systems~\cite{sm}. 
Therefore, the Onsager's reciprocity relation, being a fundamental principle for nonequilibrium thermodynamics and quantum transport, must assume a different form in the nonlinear regime, and requires further investigations~\cite{tokura2018nonreciprocal}.
Reciprocity relations at non-linear orders are also affected by the  interplay between external drive and interactions, which may give rise to new forms of non-reciprocity~\cite{Morimoto2018noncentro}.

\blue{\emph{Discussion and Conclusion}}--- 
Several remarks are in order. First, the generalized Wiedemann-Franz and Mott-type relations derived for the nonlinear thermal and thermoelectric responses vis-\`{a}-vis their electrical analogs are amenable to experimental verification~\cite{sm}.
These relations not only constrain emergent nonlinear transport phenomena in topological systems but also provide guiding principles for optimizing the thermoelectric efficiency of  topological materials exhibiting nonlinear anomalous responses~\cite{Mahan1996best, xu2017topological, he2017advances}. 

Second, various 2D materials offer potential platforms  for experimental verification of our nonlinear relations~\cite{Ghosh2025Thermopower}.
For example, our results on  bilayer graphene suggest that this system near the band bottom supports a large Berry curvature induced thermoelectric response.
Transition-metal dichalcogenides (TMDCs), with time-reversal symmetry and large Berry curvature, are also ideal for testing Berry curvature dipole effects~\cite{zhou2022isospin,You2018dipole}.
Though $C_{3z}$ symmetry forbids such dipoles, applying strain can break this symmetry and induce a Berry curvature dipole response.
Similar considerations apply to $\rm MnTe$ which is a $\cal PT$ symmetric antiferromagnet and one can verify the relations between quantum-metric-dipole induced thermoelectric  coefficients~\cite{QMDnature,QMDscience}.

Third, our results also suggest new measurements that will further elucidate topological materials.
For example, given the large spontaneous Hall response that has been measured in the Weyl-Kondo semimetal Ce$_3$Bi$_4$Pd$_3$~\cite{Dzsaber2021Giant}, one of the Wiedemann-Franz relation we have uncovered, Eq.~(\ref{WF-BCD2}), suggests that $L_{1,22}^{a, bc}$, the nonlinear Nernst response, will elucidate the Weyl-Kondo semimetal physics.
Another set of the relationships, Eqs.~(\ref{WF-BCD1},\ref{WF-BCD3}), suggest how $L_{2,22}^{a, bc}$ and $L_{2,11}^{a, bc}$, the nonlinear thermal Hall and nonlinear Ettingshausen effects, will inform the further understanding of Weyl-Kondo semimetals. Moreover, our results lead us to propose $L_{2,12}^{a, bc}$, the nonlinear response under the joint thermal and electrical drive, as a new means of probing the Weyl-Kondo semimetallicity.

Finally, our work lays the groundwork for further studies of the nonlinear thermal and thermoelectric effects. 
One promising direction concerns the behavior in the fully nonequilibrium regime, in analogy with their nonlinear charge Hall counterparts~\cite{sur2024fully}.
Another is to explore the extent to which these relations persist in the presence of interactions, as they do for the conventional Wiedemann-Franz law, and in emergent states of matter.

\blue{\emph{Acknowledgment}}---
We thank Haoyu Hu, Silke Paschen, and Fang Xie for useful discussions. 
The work has been supported by the NSF under Grant No. DMR-2220603 (conceptualization and kinetic equations analysis), by the AFOSR under Grant No. FA9550-21-1-0356 (model calculations), by the Robert A. Welch Foundation Grant No. C-1411 (Q.S.) and the Vannevar Bush Faculty Fellowship ONR-VB N00014-23-1-2870 (Q.S.).  
Y.X. was supported by the NSF CAREER DMR-2339623 and the Robert A. Welch Foundation Grant No. C-2219.
The majority of the computational calculations have been performed on the Shared University Grid at Rice funded by NSF under Grant EIA-0216467, a partnership between Rice University, Sun Microsystems, and Sigma Solutions, Inc., the Big-Data Private-Cloud Research Cyberinfrastructure MRI-award funded by NSF under Grant No. CNS-1338099, and the Extreme Science and Engineering Discovery Environment (XSEDE) by NSF under Grant No. DMR170109. 

{\it Note added}: After the completion of this manuscript, a recent work became available (X. Yang et al, arXiv:2505.00086) which studied the nonlinear effect for the case of a joint drive by a temperature gradient and an electric field. Where there is overlap, their results are compatible with ours.

\bibliography{reference.bib}

\begin{thebibliography}{74}%
\makeatletter
\providecommand \@ifxundefined [1]{%
 \@ifx{#1\undefined}
}%
\providecommand \@ifnum [1]{%
 \ifnum #1\expandafter \@firstoftwo
 \else \expandafter \@secondoftwo
 \fi
}%
\providecommand \@ifx [1]{%
 \ifx #1\expandafter \@firstoftwo
 \else \expandafter \@secondoftwo
 \fi
}%
\providecommand \natexlab [1]{#1}%
\providecommand \enquote  [1]{``#1''}%
\providecommand \bibnamefont  [1]{#1}%
\providecommand \bibfnamefont [1]{#1}%
\providecommand \citenamefont [1]{#1}%
\providecommand \href@noop [0]{\@secondoftwo}%
\providecommand \href [0]{\begingroup \@sanitize@url \@href}%
\providecommand \@href[1]{\@@startlink{#1}\@@href}%
\providecommand \@@href[1]{\endgroup#1\@@endlink}%
\providecommand \@sanitize@url [0]{\catcode `\\12\catcode `\$12\catcode
  `\&12\catcode `\#12\catcode `\^12\catcode `\_12\catcode `\%12\relax}%
\providecommand \@@startlink[1]{}%
\providecommand \@@endlink[0]{}%
\providecommand \url  [0]{\begingroup\@sanitize@url \@url }%
\providecommand \@url [1]{\endgroup\@href {#1}{\urlprefix }}%
\providecommand \urlprefix  [0]{URL }%
\providecommand \Eprint [0]{\href }%
\providecommand \doibase [0]{https://doi.org/}%
\providecommand \selectlanguage [0]{\@gobble}%
\providecommand \bibinfo  [0]{\@secondoftwo}%
\providecommand \bibfield  [0]{\@secondoftwo}%
\providecommand \translation [1]{[#1]}%
\providecommand \BibitemOpen [0]{}%
\providecommand \bibitemStop [0]{}%
\providecommand \bibitemNoStop [0]{.\EOS\space}%
\providecommand \EOS [0]{\spacefactor3000\relax}%
\providecommand \BibitemShut  [1]{\csname bibitem#1\endcsname}%
\let\auto@bib@innerbib\@empty
\bibitem [{\citenamefont {Provost}\ and\ \citenamefont
  {Vallee}(1980)}]{provost1980riemannian}%
  \BibitemOpen
  \bibfield  {author} {\bibinfo {author} {\bibfnamefont {J.~P.}\ \bibnamefont
  {Provost}}\ and\ \bibinfo {author} {\bibfnamefont {G.}~\bibnamefont
  {Vallee}},\ }\bibfield  {title} {\bibinfo {title} {Riemannian structure on
  manifolds of quantum states},\ }\href {https://doi.org/10.1007/BF02193559}
  {\bibfield  {journal} {\bibinfo  {journal} {Communications in Mathematical
  Physics}\ }\textbf {\bibinfo {volume} {76}},\ \bibinfo {pages} {289}
  (\bibinfo {year} {1980})}\BibitemShut {NoStop}%
\bibitem [{\citenamefont {Marzari}\ and\ \citenamefont
  {Vanderbilt}(1997)}]{Mar97.1}%
  \BibitemOpen
  \bibfield  {author} {\bibinfo {author} {\bibfnamefont {N.}~\bibnamefont
  {Marzari}}\ and\ \bibinfo {author} {\bibfnamefont {D.}~\bibnamefont
  {Vanderbilt}},\ }\bibfield  {title} {\bibinfo {title} {Maximally localized
  generalized wannier functions for composite energy bands},\ }\href
  {https://doi.org/10.1103/PhysRevB.56.12847} {\bibfield  {journal} {\bibinfo
  {journal} {Phys. Rev. B}\ }\textbf {\bibinfo {volume} {56}},\ \bibinfo
  {pages} {12847} (\bibinfo {year} {1997})}\BibitemShut {NoStop}%
\bibitem [{\citenamefont {Peotta}\ and\ \citenamefont
  {T{\"o}rm{\"a}}(2015)}]{Peotta2015}%
  \BibitemOpen
  \bibfield  {author} {\bibinfo {author} {\bibfnamefont {S.}~\bibnamefont
  {Peotta}}\ and\ \bibinfo {author} {\bibfnamefont {P.}~\bibnamefont
  {T{\"o}rm{\"a}}},\ }\bibfield  {title} {\bibinfo {title} {Superfluidity in
  topologically nontrivial flat bands},\ }\href
  {https://doi.org/10.1038/ncomms9944} {\bibfield  {journal} {\bibinfo
  {journal} {Nature Communications}\ }\textbf {\bibinfo {volume} {6}},\
  \bibinfo {pages} {8944} (\bibinfo {year} {2015})}\BibitemShut {NoStop}%
\bibitem [{\citenamefont {{Verma}}\ \emph {et~al.}(2025)\citenamefont
  {{Verma}}, \citenamefont {{Moll}}, \citenamefont {{Holder}},\ and\
  \citenamefont {{Queiroz}}}]{verma2025quantum}%
  \BibitemOpen
  \bibfield  {author} {\bibinfo {author} {\bibfnamefont {N.}~\bibnamefont
  {{Verma}}}, \bibinfo {author} {\bibfnamefont {P.~J.~W.}\ \bibnamefont
  {{Moll}}}, \bibinfo {author} {\bibfnamefont {T.}~\bibnamefont {{Holder}}},\
  and\ \bibinfo {author} {\bibfnamefont {R.}~\bibnamefont {{Queiroz}}},\
  }\bibfield  {title} {\bibinfo {title} {{Quantum Geometry: Revisiting
  electronic scales in quantum matter}},\ }\href
  {https://doi.org/10.48550/arXiv.2504.07173} {\bibfield  {journal} {\bibinfo
  {journal} {arXiv e-prints}\ ,\ \bibinfo {eid} {arXiv:2504.07173}} (\bibinfo
  {year} {2025})},\ \Eprint {https://arxiv.org/abs/2504.07173}
  {arXiv:2504.07173 [cond-mat.mtrl-sci]} \BibitemShut {NoStop}%
\bibitem [{\citenamefont {{Deyo}}\ \emph {et~al.}(2009)\citenamefont {{Deyo}},
  \citenamefont {{Golub}}, \citenamefont {{Ivchenko}},\ and\ \citenamefont
  {{Spivak}}}]{deyo2009semiclassical}%
  \BibitemOpen
  \bibfield  {author} {\bibinfo {author} {\bibfnamefont {E.}~\bibnamefont
  {{Deyo}}}, \bibinfo {author} {\bibfnamefont {L.~E.}\ \bibnamefont {{Golub}}},
  \bibinfo {author} {\bibfnamefont {E.~L.}\ \bibnamefont {{Ivchenko}}},\ and\
  \bibinfo {author} {\bibfnamefont {B.}~\bibnamefont {{Spivak}}},\ }\bibfield
  {title} {\bibinfo {title} {{Semiclassical theory of the photogalvanic effect
  in non-centrosymmetric systems}},\ }\href
  {https://doi.org/10.48550/arXiv.0904.1917} {\bibfield  {journal} {\bibinfo
  {journal} {arXiv e-prints}\ ,\ \bibinfo {eid} {arXiv:0904.1917}} (\bibinfo
  {year} {2009})},\ \Eprint {https://arxiv.org/abs/0904.1917} {arXiv:0904.1917
  [cond-mat.mes-hall]} \BibitemShut {NoStop}%
\bibitem [{\citenamefont {Sodemann}\ and\ \citenamefont
  {Fu}(2015)}]{Fu2015BCD}%
  \BibitemOpen
  \bibfield  {author} {\bibinfo {author} {\bibfnamefont {I.}~\bibnamefont
  {Sodemann}}\ and\ \bibinfo {author} {\bibfnamefont {L.}~\bibnamefont {Fu}},\
  }\bibfield  {title} {\bibinfo {title} {Quantum nonlinear hall effect induced
  by berry curvature dipole in time-reversal invariant materials},\ }\href
  {https://doi.org/10.1103/PhysRevLett.115.216806} {\bibfield  {journal}
  {\bibinfo  {journal} {Phys. Rev. Lett.}\ }\textbf {\bibinfo {volume} {115}},\
  \bibinfo {pages} {216806} (\bibinfo {year} {2015})}\BibitemShut {NoStop}%
\bibitem [{\citenamefont {Low}\ \emph {et~al.}(2015)\citenamefont {Low},
  \citenamefont {Jiang},\ and\ \citenamefont {Guinea}}]{low2015topological}%
  \BibitemOpen
  \bibfield  {author} {\bibinfo {author} {\bibfnamefont {T.}~\bibnamefont
  {Low}}, \bibinfo {author} {\bibfnamefont {Y.}~\bibnamefont {Jiang}},\ and\
  \bibinfo {author} {\bibfnamefont {F.}~\bibnamefont {Guinea}},\ }\bibfield
  {title} {\bibinfo {title} {Topological currents in black phosphorus with
  broken inversion symmetry},\ }\href
  {https://doi.org/10.1103/PhysRevB.92.235447} {\bibfield  {journal} {\bibinfo
  {journal} {Phys. Rev. B}\ }\textbf {\bibinfo {volume} {92}},\ \bibinfo
  {pages} {235447} (\bibinfo {year} {2015})}\BibitemShut {NoStop}%
\bibitem [{\citenamefont {Ma}\ \emph {et~al.}(2019)\citenamefont {Ma},
  \citenamefont {Xu}, \citenamefont {Shen}, \citenamefont {MacNeill},
  \citenamefont {Fatemi}, \citenamefont {Chang}, \citenamefont {Mier~Valdivia},
  \citenamefont {Wu}, \citenamefont {Du}, \citenamefont {Hsu}, \citenamefont
  {Fang}, \citenamefont {Gibson}, \citenamefont {Watanabe}, \citenamefont
  {Taniguchi}, \citenamefont {Cava}, \citenamefont {Kaxiras}, \citenamefont
  {Lu}, \citenamefont {Lin}, \citenamefont {Fu}, \citenamefont {Gedik},\ and\
  \citenamefont {Jarillo-Herrero}}]{ma2019observation}%
  \BibitemOpen
  \bibfield  {author} {\bibinfo {author} {\bibfnamefont {Q.}~\bibnamefont
  {Ma}}, \bibinfo {author} {\bibfnamefont {S.-Y.}\ \bibnamefont {Xu}}, \bibinfo
  {author} {\bibfnamefont {H.}~\bibnamefont {Shen}}, \bibinfo {author}
  {\bibfnamefont {D.}~\bibnamefont {MacNeill}}, \bibinfo {author}
  {\bibfnamefont {V.}~\bibnamefont {Fatemi}}, \bibinfo {author} {\bibfnamefont
  {T.-R.}\ \bibnamefont {Chang}}, \bibinfo {author} {\bibfnamefont {A.~M.}\
  \bibnamefont {Mier~Valdivia}}, \bibinfo {author} {\bibfnamefont
  {S.}~\bibnamefont {Wu}}, \bibinfo {author} {\bibfnamefont {Z.}~\bibnamefont
  {Du}}, \bibinfo {author} {\bibfnamefont {C.-H.}\ \bibnamefont {Hsu}},
  \bibinfo {author} {\bibfnamefont {S.}~\bibnamefont {Fang}}, \bibinfo {author}
  {\bibfnamefont {Q.~D.}\ \bibnamefont {Gibson}}, \bibinfo {author}
  {\bibfnamefont {K.}~\bibnamefont {Watanabe}}, \bibinfo {author}
  {\bibfnamefont {T.}~\bibnamefont {Taniguchi}}, \bibinfo {author}
  {\bibfnamefont {R.~J.}\ \bibnamefont {Cava}}, \bibinfo {author}
  {\bibfnamefont {E.}~\bibnamefont {Kaxiras}}, \bibinfo {author} {\bibfnamefont
  {H.-Z.}\ \bibnamefont {Lu}}, \bibinfo {author} {\bibfnamefont
  {H.}~\bibnamefont {Lin}}, \bibinfo {author} {\bibfnamefont {L.}~\bibnamefont
  {Fu}}, \bibinfo {author} {\bibfnamefont {N.}~\bibnamefont {Gedik}},\ and\
  \bibinfo {author} {\bibfnamefont {P.}~\bibnamefont {Jarillo-Herrero}},\
  }\bibfield  {title} {\bibinfo {title} {Observation of the nonlinear hall
  effect under time-reversal-symmetric conditions},\ }\href
  {https://doi.org/10.1038/s41586-018-0807-6} {\bibfield  {journal} {\bibinfo
  {journal} {Nature}\ }\textbf {\bibinfo {volume} {565}},\ \bibinfo {pages}
  {337} (\bibinfo {year} {2019})}\BibitemShut {NoStop}%
\bibitem [{\citenamefont {Kang}\ \emph {et~al.}(2019)\citenamefont {Kang},
  \citenamefont {Li}, \citenamefont {Sohn}, \citenamefont {Shan},\ and\
  \citenamefont {Mak}}]{Kang2019nonlinear}%
  \BibitemOpen
  \bibfield  {author} {\bibinfo {author} {\bibfnamefont {K.}~\bibnamefont
  {Kang}}, \bibinfo {author} {\bibfnamefont {T.}~\bibnamefont {Li}}, \bibinfo
  {author} {\bibfnamefont {E.}~\bibnamefont {Sohn}}, \bibinfo {author}
  {\bibfnamefont {J.}~\bibnamefont {Shan}},\ and\ \bibinfo {author}
  {\bibfnamefont {K.~F.}\ \bibnamefont {Mak}},\ }\bibfield  {title} {\bibinfo
  {title} {Nonlinear anomalous hall effect in few-layer wte2},\ }\href
  {https://doi.org/10.1038/s41563-019-0294-7} {\bibfield  {journal} {\bibinfo
  {journal} {Nature Materials}\ }\textbf {\bibinfo {volume} {18}},\ \bibinfo
  {pages} {324} (\bibinfo {year} {2019})}\BibitemShut {NoStop}%
\bibitem [{\citenamefont {Kumar}\ \emph {et~al.}(2021)\citenamefont {Kumar},
  \citenamefont {Hsu}, \citenamefont {Sharma}, \citenamefont {Chang},
  \citenamefont {Yu}, \citenamefont {Wang}, \citenamefont {Eda}, \citenamefont
  {Liang},\ and\ \citenamefont {Yang}}]{Kumar2021NLHE}%
  \BibitemOpen
  \bibfield  {author} {\bibinfo {author} {\bibfnamefont {D.}~\bibnamefont
  {Kumar}}, \bibinfo {author} {\bibfnamefont {C.-H.}\ \bibnamefont {Hsu}},
  \bibinfo {author} {\bibfnamefont {R.}~\bibnamefont {Sharma}}, \bibinfo
  {author} {\bibfnamefont {T.-R.}\ \bibnamefont {Chang}}, \bibinfo {author}
  {\bibfnamefont {P.}~\bibnamefont {Yu}}, \bibinfo {author} {\bibfnamefont
  {J.}~\bibnamefont {Wang}}, \bibinfo {author} {\bibfnamefont {G.}~\bibnamefont
  {Eda}}, \bibinfo {author} {\bibfnamefont {G.}~\bibnamefont {Liang}},\ and\
  \bibinfo {author} {\bibfnamefont {H.}~\bibnamefont {Yang}},\ }\bibfield
  {title} {\bibinfo {title} {Room-temperature nonlinear hall effect and
  wireless radiofrequency rectification in weyl semimetal tairte4},\ }\href
  {https://doi.org/10.1038/s41565-020-00839-3} {\bibfield  {journal} {\bibinfo
  {journal} {Nature Nanotechnology}\ }\textbf {\bibinfo {volume} {16}},\
  \bibinfo {pages} {421} (\bibinfo {year} {2021})}\BibitemShut {NoStop}%
\bibitem [{\citenamefont {{Kaplan}}\ \emph {et~al.}(2025)\citenamefont
  {{Kaplan}}, \citenamefont {{Lucht}}, \citenamefont {{Volkov}},\ and\
  \citenamefont {{Pixley}}}]{kaplan2025}%
  \BibitemOpen
  \bibfield  {author} {\bibinfo {author} {\bibfnamefont {D.}~\bibnamefont
  {{Kaplan}}}, \bibinfo {author} {\bibfnamefont {K.~P.}\ \bibnamefont
  {{Lucht}}}, \bibinfo {author} {\bibfnamefont {P.~A.}\ \bibnamefont
  {{Volkov}}},\ and\ \bibinfo {author} {\bibfnamefont {J.~H.}\ \bibnamefont
  {{Pixley}}},\ }\bibfield  {title} {\bibinfo {title} {{Quantum geometric
  photocurrents of quasiparticles in superconductors}},\ }\href
  {https://doi.org/10.48550/arXiv.2502.12265} {\bibfield  {journal} {\bibinfo
  {journal} {arXiv e-prints}\ ,\ \bibinfo {eid} {arXiv:2502.12265}} (\bibinfo
  {year} {2025})},\ \Eprint {https://arxiv.org/abs/2502.12265}
  {arXiv:2502.12265 [cond-mat.supr-con]} \BibitemShut {NoStop}%
\bibitem [{\citenamefont {Lai}\ \emph {et~al.}(2018)\citenamefont {Lai},
  \citenamefont {Grefe}, \citenamefont {Paschen},\ and\ \citenamefont
  {Si}}]{lai2018weyl}%
  \BibitemOpen
  \bibfield  {author} {\bibinfo {author} {\bibfnamefont {H.-H.}\ \bibnamefont
  {Lai}}, \bibinfo {author} {\bibfnamefont {S.~E.}\ \bibnamefont {Grefe}},
  \bibinfo {author} {\bibfnamefont {S.}~\bibnamefont {Paschen}},\ and\ \bibinfo
  {author} {\bibfnamefont {Q.}~\bibnamefont {Si}},\ }\bibfield  {title}
  {\bibinfo {title} {Weyl–kondo semimetal in heavy-fermion systems},\ }\href
  {https://doi.org/10.1073/pnas.1715851115} {\bibfield  {journal} {\bibinfo
  {journal} {PNAS}\ }\textbf {\bibinfo {volume} {115}},\ \bibinfo {pages} {93}
  (\bibinfo {year} {2018})}\BibitemShut {NoStop}%
\bibitem [{\citenamefont {Grefe}\ \emph {et~al.}(2020)\citenamefont {Grefe},
  \citenamefont {Lai}, \citenamefont {Paschen},\ and\ \citenamefont
  {Si}}]{grefe2020weyl}%
  \BibitemOpen
  \bibfield  {author} {\bibinfo {author} {\bibfnamefont {S.~E.}\ \bibnamefont
  {Grefe}}, \bibinfo {author} {\bibfnamefont {H.-H.}\ \bibnamefont {Lai}},
  \bibinfo {author} {\bibfnamefont {S.}~\bibnamefont {Paschen}},\ and\ \bibinfo
  {author} {\bibfnamefont {Q.}~\bibnamefont {Si}},\ }\bibfield  {title}
  {\bibinfo {title} {Weyl-kondo semimetals in nonsymmorphic systems},\ }\href
  {https://doi.org/10.1103/PhysRevB.101.075138} {\bibfield  {journal} {\bibinfo
   {journal} {Phys. Rev. B}\ }\textbf {\bibinfo {volume} {101}},\ \bibinfo
  {pages} {075138} (\bibinfo {year} {2020})}\BibitemShut {NoStop}%
\bibitem [{\citenamefont {Dzsaber}\ \emph {et~al.}(2017)\citenamefont
  {Dzsaber}, \citenamefont {Prochaska}, \citenamefont {Sidorenko},
  \citenamefont {Eguchi}, \citenamefont {Svagera}, \citenamefont {Waas},
  \citenamefont {Prokofiev}, \citenamefont {Si},\ and\ \citenamefont
  {Paschen}}]{dzsaber2017kondo}%
  \BibitemOpen
  \bibfield  {author} {\bibinfo {author} {\bibfnamefont {S.}~\bibnamefont
  {Dzsaber}}, \bibinfo {author} {\bibfnamefont {L.}~\bibnamefont {Prochaska}},
  \bibinfo {author} {\bibfnamefont {A.}~\bibnamefont {Sidorenko}}, \bibinfo
  {author} {\bibfnamefont {G.}~\bibnamefont {Eguchi}}, \bibinfo {author}
  {\bibfnamefont {R.}~\bibnamefont {Svagera}}, \bibinfo {author} {\bibfnamefont
  {M.}~\bibnamefont {Waas}}, \bibinfo {author} {\bibfnamefont {A.}~\bibnamefont
  {Prokofiev}}, \bibinfo {author} {\bibfnamefont {Q.}~\bibnamefont {Si}},\ and\
  \bibinfo {author} {\bibfnamefont {S.}~\bibnamefont {Paschen}},\ }\bibfield
  {title} {\bibinfo {title} {Kondo insulator to semimetal transformation tuned
  by spin-orbit coupling},\ }\href
  {https://doi.org/10.1103/PhysRevLett.118.246601} {\bibfield  {journal}
  {\bibinfo  {journal} {Phys. Rev. Lett.}\ }\textbf {\bibinfo {volume} {118}},\
  \bibinfo {pages} {246601} (\bibinfo {year} {2017})}\BibitemShut {NoStop}%
\bibitem [{\citenamefont {{Sur}}\ \emph {et~al.}(2024)\citenamefont {{Sur}},
  \citenamefont {{Chen}}, \citenamefont {{Wang}}, \citenamefont {{Setty}},
  \citenamefont {{Paschen}},\ and\ \citenamefont {{Si}}}]{sur2024fully}%
  \BibitemOpen
  \bibfield  {author} {\bibinfo {author} {\bibfnamefont {S.}~\bibnamefont
  {{Sur}}}, \bibinfo {author} {\bibfnamefont {L.}~\bibnamefont {{Chen}}},
  \bibinfo {author} {\bibfnamefont {Y.}~\bibnamefont {{Wang}}}, \bibinfo
  {author} {\bibfnamefont {C.}~\bibnamefont {{Setty}}}, \bibinfo {author}
  {\bibfnamefont {S.}~\bibnamefont {{Paschen}}},\ and\ \bibinfo {author}
  {\bibfnamefont {Q.}~\bibnamefont {{Si}}},\ }\bibfield  {title} {\bibinfo
  {title} {{Fully nonequilibrium Hall response from Berry curvature}},\ }\href
  {https://doi.org/10.48550/arXiv.2411.16675} {\bibfield  {journal} {\bibinfo
  {journal} {arXiv e-prints}\ ,\ \bibinfo {eid} {arXiv:2411.16675}} (\bibinfo
  {year} {2024})},\ \Eprint {https://arxiv.org/abs/2411.16675}
  {arXiv:2411.16675 [cond-mat.mes-hall]} \BibitemShut {NoStop}%
\bibitem [{\citenamefont {Dzsaber}\ \emph {et~al.}(2021)\citenamefont
  {Dzsaber}, \citenamefont {Yan}, \citenamefont {Taupin}, \citenamefont
  {Eguchi}, \citenamefont {Prokofiev}, \citenamefont {Shiroka}, \citenamefont
  {Blaha}, \citenamefont {Rubel}, \citenamefont {Grefe}, \citenamefont {Lai},
  \citenamefont {Si},\ and\ \citenamefont {Paschen}}]{Dzsaber2021Giant}%
  \BibitemOpen
  \bibfield  {author} {\bibinfo {author} {\bibfnamefont {S.}~\bibnamefont
  {Dzsaber}}, \bibinfo {author} {\bibfnamefont {X.}~\bibnamefont {Yan}},
  \bibinfo {author} {\bibfnamefont {M.}~\bibnamefont {Taupin}}, \bibinfo
  {author} {\bibfnamefont {G.}~\bibnamefont {Eguchi}}, \bibinfo {author}
  {\bibfnamefont {A.}~\bibnamefont {Prokofiev}}, \bibinfo {author}
  {\bibfnamefont {T.}~\bibnamefont {Shiroka}}, \bibinfo {author} {\bibfnamefont
  {P.}~\bibnamefont {Blaha}}, \bibinfo {author} {\bibfnamefont
  {O.}~\bibnamefont {Rubel}}, \bibinfo {author} {\bibfnamefont {S.~E.}\
  \bibnamefont {Grefe}}, \bibinfo {author} {\bibfnamefont {H.-H.}\ \bibnamefont
  {Lai}}, \bibinfo {author} {\bibfnamefont {Q.}~\bibnamefont {Si}},\ and\
  \bibinfo {author} {\bibfnamefont {S.}~\bibnamefont {Paschen}},\ }\bibfield
  {title} {\bibinfo {title} {Giant spontaneous hall effect in a nonmagnetic
  weyl–kondo semimetal},\ }\href {https://doi.org/10.1073/pnas.2013386118}
  {\bibfield  {journal} {\bibinfo  {journal} {Proceedings of the National
  Academy of Sciences}\ }\textbf {\bibinfo {volume} {118}},\ \bibinfo {pages}
  {e2013386118} (\bibinfo {year} {2021})}\BibitemShut {NoStop}%
\bibitem [{\citenamefont {Gao}\ \emph {et~al.}(2014)\citenamefont {Gao},
  \citenamefont {Yang},\ and\ \citenamefont {Niu}}]{Gao2014Field}%
  \BibitemOpen
  \bibfield  {author} {\bibinfo {author} {\bibfnamefont {Y.}~\bibnamefont
  {Gao}}, \bibinfo {author} {\bibfnamefont {S.~A.}\ \bibnamefont {Yang}},\ and\
  \bibinfo {author} {\bibfnamefont {Q.}~\bibnamefont {Niu}},\ }\bibfield
  {title} {\bibinfo {title} {Field induced positional shift of bloch electrons
  and its dynamical implications},\ }\href
  {https://doi.org/10.1103/PhysRevLett.112.166601} {\bibfield  {journal}
  {\bibinfo  {journal} {Phys. Rev. Lett.}\ }\textbf {\bibinfo {volume} {112}},\
  \bibinfo {pages} {166601} (\bibinfo {year} {2014})}\BibitemShut {NoStop}%
\bibitem [{\citenamefont {Wang}\ \emph {et~al.}(2021)\citenamefont {Wang},
  \citenamefont {Gao},\ and\ \citenamefont {Xiao}}]{Wang2021Intrinsic}%
  \BibitemOpen
  \bibfield  {author} {\bibinfo {author} {\bibfnamefont {C.}~\bibnamefont
  {Wang}}, \bibinfo {author} {\bibfnamefont {Y.}~\bibnamefont {Gao}},\ and\
  \bibinfo {author} {\bibfnamefont {D.}~\bibnamefont {Xiao}},\ }\bibfield
  {title} {\bibinfo {title} {Intrinsic nonlinear hall effect in
  antiferromagnetic tetragonal cumnas},\ }\href
  {https://doi.org/10.1103/PhysRevLett.127.277201} {\bibfield  {journal}
  {\bibinfo  {journal} {Phys. Rev. Lett.}\ }\textbf {\bibinfo {volume} {127}},\
  \bibinfo {pages} {277201} (\bibinfo {year} {2021})}\BibitemShut {NoStop}%
\bibitem [{\citenamefont {Liu}\ \emph {et~al.}(2021{\natexlab{a}})\citenamefont
  {Liu}, \citenamefont {Zhao}, \citenamefont {Huang}, \citenamefont {Wu},
  \citenamefont {Sheng}, \citenamefont {Xiao},\ and\ \citenamefont
  {Yang}}]{Liu2021Intrinsic}%
  \BibitemOpen
  \bibfield  {author} {\bibinfo {author} {\bibfnamefont {H.}~\bibnamefont
  {Liu}}, \bibinfo {author} {\bibfnamefont {J.}~\bibnamefont {Zhao}}, \bibinfo
  {author} {\bibfnamefont {Y.-X.}\ \bibnamefont {Huang}}, \bibinfo {author}
  {\bibfnamefont {W.}~\bibnamefont {Wu}}, \bibinfo {author} {\bibfnamefont
  {X.-L.}\ \bibnamefont {Sheng}}, \bibinfo {author} {\bibfnamefont
  {C.}~\bibnamefont {Xiao}},\ and\ \bibinfo {author} {\bibfnamefont {S.~A.}\
  \bibnamefont {Yang}},\ }\bibfield  {title} {\bibinfo {title} {Intrinsic
  second-order anomalous hall effect and its application in compensated
  antiferromagnets},\ }\href {https://doi.org/10.1103/PhysRevLett.127.277202}
  {\bibfield  {journal} {\bibinfo  {journal} {Phys. Rev. Lett.}\ }\textbf
  {\bibinfo {volume} {127}},\ \bibinfo {pages} {277202} (\bibinfo {year}
  {2021}{\natexlab{a}})}\BibitemShut {NoStop}%
\bibitem [{\citenamefont {Liu}\ \emph {et~al.}(2021{\natexlab{b}})\citenamefont
  {Liu}, \citenamefont {Zhao}, \citenamefont {Huang}, \citenamefont {Wu},
  \citenamefont {Sheng}, \citenamefont {Xiao},\ and\ \citenamefont
  {Yang}}]{PhysRevLett.127.277202}%
  \BibitemOpen
  \bibfield  {author} {\bibinfo {author} {\bibfnamefont {H.}~\bibnamefont
  {Liu}}, \bibinfo {author} {\bibfnamefont {J.}~\bibnamefont {Zhao}}, \bibinfo
  {author} {\bibfnamefont {Y.-X.}\ \bibnamefont {Huang}}, \bibinfo {author}
  {\bibfnamefont {W.}~\bibnamefont {Wu}}, \bibinfo {author} {\bibfnamefont
  {X.-L.}\ \bibnamefont {Sheng}}, \bibinfo {author} {\bibfnamefont
  {C.}~\bibnamefont {Xiao}},\ and\ \bibinfo {author} {\bibfnamefont {S.~A.}\
  \bibnamefont {Yang}},\ }\bibfield  {title} {\bibinfo {title} {Intrinsic
  second-order anomalous hall effect and its application in compensated
  antiferromagnets},\ }\href {https://doi.org/10.1103/PhysRevLett.127.277202}
  {\bibfield  {journal} {\bibinfo  {journal} {Phys. Rev. Lett.}\ }\textbf
  {\bibinfo {volume} {127}},\ \bibinfo {pages} {277202} (\bibinfo {year}
  {2021}{\natexlab{b}})}\BibitemShut {NoStop}%
\bibitem [{\citenamefont {Kaplan}\ \emph {et~al.}(2024)\citenamefont {Kaplan},
  \citenamefont {Holder},\ and\ \citenamefont {Yan}}]{kaplan2024unification}%
  \BibitemOpen
  \bibfield  {author} {\bibinfo {author} {\bibfnamefont {D.}~\bibnamefont
  {Kaplan}}, \bibinfo {author} {\bibfnamefont {T.}~\bibnamefont {Holder}},\
  and\ \bibinfo {author} {\bibfnamefont {B.}~\bibnamefont {Yan}},\ }\bibfield
  {title} {\bibinfo {title} {Unification of nonlinear anomalous hall effect and
  nonreciprocal magnetoresistance in metals by the quantum geometry},\ }\href
  {https://doi.org/10.1103/PhysRevLett.132.026301} {\bibfield  {journal}
  {\bibinfo  {journal} {Phys. Rev. Lett.}\ }\textbf {\bibinfo {volume} {132}},\
  \bibinfo {pages} {026301} (\bibinfo {year} {2024})}\BibitemShut {NoStop}%
\bibitem [{\citenamefont {Murakami}\ \emph {et~al.}(2003)\citenamefont
  {Murakami}, \citenamefont {Nagaosa},\ and\ \citenamefont
  {Zhang}}]{murakami2003dissipationless}%
  \BibitemOpen
  \bibfield  {author} {\bibinfo {author} {\bibfnamefont {S.}~\bibnamefont
  {Murakami}}, \bibinfo {author} {\bibfnamefont {N.}~\bibnamefont {Nagaosa}},\
  and\ \bibinfo {author} {\bibfnamefont {S.-C.}\ \bibnamefont {Zhang}},\
  }\bibfield  {title} {\bibinfo {title} {Dissipationless quantum spin current
  at room temperature},\ }\href {https://doi.org/10.1126/science.1087128}
  {\bibfield  {journal} {\bibinfo  {journal} {Science}\ }\textbf {\bibinfo
  {volume} {301}},\ \bibinfo {pages} {1348} (\bibinfo {year}
  {2003})}\BibitemShut {NoStop}%
\bibitem [{\citenamefont {Wang}\ \emph {et~al.}(2023)\citenamefont {Wang},
  \citenamefont {Kaplan}, \citenamefont {Zhang}, \citenamefont {Holder},
  \citenamefont {Cao}, \citenamefont {Wang}, \citenamefont {Zhou},
  \citenamefont {Zhou}, \citenamefont {Jiang}, \citenamefont {Zhang} \emph
  {et~al.}}]{QMDnature}%
  \BibitemOpen
  \bibfield  {author} {\bibinfo {author} {\bibfnamefont {N.}~\bibnamefont
  {Wang}}, \bibinfo {author} {\bibfnamefont {D.}~\bibnamefont {Kaplan}},
  \bibinfo {author} {\bibfnamefont {Z.}~\bibnamefont {Zhang}}, \bibinfo
  {author} {\bibfnamefont {T.}~\bibnamefont {Holder}}, \bibinfo {author}
  {\bibfnamefont {N.}~\bibnamefont {Cao}}, \bibinfo {author} {\bibfnamefont
  {A.}~\bibnamefont {Wang}}, \bibinfo {author} {\bibfnamefont {X.}~\bibnamefont
  {Zhou}}, \bibinfo {author} {\bibfnamefont {F.}~\bibnamefont {Zhou}}, \bibinfo
  {author} {\bibfnamefont {Z.}~\bibnamefont {Jiang}}, \bibinfo {author}
  {\bibfnamefont {C.}~\bibnamefont {Zhang}}, \emph {et~al.},\ }\bibfield
  {title} {\bibinfo {title} {Quantum metric-induced nonlinear transport in a
  topological antiferromagnet},\ }\href
  {https://doi.org/10.1038/s41586-023-06363-3} {\bibfield  {journal} {\bibinfo
  {journal} {Nature}\ }\textbf {\bibinfo {volume} {621}},\ \bibinfo {pages}
  {487} (\bibinfo {year} {2023})}\BibitemShut {NoStop}%
\bibitem [{\citenamefont {Gao}\ \emph {et~al.}(2023)\citenamefont {Gao},
  \citenamefont {Liu}, \citenamefont {Qiu}, \citenamefont {Ghosh},
  \citenamefont {V.~Trevisan}, \citenamefont {Onishi}, \citenamefont {Hu},
  \citenamefont {Qian}, \citenamefont {Tien}, \citenamefont {Chen} \emph
  {et~al.}}]{QMDscience}%
  \BibitemOpen
  \bibfield  {author} {\bibinfo {author} {\bibfnamefont {A.}~\bibnamefont
  {Gao}}, \bibinfo {author} {\bibfnamefont {Y.-F.}\ \bibnamefont {Liu}},
  \bibinfo {author} {\bibfnamefont {J.-X.}\ \bibnamefont {Qiu}}, \bibinfo
  {author} {\bibfnamefont {B.}~\bibnamefont {Ghosh}}, \bibinfo {author}
  {\bibfnamefont {T.}~\bibnamefont {V.~Trevisan}}, \bibinfo {author}
  {\bibfnamefont {Y.}~\bibnamefont {Onishi}}, \bibinfo {author} {\bibfnamefont
  {C.}~\bibnamefont {Hu}}, \bibinfo {author} {\bibfnamefont {T.}~\bibnamefont
  {Qian}}, \bibinfo {author} {\bibfnamefont {H.-J.}\ \bibnamefont {Tien}},
  \bibinfo {author} {\bibfnamefont {S.-W.}\ \bibnamefont {Chen}}, \emph
  {et~al.},\ }\bibfield  {title} {\bibinfo {title} {Quantum metric nonlinear
  hall effect in a topological antiferromagnetic heterostructure},\ }\href
  {https://doi.org/10.1126/science.adf1506} {\bibfield  {journal} {\bibinfo
  {journal} {Science}\ ,\ \bibinfo {pages} {eadf1506}} (\bibinfo {year}
  {2023})}\BibitemShut {NoStop}%
\bibitem [{\citenamefont {Neupert}\ \emph {et~al.}(2013)\citenamefont
  {Neupert}, \citenamefont {Chamon},\ and\ \citenamefont
  {Mudry}}]{Neupert2013Measuring}%
  \BibitemOpen
  \bibfield  {author} {\bibinfo {author} {\bibfnamefont {T.}~\bibnamefont
  {Neupert}}, \bibinfo {author} {\bibfnamefont {C.}~\bibnamefont {Chamon}},\
  and\ \bibinfo {author} {\bibfnamefont {C.}~\bibnamefont {Mudry}},\ }\bibfield
   {title} {\bibinfo {title} {Measuring the quantum geometry of bloch bands
  with current noise},\ }\href {https://doi.org/10.1103/PhysRevB.87.245103}
  {\bibfield  {journal} {\bibinfo  {journal} {Phys. Rev. B}\ }\textbf {\bibinfo
  {volume} {87}},\ \bibinfo {pages} {245103} (\bibinfo {year}
  {2013})}\BibitemShut {NoStop}%
\bibitem [{\citenamefont {Kang}\ \emph {et~al.}(2025)\citenamefont {Kang},
  \citenamefont {Kim}, \citenamefont {Qian}, \citenamefont {Neves},
  \citenamefont {Ye}, \citenamefont {Jung}, \citenamefont {Puntel},
  \citenamefont {Mazzola}, \citenamefont {Fang}, \citenamefont {Jozwiak},
  \citenamefont {Bostwick}, \citenamefont {Rotenberg}, \citenamefont {Fuji},
  \citenamefont {Vobornik}, \citenamefont {Park}, \citenamefont {Checkelsky},
  \citenamefont {Yang},\ and\ \citenamefont {Comin}}]{Kang2025}%
  \BibitemOpen
  \bibfield  {author} {\bibinfo {author} {\bibfnamefont {M.}~\bibnamefont
  {Kang}}, \bibinfo {author} {\bibfnamefont {S.}~\bibnamefont {Kim}}, \bibinfo
  {author} {\bibfnamefont {Y.}~\bibnamefont {Qian}}, \bibinfo {author}
  {\bibfnamefont {P.~M.}\ \bibnamefont {Neves}}, \bibinfo {author}
  {\bibfnamefont {L.}~\bibnamefont {Ye}}, \bibinfo {author} {\bibfnamefont
  {J.}~\bibnamefont {Jung}}, \bibinfo {author} {\bibfnamefont {D.}~\bibnamefont
  {Puntel}}, \bibinfo {author} {\bibfnamefont {F.}~\bibnamefont {Mazzola}},
  \bibinfo {author} {\bibfnamefont {S.}~\bibnamefont {Fang}}, \bibinfo {author}
  {\bibfnamefont {C.}~\bibnamefont {Jozwiak}}, \bibinfo {author} {\bibfnamefont
  {A.}~\bibnamefont {Bostwick}}, \bibinfo {author} {\bibfnamefont
  {E.}~\bibnamefont {Rotenberg}}, \bibinfo {author} {\bibfnamefont
  {J.}~\bibnamefont {Fuji}}, \bibinfo {author} {\bibfnamefont {I.}~\bibnamefont
  {Vobornik}}, \bibinfo {author} {\bibfnamefont {J.-H.}\ \bibnamefont {Park}},
  \bibinfo {author} {\bibfnamefont {J.~G.}\ \bibnamefont {Checkelsky}},
  \bibinfo {author} {\bibfnamefont {B.-J.}\ \bibnamefont {Yang}},\ and\
  \bibinfo {author} {\bibfnamefont {R.}~\bibnamefont {Comin}},\ }\bibfield
  {title} {\bibinfo {title} {Measurements of the quantum geometric tensor in
  solids},\ }\href {https://doi.org/10.1038/s41567-024-02678-8} {\bibfield
  {journal} {\bibinfo  {journal} {Nature Physics}\ }\textbf {\bibinfo {volume}
  {21}},\ \bibinfo {pages} {110} (\bibinfo {year} {2025})}\BibitemShut
  {NoStop}%
\bibitem [{\citenamefont {Zeng}\ \emph {et~al.}(2020)\citenamefont {Zeng},
  \citenamefont {Nandy},\ and\ \citenamefont {Tewari}}]{Zeng2020Fundamental}%
  \BibitemOpen
  \bibfield  {author} {\bibinfo {author} {\bibfnamefont {C.}~\bibnamefont
  {Zeng}}, \bibinfo {author} {\bibfnamefont {S.}~\bibnamefont {Nandy}},\ and\
  \bibinfo {author} {\bibfnamefont {S.}~\bibnamefont {Tewari}},\ }\bibfield
  {title} {\bibinfo {title} {Fundamental relations for anomalous thermoelectric
  transport coefficients in the nonlinear regime},\ }\href
  {https://doi.org/10.1103/PhysRevResearch.2.032066} {\bibfield  {journal}
  {\bibinfo  {journal} {Phys. Rev. Res.}\ }\textbf {\bibinfo {volume} {2}},\
  \bibinfo {pages} {032066} (\bibinfo {year} {2020})}\BibitemShut {NoStop}%
\bibitem [{\citenamefont {Franz}\ and\ \citenamefont
  {Wiedemann}(1853)}]{franz1853ueber}%
  \BibitemOpen
  \bibfield  {author} {\bibinfo {author} {\bibfnamefont {R.}~\bibnamefont
  {Franz}}\ and\ \bibinfo {author} {\bibfnamefont {G.}~\bibnamefont
  {Wiedemann}},\ }\bibfield  {title} {\bibinfo {title} {Ueber die
  w{\"a}rme-leitungsf{\"a}higkeit der metalle},\ }\href@noop {} {\bibfield
  {journal} {\bibinfo  {journal} {Annalen der Physik}\ }\textbf {\bibinfo
  {volume} {165}},\ \bibinfo {pages} {497} (\bibinfo {year}
  {1853})}\BibitemShut {NoStop}%
\bibitem [{\citenamefont {Jonson}\ and\ \citenamefont
  {Mahan}(1980)}]{jonson1980mott}%
  \BibitemOpen
  \bibfield  {author} {\bibinfo {author} {\bibfnamefont {M.}~\bibnamefont
  {Jonson}}\ and\ \bibinfo {author} {\bibfnamefont {G.~D.}\ \bibnamefont
  {Mahan}},\ }\bibfield  {title} {\bibinfo {title} {Mott's formula for the
  thermopower and the wiedemann-franz law},\ }\href
  {https://doi.org/10.1103/PhysRevB.21.4223} {\bibfield  {journal} {\bibinfo
  {journal} {Phys. Rev. B}\ }\textbf {\bibinfo {volume} {21}},\ \bibinfo
  {pages} {4223} (\bibinfo {year} {1980})}\BibitemShut {NoStop}%
\bibitem [{\citenamefont {Cutler}\ and\ \citenamefont {Mott}(1969)}]{Mott1969}%
  \BibitemOpen
  \bibfield  {author} {\bibinfo {author} {\bibfnamefont {M.}~\bibnamefont
  {Cutler}}\ and\ \bibinfo {author} {\bibfnamefont {N.~F.}\ \bibnamefont
  {Mott}},\ }\bibfield  {title} {\bibinfo {title} {Observation of anderson
  localization in an electron gas},\ }\href
  {https://doi.org/10.1103/PhysRev.181.1336} {\bibfield  {journal} {\bibinfo
  {journal} {Phys. Rev.}\ }\textbf {\bibinfo {volume} {181}},\ \bibinfo {pages}
  {1336} (\bibinfo {year} {1969})}\BibitemShut {NoStop}%
\bibitem [{\citenamefont {Smrcka}\ and\ \citenamefont
  {Streda}(1977)}]{smrcka1977transport}%
  \BibitemOpen
  \bibfield  {author} {\bibinfo {author} {\bibfnamefont {L.}~\bibnamefont
  {Smrcka}}\ and\ \bibinfo {author} {\bibfnamefont {P.}~\bibnamefont
  {Streda}},\ }\bibfield  {title} {\bibinfo {title} {Transport coefficients in
  strong magnetic fields},\ }\href
  {https://doi.org/10.1088/0022-3719/10/12/021} {\bibfield  {journal} {\bibinfo
   {journal} {Journal of Physics C: Solid State Physics}\ }\textbf {\bibinfo
  {volume} {10}},\ \bibinfo {pages} {2153} (\bibinfo {year}
  {1977})}\BibitemShut {NoStop}%
\bibitem [{\citenamefont {Onsager}(1931)}]{Onsager1931}%
  \BibitemOpen
  \bibfield  {author} {\bibinfo {author} {\bibfnamefont {L.}~\bibnamefont
  {Onsager}},\ }\bibfield  {title} {\bibinfo {title} {Reciprocal relations in
  irreversible processes. i.},\ }\href {https://doi.org/10.1103/PhysRev.37.405}
  {\bibfield  {journal} {\bibinfo  {journal} {Phys. Rev.}\ }\textbf {\bibinfo
  {volume} {37}},\ \bibinfo {pages} {405} (\bibinfo {year} {1931})}\BibitemShut
  {NoStop}%
\bibitem [{\citenamefont {Callen}(1948)}]{callen1948application}%
  \BibitemOpen
  \bibfield  {author} {\bibinfo {author} {\bibfnamefont {H.~B.}\ \bibnamefont
  {Callen}},\ }\bibfield  {title} {\bibinfo {title} {The application of
  onsager's reciprocal relations to thermoelectric, thermomagnetic, and
  galvanomagnetic effects},\ }\href {https://doi.org/10.1103/PhysRev.73.1349}
  {\bibfield  {journal} {\bibinfo  {journal} {Phys. Rev.}\ }\textbf {\bibinfo
  {volume} {73}},\ \bibinfo {pages} {1349} (\bibinfo {year}
  {1948})}\BibitemShut {NoStop}%
\bibitem [{\citenamefont {Ashcroft}\ and\ \citenamefont
  {Mermin}(1976)}]{ashcroft1976solid}%
  \BibitemOpen
  \bibfield  {author} {\bibinfo {author} {\bibfnamefont {N.~W.}\ \bibnamefont
  {Ashcroft}}\ and\ \bibinfo {author} {\bibfnamefont {N.~D.}\ \bibnamefont
  {Mermin}},\ }\href@noop {} {\emph {\bibinfo {title} {{S}olid {S}tate
  {P}hysics}}}\ (\bibinfo  {publisher} {Holt-Saunders},\ \bibinfo {year}
  {1976})\BibitemShut {NoStop}%
\bibitem [{\citenamefont {Girvin}\ and\ \citenamefont
  {Yang}(2019)}]{Girvin_Yang_2019}%
  \BibitemOpen
  \bibfield  {author} {\bibinfo {author} {\bibfnamefont {S.~M.}\ \bibnamefont
  {Girvin}}\ and\ \bibinfo {author} {\bibfnamefont {K.}~\bibnamefont {Yang}},\
  }\href@noop {} {\emph {\bibinfo {title} {Modern Condensed Matter Physics}}}\
  (\bibinfo  {publisher} {Cambridge University Press},\ \bibinfo {year}
  {2019})\BibitemShut {NoStop}%
\bibitem [{\citenamefont {Mahan}(2013)}]{mahan2013many}%
  \BibitemOpen
  \bibfield  {author} {\bibinfo {author} {\bibfnamefont {G.~D.}\ \bibnamefont
  {Mahan}},\ }\href@noop {} {\emph {\bibinfo {title} {Many-particle physics}}}\
  (\bibinfo  {publisher} {Springer Science \& Business Media},\ \bibinfo {year}
  {2013})\BibitemShut {NoStop}%
\bibitem [{\citenamefont {Pfau}\ \emph {et~al.}(2012)\citenamefont {Pfau},
  \citenamefont {Hartmann}, \citenamefont {Stockert}, \citenamefont {Sun},
  \citenamefont {Lausberg}, \citenamefont {Brando}, \citenamefont {Friedemann},
  \citenamefont {Krellner}, \citenamefont {Geibel}, \citenamefont {Wirth},
  \citenamefont {Kirchner}, \citenamefont {Abrahams}, \citenamefont {Si},\ and\
  \citenamefont {Steglich}}]{Pfa12.1}%
  \BibitemOpen
  \bibfield  {author} {\bibinfo {author} {\bibfnamefont {H.}~\bibnamefont
  {Pfau}}, \bibinfo {author} {\bibfnamefont {S.}~\bibnamefont {Hartmann}},
  \bibinfo {author} {\bibfnamefont {U.}~\bibnamefont {Stockert}}, \bibinfo
  {author} {\bibfnamefont {P.}~\bibnamefont {Sun}}, \bibinfo {author}
  {\bibfnamefont {S.}~\bibnamefont {Lausberg}}, \bibinfo {author}
  {\bibfnamefont {M.}~\bibnamefont {Brando}}, \bibinfo {author} {\bibfnamefont
  {S.}~\bibnamefont {Friedemann}}, \bibinfo {author} {\bibfnamefont
  {C.}~\bibnamefont {Krellner}}, \bibinfo {author} {\bibfnamefont
  {C.}~\bibnamefont {Geibel}}, \bibinfo {author} {\bibfnamefont
  {S.}~\bibnamefont {Wirth}}, \bibinfo {author} {\bibfnamefont
  {S.}~\bibnamefont {Kirchner}}, \bibinfo {author} {\bibfnamefont
  {E.}~\bibnamefont {Abrahams}}, \bibinfo {author} {\bibfnamefont
  {Q.}~\bibnamefont {Si}},\ and\ \bibinfo {author} {\bibfnamefont
  {F.}~\bibnamefont {Steglich}},\ }\bibfield  {title} {\bibinfo {title}
  {Thermal and electrical transport across a magnetic quantum critical point},\
  }\href {https://doi.org/10.1038/nature11072} {\bibfield  {journal} {\bibinfo
  {journal} {Nature}\ }\textbf {\bibinfo {volume} {484}},\ \bibinfo {pages}
  {493} (\bibinfo {year} {2012})}\BibitemShut {NoStop}%
\bibitem [{\citenamefont {Ghawri}\ \emph {et~al.}(2022)\citenamefont {Ghawri},
  \citenamefont {Mahapatra}, \citenamefont {Garg}, \citenamefont {Mandal},
  \citenamefont {Bhowmik}, \citenamefont {Jayaraman}, \citenamefont {Soni},
  \citenamefont {Watanabe}, \citenamefont {Taniguchi}, \citenamefont
  {Krishnamurthy}, \citenamefont {Jain}, \citenamefont {Banerjee},
  \citenamefont {Chandni},\ and\ \citenamefont {Ghosh}}]{ghawri2022breakdown}%
  \BibitemOpen
  \bibfield  {author} {\bibinfo {author} {\bibfnamefont {B.}~\bibnamefont
  {Ghawri}}, \bibinfo {author} {\bibfnamefont {P.~S.}\ \bibnamefont
  {Mahapatra}}, \bibinfo {author} {\bibfnamefont {M.}~\bibnamefont {Garg}},
  \bibinfo {author} {\bibfnamefont {S.}~\bibnamefont {Mandal}}, \bibinfo
  {author} {\bibfnamefont {S.}~\bibnamefont {Bhowmik}}, \bibinfo {author}
  {\bibfnamefont {A.}~\bibnamefont {Jayaraman}}, \bibinfo {author}
  {\bibfnamefont {R.}~\bibnamefont {Soni}}, \bibinfo {author} {\bibfnamefont
  {K.}~\bibnamefont {Watanabe}}, \bibinfo {author} {\bibfnamefont
  {T.}~\bibnamefont {Taniguchi}}, \bibinfo {author} {\bibfnamefont {H.~R.}\
  \bibnamefont {Krishnamurthy}}, \bibinfo {author} {\bibfnamefont
  {M.}~\bibnamefont {Jain}}, \bibinfo {author} {\bibfnamefont {S.}~\bibnamefont
  {Banerjee}}, \bibinfo {author} {\bibfnamefont {U.}~\bibnamefont {Chandni}},\
  and\ \bibinfo {author} {\bibfnamefont {A.}~\bibnamefont {Ghosh}},\ }\bibfield
   {title} {\bibinfo {title} {Breakdown of semiclassical description of
  thermoelectricity in near-magic angle twisted bilayer graphene},\ }\href
  {https://doi.org/10.1038/s41467-022-29198-4} {\bibfield  {journal} {\bibinfo
  {journal} {Nature Communications}\ }\textbf {\bibinfo {volume} {13}},\
  \bibinfo {pages} {1522} (\bibinfo {year} {2022})}\BibitemShut {NoStop}%
\bibitem [{\citenamefont {Xiao}\ \emph {et~al.}(2006)\citenamefont {Xiao},
  \citenamefont {Yao}, \citenamefont {Fang},\ and\ \citenamefont
  {Niu}}]{Xiao2006Berry}%
  \BibitemOpen
  \bibfield  {author} {\bibinfo {author} {\bibfnamefont {D.}~\bibnamefont
  {Xiao}}, \bibinfo {author} {\bibfnamefont {Y.}~\bibnamefont {Yao}}, \bibinfo
  {author} {\bibfnamefont {Z.}~\bibnamefont {Fang}},\ and\ \bibinfo {author}
  {\bibfnamefont {Q.}~\bibnamefont {Niu}},\ }\bibfield  {title} {\bibinfo
  {title} {Berry-phase effect in anomalous thermoelectric transport},\ }\href
  {https://doi.org/10.1103/PhysRevLett.97.026603} {\bibfield  {journal}
  {\bibinfo  {journal} {Phys. Rev. Lett.}\ }\textbf {\bibinfo {volume} {97}},\
  \bibinfo {pages} {026603} (\bibinfo {year} {2006})}\BibitemShut {NoStop}%
\bibitem [{\citenamefont {Onoda}\ \emph {et~al.}(2008)\citenamefont {Onoda},
  \citenamefont {Sugimoto},\ and\ \citenamefont {Nagaosa}}]{onoda2008quantum}%
  \BibitemOpen
  \bibfield  {author} {\bibinfo {author} {\bibfnamefont {S.}~\bibnamefont
  {Onoda}}, \bibinfo {author} {\bibfnamefont {N.}~\bibnamefont {Sugimoto}},\
  and\ \bibinfo {author} {\bibfnamefont {N.}~\bibnamefont {Nagaosa}},\
  }\bibfield  {title} {\bibinfo {title} {Quantum transport theory of anomalous
  electric, thermoelectric, and thermal hall effects in ferromagnets},\ }\href
  {https://doi.org/10.1103/PhysRevB.77.165103} {\bibfield  {journal} {\bibinfo
  {journal} {Phys. Rev. B}\ }\textbf {\bibinfo {volume} {77}},\ \bibinfo
  {pages} {165103} (\bibinfo {year} {2008})}\BibitemShut {NoStop}%
\bibitem [{\citenamefont {Pu}\ \emph {et~al.}(2008)\citenamefont {Pu},
  \citenamefont {Chiba}, \citenamefont {Matsukura}, \citenamefont {Ohno},\ and\
  \citenamefont {Shi}}]{pu2008mott}%
  \BibitemOpen
  \bibfield  {author} {\bibinfo {author} {\bibfnamefont {Y.}~\bibnamefont
  {Pu}}, \bibinfo {author} {\bibfnamefont {D.}~\bibnamefont {Chiba}}, \bibinfo
  {author} {\bibfnamefont {F.}~\bibnamefont {Matsukura}}, \bibinfo {author}
  {\bibfnamefont {H.}~\bibnamefont {Ohno}},\ and\ \bibinfo {author}
  {\bibfnamefont {J.}~\bibnamefont {Shi}},\ }\bibfield  {title} {\bibinfo
  {title} {Mott relation for anomalous hall and nernst effects in
  ${\mathrm{ga}}_{1\ensuremath{-}x}{\mathrm{mn}}_{x}\mathrm{As}$ ferromagnetic
  semiconductors},\ }\href {https://doi.org/10.1103/PhysRevLett.101.117208}
  {\bibfield  {journal} {\bibinfo  {journal} {Phys. Rev. Lett.}\ }\textbf
  {\bibinfo {volume} {101}},\ \bibinfo {pages} {117208} (\bibinfo {year}
  {2008})}\BibitemShut {NoStop}%
\bibitem [{sm()}]{sm}%
  \BibitemOpen
  \href@noop {} {\bibinfo  {journal} {See `Supplementary Materials' for (A)
  additional details on the semiclassical formalism for nonlinear transport;
  (B) additional details of the computation of response coefficients; (C)
  computational details that inform our results on Bernal stacked bilayer
  graphene; (D) potential experimental protocols.}\ }\BibitemShut {NoStop}%
\bibitem [{\citenamefont {Fang}\ \emph {et~al.}(2024)\citenamefont {Fang},
  \citenamefont {Cano},\ and\ \citenamefont {Ghorashi}}]{Fang2024Quantum}%
  \BibitemOpen
\bibfield  {journal} {  }\bibfield  {author} {\bibinfo {author} {\bibfnamefont
  {Y.}~\bibnamefont {Fang}}, \bibinfo {author} {\bibfnamefont {J.}~\bibnamefont
  {Cano}},\ and\ \bibinfo {author} {\bibfnamefont {S.~A.~A.}\ \bibnamefont
  {Ghorashi}},\ }\bibfield  {title} {\bibinfo {title} {Quantum geometry induced
  nonlinear transport in altermagnets},\ }\href
  {https://doi.org/10.1103/PhysRevLett.133.106701} {\bibfield  {journal}
  {\bibinfo  {journal} {Phys. Rev. Lett.}\ }\textbf {\bibinfo {volume} {133}},\
  \bibinfo {pages} {106701} (\bibinfo {year} {2024})}\BibitemShut {NoStop}%
\bibitem [{\citenamefont {{Wen}}\ \emph {et~al.}(2025)\citenamefont {{Wen}},
  \citenamefont {{Xie}}, \citenamefont {{Auerbach}},\ and\ \citenamefont
  {{Uchoa}}}]{Wen2025Thermal}%
  \BibitemOpen
  \bibfield  {author} {\bibinfo {author} {\bibfnamefont {K.}~\bibnamefont
  {{Wen}}}, \bibinfo {author} {\bibfnamefont {H.-Y.}\ \bibnamefont {{Xie}}},
  \bibinfo {author} {\bibfnamefont {A.}~\bibnamefont {{Auerbach}}},\ and\
  \bibinfo {author} {\bibfnamefont {B.}~\bibnamefont {{Uchoa}}},\ }\bibfield
  {title} {\bibinfo {title} {{Thermal and thermoelectric transport in flat
  bands with non-trivial quantum geometry}},\ }\href
  {https://doi.org/10.48550/arXiv.2502.10504} {\bibfield  {journal} {\bibinfo
  {journal} {arXiv e-prints}\ ,\ \bibinfo {eid} {arXiv:2502.10504}} (\bibinfo
  {year} {2025})},\ \Eprint {https://arxiv.org/abs/2502.10504}
  {arXiv:2502.10504 [cond-mat.mes-hall]} \BibitemShut {NoStop}%
\bibitem [{\citenamefont {Luttinger}(1964)}]{Luttinger1964}%
  \BibitemOpen
  \bibfield  {author} {\bibinfo {author} {\bibfnamefont {J.~M.}\ \bibnamefont
  {Luttinger}},\ }\bibfield  {title} {\bibinfo {title} {Theory of thermal
  transport coefficients},\ }\href {https://doi.org/10.1103/PhysRev.135.A1505}
  {\bibfield  {journal} {\bibinfo  {journal} {Phys. Rev.}\ }\textbf {\bibinfo
  {volume} {135}},\ \bibinfo {pages} {A1505} (\bibinfo {year}
  {1964})}\BibitemShut {NoStop}%
\bibitem [{Note1()}]{Note1}%
  \BibitemOpen
  \bibinfo {note} {The contributions from the Drude terms to transverse
  response can be eliminated by a suitable (anti-)symmetrization.}\BibitemShut
  {Stop}%
\bibitem [{\citenamefont {Matsumoto}\ and\ \citenamefont
  {Murakami}(2011)}]{Matsumoto2011Theoretical}%
  \BibitemOpen
  \bibfield  {author} {\bibinfo {author} {\bibfnamefont {R.}~\bibnamefont
  {Matsumoto}}\ and\ \bibinfo {author} {\bibfnamefont {S.}~\bibnamefont
  {Murakami}},\ }\bibfield  {title} {\bibinfo {title} {Theoretical prediction
  of a rotating magnon wave packet in ferromagnets},\ }\href
  {https://doi.org/10.1103/PhysRevLett.106.197202} {\bibfield  {journal}
  {\bibinfo  {journal} {Phys. Rev. Lett.}\ }\textbf {\bibinfo {volume} {106}},\
  \bibinfo {pages} {197202} (\bibinfo {year} {2011})}\BibitemShut {NoStop}%
\bibitem [{Note2()}]{Note2}%
  \BibitemOpen
  \bibinfo {note} {An equivalent approach is to consider the orbital
  magnetization current up to magnetic quadrupole moment induced by Berry
  curvature~\cite
  {Xiao2010RMP,Xiao2005Berry,Xiao2006Berry,Gao2018Orbital,Oji1985,Nakai2019Nonreciprocal}.}\BibitemShut
  {Stop}%
\bibitem [{Note3()}]{Note3}%
  \BibitemOpen
  \bibinfo {note} {These integrals can be expressed in terms of polylogarithm
  functions~\cite {Zagier2007}.}\BibitemShut {Stop}%
\bibitem [{\citenamefont {Wang}\ \emph {et~al.}(2022)\citenamefont {Wang},
  \citenamefont {Zhu},\ and\ \citenamefont {Su}}]{wang2022quantum}%
  \BibitemOpen
  \bibfield  {author} {\bibinfo {author} {\bibfnamefont {Y.}~\bibnamefont
  {Wang}}, \bibinfo {author} {\bibfnamefont {Z.-G.}\ \bibnamefont {Zhu}},\ and\
  \bibinfo {author} {\bibfnamefont {G.}~\bibnamefont {Su}},\ }\bibfield
  {title} {\bibinfo {title} {Quantum theory of nonlinear thermal response},\
  }\href {https://doi.org/10.1103/PhysRevB.106.035148} {\bibfield  {journal}
  {\bibinfo  {journal} {Phys. Rev. B}\ }\textbf {\bibinfo {volume} {106}},\
  \bibinfo {pages} {035148} (\bibinfo {year} {2022})}\BibitemShut {NoStop}%
\bibitem [{Note4()}]{Note4}%
  \BibitemOpen
  \bibinfo {note} {We note that Eqs.~(\ref {WF-BCD3}) and (\ref {Mott-BCD2})
  require $\protect \bm {E} \times \protect \bm {\nabla }T = 0$}\BibitemShut
  {NoStop}%
\bibitem [{\citenamefont {Das}\ \emph {et~al.}(2023)\citenamefont {Das},
  \citenamefont {Lahiri}, \citenamefont {Atencia}, \citenamefont {Culcer},\
  and\ \citenamefont {Agarwal}}]{das2023intrinsic}%
  \BibitemOpen
  \bibfield  {author} {\bibinfo {author} {\bibfnamefont {K.}~\bibnamefont
  {Das}}, \bibinfo {author} {\bibfnamefont {S.}~\bibnamefont {Lahiri}},
  \bibinfo {author} {\bibfnamefont {R.~B.}\ \bibnamefont {Atencia}}, \bibinfo
  {author} {\bibfnamefont {D.}~\bibnamefont {Culcer}},\ and\ \bibinfo {author}
  {\bibfnamefont {A.}~\bibnamefont {Agarwal}},\ }\bibfield  {title} {\bibinfo
  {title} {Intrinsic nonlinear conductivities induced by the quantum metric},\
  }\href {https://doi.org/10.1103/PhysRevB.108.L201405} {\bibfield  {journal}
  {\bibinfo  {journal} {Phys. Rev. B}\ }\textbf {\bibinfo {volume} {108}},\
  \bibinfo {pages} {L201405} (\bibinfo {year} {2023})}\BibitemShut {NoStop}%
\bibitem [{\citenamefont {Oh}\ \emph {et~al.}(2024)\citenamefont {Oh},
  \citenamefont {Kim},\ and\ \citenamefont {Rhim}}]{oh2024thermoelectric}%
  \BibitemOpen
  \bibfield  {author} {\bibinfo {author} {\bibfnamefont {C.-g.}\ \bibnamefont
  {Oh}}, \bibinfo {author} {\bibfnamefont {K.~W.}\ \bibnamefont {Kim}},\ and\
  \bibinfo {author} {\bibfnamefont {J.-W.}\ \bibnamefont {Rhim}},\ }\bibfield
  {title} {\bibinfo {title} {Thermoelectric transport driven by the
  hilbert–schmidt distance},\ }\href
  {https://doi.org/https://doi.org/10.1002/advs.202411313} {\bibfield
  {journal} {\bibinfo  {journal} {Advanced Science}\ }\textbf {\bibinfo
  {volume} {11}},\ \bibinfo {pages} {2411313} (\bibinfo {year}
  {2024})}\BibitemShut {NoStop}%
\bibitem [{Note5()}]{Note5}%
  \BibitemOpen
  \bibinfo {note} {Note that the Drude contributions mix with the contributions
  from the quantum metric. However, these contributions can be distinguished,
  as discussed in section B.2 of the SM~\cite {sm}.}\BibitemShut {Stop}%
\bibitem [{\citenamefont {Zhou}\ \emph {et~al.}(2022)\citenamefont {Zhou},
  \citenamefont {Holleis}, \citenamefont {Saito}, \citenamefont {Cohen},
  \citenamefont {Huynh}, \citenamefont {Patterson}, \citenamefont {Yang},
  \citenamefont {Taniguchi}, \citenamefont {Watanabe},\ and\ \citenamefont
  {Young}}]{zhou2022isospin}%
  \BibitemOpen
  \bibfield  {author} {\bibinfo {author} {\bibfnamefont {H.}~\bibnamefont
  {Zhou}}, \bibinfo {author} {\bibfnamefont {L.}~\bibnamefont {Holleis}},
  \bibinfo {author} {\bibfnamefont {Y.}~\bibnamefont {Saito}}, \bibinfo
  {author} {\bibfnamefont {L.}~\bibnamefont {Cohen}}, \bibinfo {author}
  {\bibfnamefont {W.}~\bibnamefont {Huynh}}, \bibinfo {author} {\bibfnamefont
  {C.~L.}\ \bibnamefont {Patterson}}, \bibinfo {author} {\bibfnamefont
  {F.}~\bibnamefont {Yang}}, \bibinfo {author} {\bibfnamefont {T.}~\bibnamefont
  {Taniguchi}}, \bibinfo {author} {\bibfnamefont {K.}~\bibnamefont
  {Watanabe}},\ and\ \bibinfo {author} {\bibfnamefont {A.~F.}\ \bibnamefont
  {Young}},\ }\bibfield  {title} {\bibinfo {title} {Isospin magnetism and
  spin-polarized superconductivity in bernal bilayer graphene},\ }\href
  {https://doi.org/10.1126/science.abm8386} {\bibfield  {journal} {\bibinfo
  {journal} {Science}\ }\textbf {\bibinfo {volume} {375}},\ \bibinfo {pages}
  {774} (\bibinfo {year} {2022})}\BibitemShut {NoStop}%
\bibitem [{\citenamefont {Seiler}\ \emph {et~al.}(2022)\citenamefont {Seiler},
  \citenamefont {Geisenhof}, \citenamefont {Winterer}, \citenamefont
  {Watanabe}, \citenamefont {Taniguchi}, \citenamefont {Xu}, \citenamefont
  {Zhang},\ and\ \citenamefont {Weitz}}]{seiler2022quantum}%
  \BibitemOpen
  \bibfield  {author} {\bibinfo {author} {\bibfnamefont {A.~M.}\ \bibnamefont
  {Seiler}}, \bibinfo {author} {\bibfnamefont {F.~R.}\ \bibnamefont
  {Geisenhof}}, \bibinfo {author} {\bibfnamefont {F.}~\bibnamefont {Winterer}},
  \bibinfo {author} {\bibfnamefont {K.}~\bibnamefont {Watanabe}}, \bibinfo
  {author} {\bibfnamefont {T.}~\bibnamefont {Taniguchi}}, \bibinfo {author}
  {\bibfnamefont {T.}~\bibnamefont {Xu}}, \bibinfo {author} {\bibfnamefont
  {F.}~\bibnamefont {Zhang}},\ and\ \bibinfo {author} {\bibfnamefont {R.~T.}\
  \bibnamefont {Weitz}},\ }\bibfield  {title} {\bibinfo {title} {Quantum
  cascade of correlated phases in trigonally warped bilayer graphene},\ }\href
  {https://doi.org/10.1038/s41586-022-04937-1} {\bibfield  {journal} {\bibinfo
  {journal} {Nature}\ }\textbf {\bibinfo {volume} {608}},\ \bibinfo {pages}
  {298} (\bibinfo {year} {2022})}\BibitemShut {NoStop}%
\bibitem [{\citenamefont {de~la Barrera}\ \emph {et~al.}(2022)\citenamefont
  {de~la Barrera}, \citenamefont {Aronson}, \citenamefont {Zheng},
  \citenamefont {Watanabe}, \citenamefont {Taniguchi}, \citenamefont {Ma},
  \citenamefont {Jarillo-Herrero},\ and\ \citenamefont
  {Ashoori}}]{de2022cascade}%
  \BibitemOpen
  \bibfield  {author} {\bibinfo {author} {\bibfnamefont {S.~C.}\ \bibnamefont
  {de~la Barrera}}, \bibinfo {author} {\bibfnamefont {S.}~\bibnamefont
  {Aronson}}, \bibinfo {author} {\bibfnamefont {Z.}~\bibnamefont {Zheng}},
  \bibinfo {author} {\bibfnamefont {K.}~\bibnamefont {Watanabe}}, \bibinfo
  {author} {\bibfnamefont {T.}~\bibnamefont {Taniguchi}}, \bibinfo {author}
  {\bibfnamefont {Q.}~\bibnamefont {Ma}}, \bibinfo {author} {\bibfnamefont
  {P.}~\bibnamefont {Jarillo-Herrero}},\ and\ \bibinfo {author} {\bibfnamefont
  {R.}~\bibnamefont {Ashoori}},\ }\bibfield  {title} {\bibinfo {title} {Cascade
  of isospin phase transitions in bernal-stacked bilayer graphene at zero
  magnetic field},\ }\href {https://doi.org/10.1038/s41567-022-01616-w}
  {\bibfield  {journal} {\bibinfo  {journal} {Nature Physics}\ }\textbf
  {\bibinfo {volume} {18}},\ \bibinfo {pages} {771} (\bibinfo {year}
  {2022})}\BibitemShut {NoStop}%
\bibitem [{\citenamefont {{Chichinadze}}\ \emph {et~al.}(2024)\citenamefont
  {{Chichinadze}}, \citenamefont {{Zhang}}, \citenamefont {{Lin}},
  \citenamefont {{Wang}}, \citenamefont {{Watanabe}}, \citenamefont
  {{Taniguchi}}, \citenamefont {{Vafek}},\ and\ \citenamefont
  {{Li}}}]{chichinadze2024observation}%
  \BibitemOpen
  \bibfield  {author} {\bibinfo {author} {\bibfnamefont {D.~V.}\ \bibnamefont
  {{Chichinadze}}}, \bibinfo {author} {\bibfnamefont {N.~J.}\ \bibnamefont
  {{Zhang}}}, \bibinfo {author} {\bibfnamefont {J.-X.}\ \bibnamefont {{Lin}}},
  \bibinfo {author} {\bibfnamefont {X.}~\bibnamefont {{Wang}}}, \bibinfo
  {author} {\bibfnamefont {K.}~\bibnamefont {{Watanabe}}}, \bibinfo {author}
  {\bibfnamefont {T.}~\bibnamefont {{Taniguchi}}}, \bibinfo {author}
  {\bibfnamefont {O.}~\bibnamefont {{Vafek}}},\ and\ \bibinfo {author}
  {\bibfnamefont {J.~I.~A.}\ \bibnamefont {{Li}}},\ }\bibfield  {title}
  {\bibinfo {title} {{Observation of giant nonlinear Hall conductivity in
  Bernal bilayer graphene}},\ }\href
  {https://doi.org/10.48550/arXiv.2411.11156} {\bibfield  {journal} {\bibinfo
  {journal} {arXiv e-prints}\ ,\ \bibinfo {eid} {arXiv:2411.11156}} (\bibinfo
  {year} {2024})},\ \Eprint {https://arxiv.org/abs/2411.11156}
  {arXiv:2411.11156 [cond-mat.mes-hall]} \BibitemShut {NoStop}%
\bibitem [{\citenamefont {Koshino}\ and\ \citenamefont
  {McCann}(2011)}]{Koshino2011Landau}%
  \BibitemOpen
  \bibfield  {author} {\bibinfo {author} {\bibfnamefont {M.}~\bibnamefont
  {Koshino}}\ and\ \bibinfo {author} {\bibfnamefont {E.}~\bibnamefont
  {McCann}},\ }\bibfield  {title} {\bibinfo {title} {Landau level spectra and
  the quantum hall effect of multilayer graphene},\ }\href
  {https://doi.org/10.1103/PhysRevB.83.165443} {\bibfield  {journal} {\bibinfo
  {journal} {Phys. Rev. B}\ }\textbf {\bibinfo {volume} {83}},\ \bibinfo
  {pages} {165443} (\bibinfo {year} {2011})}\BibitemShut {NoStop}%
\bibitem [{\citenamefont {Bradley}\ and\ \citenamefont
  {Cracknell}(2010)}]{bradley2010mathematical}%
  \BibitemOpen
  \bibfield  {author} {\bibinfo {author} {\bibfnamefont {C.}~\bibnamefont
  {Bradley}}\ and\ \bibinfo {author} {\bibfnamefont {A.}~\bibnamefont
  {Cracknell}},\ }\href@noop {} {\emph {\bibinfo {title} {The mathematical
  theory of symmetry in solids: representation theory for point groups and
  space groups}}}\ (\bibinfo  {publisher} {Oxford University Press},\ \bibinfo
  {year} {2010})\BibitemShut {NoStop}%
\bibitem [{\citenamefont {Wu}\ \emph {et~al.}(2021)\citenamefont {Wu},
  \citenamefont {Zhu},\ and\ \citenamefont {Yu}}]{Wu2021nernst}%
  \BibitemOpen
  \bibfield  {author} {\bibinfo {author} {\bibfnamefont {Y.-L.}\ \bibnamefont
  {Wu}}, \bibinfo {author} {\bibfnamefont {G.-H.}\ \bibnamefont {Zhu}},\ and\
  \bibinfo {author} {\bibfnamefont {X.-Q.}\ \bibnamefont {Yu}},\ }\bibfield
  {title} {\bibinfo {title} {Nonlinear anomalous nernst effect in strained
  graphene induced by trigonal warping},\ }\href
  {https://doi.org/10.1103/PhysRevB.104.195427} {\bibfield  {journal} {\bibinfo
   {journal} {Phys. Rev. B}\ }\textbf {\bibinfo {volume} {104}},\ \bibinfo
  {pages} {195427} (\bibinfo {year} {2021})}\BibitemShut {NoStop}%
\bibitem [{\citenamefont {Tokura}\ and\ \citenamefont
  {Nagaosa}(2018)}]{tokura2018nonreciprocal}%
  \BibitemOpen
  \bibfield  {author} {\bibinfo {author} {\bibfnamefont {Y.}~\bibnamefont
  {Tokura}}\ and\ \bibinfo {author} {\bibfnamefont {N.}~\bibnamefont
  {Nagaosa}},\ }\bibfield  {title} {\bibinfo {title} {Nonreciprocal responses
  from non-centrosymmetric quantum materials},\ }\href
  {https://doi.org/10.1038/s41467-018-05759-4} {\bibfield  {journal} {\bibinfo
  {journal} {Nature Communications}\ }\textbf {\bibinfo {volume} {9}},\
  \bibinfo {pages} {3740} (\bibinfo {year} {2018})}\BibitemShut {NoStop}%
\bibitem [{\citenamefont {Morimoto}\ and\ \citenamefont
  {Nagaosa}(2018)}]{Morimoto2018noncentro}%
  \BibitemOpen
  \bibfield  {author} {\bibinfo {author} {\bibfnamefont {T.}~\bibnamefont
  {Morimoto}}\ and\ \bibinfo {author} {\bibfnamefont {N.}~\bibnamefont
  {Nagaosa}},\ }\bibfield  {title} {\bibinfo {title} {Nonreciprocal current
  from electron interactions in noncentrosymmetric crystals: roles of time
  reversal symmetry and dissipation},\ }\href
  {https://doi.org/10.1038/s41598-018-20539-2} {\bibfield  {journal} {\bibinfo
  {journal} {Scientific Reports}\ }\textbf {\bibinfo {volume} {8}},\ \bibinfo
  {pages} {2973} (\bibinfo {year} {2018})}\BibitemShut {NoStop}%
\bibitem [{\citenamefont {Mahan}\ and\ \citenamefont
  {Sofo}(1996)}]{Mahan1996best}%
  \BibitemOpen
  \bibfield  {author} {\bibinfo {author} {\bibfnamefont {G.~D.}\ \bibnamefont
  {Mahan}}\ and\ \bibinfo {author} {\bibfnamefont {J.~O.}\ \bibnamefont
  {Sofo}},\ }\bibfield  {title} {\bibinfo {title} {The best thermoelectric.},\
  }\href {https://doi.org/10.1073/pnas.93.15.7436} {\bibfield  {journal}
  {\bibinfo  {journal} {Proceedings of the National Academy of Sciences}\
  }\textbf {\bibinfo {volume} {93}},\ \bibinfo {pages} {7436} (\bibinfo {year}
  {1996})}\BibitemShut {NoStop}%
\bibitem [{\citenamefont {Xu}\ \emph {et~al.}(2017)\citenamefont {Xu},
  \citenamefont {Xu},\ and\ \citenamefont {Zhu}}]{xu2017topological}%
  \BibitemOpen
  \bibfield  {author} {\bibinfo {author} {\bibfnamefont {N.}~\bibnamefont
  {Xu}}, \bibinfo {author} {\bibfnamefont {Y.}~\bibnamefont {Xu}},\ and\
  \bibinfo {author} {\bibfnamefont {J.}~\bibnamefont {Zhu}},\ }\bibfield
  {title} {\bibinfo {title} {Topological insulators for thermoelectrics},\
  }\href {https://doi.org/10.1038/s41535-017-0054-3} {\bibfield  {journal}
  {\bibinfo  {journal} {npj Quantum Materials}\ }\textbf {\bibinfo {volume}
  {2}},\ \bibinfo {pages} {51} (\bibinfo {year} {2017})}\BibitemShut {NoStop}%
\bibitem [{\citenamefont {He}\ and\ \citenamefont
  {Tritt}(2017)}]{he2017advances}%
  \BibitemOpen
  \bibfield  {author} {\bibinfo {author} {\bibfnamefont {J.}~\bibnamefont
  {He}}\ and\ \bibinfo {author} {\bibfnamefont {T.~M.}\ \bibnamefont {Tritt}},\
  }\bibfield  {title} {\bibinfo {title} {Advances in thermoelectric materials
  research: Looking back and moving forward},\ }\href
  {https://doi.org/10.1126/science.aak9997} {\bibfield  {journal} {\bibinfo
  {journal} {Science}\ }\textbf {\bibinfo {volume} {357}},\ \bibinfo {pages}
  {eaak9997} (\bibinfo {year} {2017})}\BibitemShut {NoStop}%
\bibitem [{\citenamefont {Ghosh}\ \emph {et~al.}(2025)\citenamefont {Ghosh},
  \citenamefont {Chakraborty}, \citenamefont {Dutta}, \citenamefont {Agarwala},
  \citenamefont {Watanabe}, \citenamefont {Taniguchi}, \citenamefont
  {Banerjee}, \citenamefont {Trivedi}, \citenamefont {Mukerjee},\ and\
  \citenamefont {Das}}]{Ghosh2025Thermopower}%
  \BibitemOpen
  \bibfield  {author} {\bibinfo {author} {\bibfnamefont {A.}~\bibnamefont
  {Ghosh}}, \bibinfo {author} {\bibfnamefont {S.}~\bibnamefont {Chakraborty}},
  \bibinfo {author} {\bibfnamefont {R.}~\bibnamefont {Dutta}}, \bibinfo
  {author} {\bibfnamefont {A.}~\bibnamefont {Agarwala}}, \bibinfo {author}
  {\bibfnamefont {K.}~\bibnamefont {Watanabe}}, \bibinfo {author}
  {\bibfnamefont {T.}~\bibnamefont {Taniguchi}}, \bibinfo {author}
  {\bibfnamefont {S.}~\bibnamefont {Banerjee}}, \bibinfo {author}
  {\bibfnamefont {N.}~\bibnamefont {Trivedi}}, \bibinfo {author} {\bibfnamefont
  {S.}~\bibnamefont {Mukerjee}},\ and\ \bibinfo {author} {\bibfnamefont
  {A.}~\bibnamefont {Das}},\ }\bibfield  {title} {\bibinfo {title} {Thermopower
  probes of emergent local moments in magic-angle twisted bilayer graphene},\
  }\href {https://doi.org/10.1038/s41567-025-02849-1} {\bibfield  {journal}
  {\bibinfo  {journal} {Nature Physics}\ }\textbf {\bibinfo {volume} {21}},\
  \bibinfo {pages} {732} (\bibinfo {year} {2025})}\BibitemShut {NoStop}%
\bibitem [{\citenamefont {You}\ \emph {et~al.}(2018)\citenamefont {You},
  \citenamefont {Fang}, \citenamefont {Xu}, \citenamefont {Kaxiras},\ and\
  \citenamefont {Low}}]{You2018dipole}%
  \BibitemOpen
  \bibfield  {author} {\bibinfo {author} {\bibfnamefont {J.-S.}\ \bibnamefont
  {You}}, \bibinfo {author} {\bibfnamefont {S.}~\bibnamefont {Fang}}, \bibinfo
  {author} {\bibfnamefont {S.-Y.}\ \bibnamefont {Xu}}, \bibinfo {author}
  {\bibfnamefont {E.}~\bibnamefont {Kaxiras}},\ and\ \bibinfo {author}
  {\bibfnamefont {T.}~\bibnamefont {Low}},\ }\bibfield  {title} {\bibinfo
  {title} {Berry curvature dipole current in the transition metal
  dichalcogenides family},\ }\href {https://doi.org/10.1103/PhysRevB.98.121109}
  {\bibfield  {journal} {\bibinfo  {journal} {Phys. Rev. B}\ }\textbf {\bibinfo
  {volume} {98}},\ \bibinfo {pages} {121109} (\bibinfo {year}
  {2018})}\BibitemShut {NoStop}%
\bibitem [{\citenamefont {Xiao}\ \emph {et~al.}(2010)\citenamefont {Xiao},
  \citenamefont {Chang},\ and\ \citenamefont {Niu}}]{Xiao2010RMP}%
  \BibitemOpen
  \bibfield  {author} {\bibinfo {author} {\bibfnamefont {D.}~\bibnamefont
  {Xiao}}, \bibinfo {author} {\bibfnamefont {M.-C.}\ \bibnamefont {Chang}},\
  and\ \bibinfo {author} {\bibfnamefont {Q.}~\bibnamefont {Niu}},\ }\bibfield
  {title} {\bibinfo {title} {Berry phase effects on electronic properties},\
  }\href {https://doi.org/10.1103/RevModPhys.82.1959} {\bibfield  {journal}
  {\bibinfo  {journal} {Rev. Mod. Phys.}\ }\textbf {\bibinfo {volume} {82}},\
  \bibinfo {pages} {1959} (\bibinfo {year} {2010})}\BibitemShut {NoStop}%
\bibitem [{\citenamefont {Xiao}\ \emph {et~al.}(2005)\citenamefont {Xiao},
  \citenamefont {Shi},\ and\ \citenamefont {Niu}}]{Xiao2005Berry}%
  \BibitemOpen
  \bibfield  {author} {\bibinfo {author} {\bibfnamefont {D.}~\bibnamefont
  {Xiao}}, \bibinfo {author} {\bibfnamefont {J.}~\bibnamefont {Shi}},\ and\
  \bibinfo {author} {\bibfnamefont {Q.}~\bibnamefont {Niu}},\ }\bibfield
  {title} {\bibinfo {title} {Berry phase correction to electron density of
  states in solids},\ }\href {https://doi.org/10.1103/PhysRevLett.95.137204}
  {\bibfield  {journal} {\bibinfo  {journal} {Phys. Rev. Lett.}\ }\textbf
  {\bibinfo {volume} {95}},\ \bibinfo {pages} {137204} (\bibinfo {year}
  {2005})}\BibitemShut {NoStop}%
\bibitem [{\citenamefont {Gao}\ and\ \citenamefont
  {Xiao}(2018)}]{Gao2018Orbital}%
  \BibitemOpen
  \bibfield  {author} {\bibinfo {author} {\bibfnamefont {Y.}~\bibnamefont
  {Gao}}\ and\ \bibinfo {author} {\bibfnamefont {D.}~\bibnamefont {Xiao}},\
  }\bibfield  {title} {\bibinfo {title} {Orbital magnetic quadrupole moment and
  nonlinear anomalous thermoelectric transport},\ }\href
  {https://doi.org/10.1103/PhysRevB.98.060402} {\bibfield  {journal} {\bibinfo
  {journal} {Phys. Rev. B}\ }\textbf {\bibinfo {volume} {98}},\ \bibinfo
  {pages} {060402} (\bibinfo {year} {2018})}\BibitemShut {NoStop}%
\bibitem [{\citenamefont {Oji}\ and\ \citenamefont {Streda}(1985)}]{Oji1985}%
  \BibitemOpen
  \bibfield  {author} {\bibinfo {author} {\bibfnamefont {H.}~\bibnamefont
  {Oji}}\ and\ \bibinfo {author} {\bibfnamefont {P.}~\bibnamefont {Streda}},\
  }\bibfield  {title} {\bibinfo {title} {Theory of electronic thermal
  transport: Magnetoquantum corrections to the thermal transport
  coefficients},\ }\href {https://doi.org/10.1103/PhysRevB.31.7291} {\bibfield
  {journal} {\bibinfo  {journal} {Phys. Rev. B}\ }\textbf {\bibinfo {volume}
  {31}},\ \bibinfo {pages} {7291} (\bibinfo {year} {1985})}\BibitemShut
  {NoStop}%
\bibitem [{\citenamefont {Nakai}\ and\ \citenamefont
  {Nagaosa}(2019)}]{Nakai2019Nonreciprocal}%
  \BibitemOpen
  \bibfield  {author} {\bibinfo {author} {\bibfnamefont {R.}~\bibnamefont
  {Nakai}}\ and\ \bibinfo {author} {\bibfnamefont {N.}~\bibnamefont
  {Nagaosa}},\ }\bibfield  {title} {\bibinfo {title} {Nonreciprocal thermal and
  thermoelectric transport of electrons in noncentrosymmetric crystals},\
  }\href {https://doi.org/10.1103/PhysRevB.99.115201} {\bibfield  {journal}
  {\bibinfo  {journal} {Phys. Rev. B}\ }\textbf {\bibinfo {volume} {99}},\
  \bibinfo {pages} {115201} (\bibinfo {year} {2019})}\BibitemShut {NoStop}%
\bibitem [{\citenamefont {Zagier}(2007)}]{Zagier2007}%
  \BibitemOpen
  \bibfield  {author} {\bibinfo {author} {\bibfnamefont {D.}~\bibnamefont
  {Zagier}},\ }\bibinfo {title} {The dilogarithm function},\ in\ \href
  {https://doi.org/10.1007/978-3-540-30308-4_1} {\emph {\bibinfo {booktitle}
  {Frontiers in Number Theory, Physics, and Geometry II: On Conformal Field
  Theories, Discrete Groups and Renormalization}}},\ \bibinfo {editor} {edited
  by\ \bibinfo {editor} {\bibfnamefont {P.}~\bibnamefont {Cartier}}, \bibinfo
  {editor} {\bibfnamefont {P.}~\bibnamefont {Moussa}}, \bibinfo {editor}
  {\bibfnamefont {B.}~\bibnamefont {Julia}},\ and\ \bibinfo {editor}
  {\bibfnamefont {P.}~\bibnamefont {Vanhove}}}\ (\bibinfo  {publisher}
  {Springer Berlin Heidelberg},\ \bibinfo {address} {Berlin, Heidelberg},\
  \bibinfo {year} {2007})\ pp.\ \bibinfo {pages} {3--65}\BibitemShut {NoStop}%
\end{thebibliography}%

\clearpage


\onecolumngrid

\setcounter{secnumdepth}{3}

\onecolumngrid
\newpage
\beginsupplement

\section*{Supplemental Materials}

\tableofcontents

\section{Further details on the semiclassical formalism for nonlinear transport}\label{sec:formalism}
Understanding transport phenomena is central to the study of condensed matter systems, as it provides the link between microscopic particle dynamics and macroscopic observables such as electrical and thermal conductivity. 
In particular, charge transport involves the response of mobile electrons to external perturbations, typically electric fields or temperature gradient. 
A more systematic and microscopic model arises from the semiclassical Boltzmann transport equation. 
By treating the electron distribution function in phase space and incorporating scattering via a collision integral, the Boltzmann equation formalism provides a general framework for transport in solids. 
In the relaxation-time approximation, the solution to the Boltzmann equation under an applied electric field reproduces the classical Drude formula for conductivity.

This approach is extended to include the effects of Berry curvature and quantum metric, which are important in systems with nontrivial band topology.
The anomalous Hall effect, for example, arises from the Berry curvature of Bloch states, which modifies the semiclassical equations of motion for electrons with the anomalous velocity term.

\subsection{Boltzmann equation}\label{sec:formalism_Bolzmann}
To start with, we use the semiclassical Boltzmann equation to derive the perturbation of distribution function $f$ in the presence of electric field ${\bs E}$ and temperature gradient $\Nabla T$.

The Boltzmann equation describes the time evolution of the distribution function $f(\rr,\kk,t)$ of our wave packet in phase space, which is a function of position $\rr$, momentum $\kk$ and time $t$. 
The equation is given by
\begin{align} 
    \{ \partial_t + \dot \rr \cdot \nabla_{\rr} + \dot \kk \cdot \nabla_{\kk} \} f(\rr,\kk,t) = I_{\text{coll}} \{f(\rr,\kk,t)\}
    \approx -\frac{f(\rr,\kk,t)-f_0(\kk)}{\tau}
\end{align}
where we have used relaxation time approximation and $f_0(\kk)$ is the Fermi-Dirac distribution function at equilibrium.
Here we use the semiclassical equations of motion without magnetic field:
\begin{align}
    \dot \rr &= \frac{1}{\hbar} \frac{\partial \epsilon_n(\kk)}{\partial \kk} - e {\bs E} \times{\bs \Omega} \,, \label{smeqn:vr}\\
    \dot \kk &= -\frac{e}{\hbar} {\bs E} \,.  \label{smeqn:vk}
\end{align}

We can expand $f$ perturbatively
\begin{equation}
    f = f_0 + \sum_{n,m} f^{(n,m)} E^n (\nabla T/T)^m \,.
\end{equation}
We are interested in the direct current (DC) limit, i.e. $\partial_t f = 0$. We have
\begin{align}
    \left\{ \frac{1}{\hbar} \frac{\partial \epsilon_n(\kk)}{\partial \kk} \cdot \left(\frac{\nabla_\rr T}{T}\right)\left(T\frac{\partial}{\partial T}\right) + (-e{\bs E} \times \mathbf\Omega)\left(\frac{\nabla_\rr T}{T}\right)\left(T\frac{\partial}{\partial T}\right)+\frac{-e}{\hbar} {\bs E} \cdot \nabla_\kk \right\} f = -\frac{f-f_0}{\tau} \,.
\end{align}
This leads to the recursive equation for the distribution function $f^{(n,m)}$:
\begin{align}
    \sum_l\left(\frac{1}{\hbar} \frac{\partial \epsilon_n(\kk)}{\partial \kk}\right)^{(l)} \cdot \left(\frac{\nabla_\rr T}{T}\right)\left(T\frac{\partial}{\partial T}\right) f^{(n-l,m-1)}
    + \sum_l (-e{\bs E} \times \mathbf\Omega)^{(l)} \left(\frac{\nabla_\rr T}{T}\right)\left(T\frac{\partial}{\partial T}\right)f^{(n-l,m-1)} \nonumber \\
    +\frac{-e}{\hbar} {\bs E} \cdot \nabla_\kk  f^{(n-1,m)} = -\frac{f^{(n,m)}}{\tau} \,.
\end{align}
This equation can be solved order by order in $\bs E$ and $\Nabla T$.

Here we list the results for the leading orders, which will be used in the later subsections:
\begin{align}
    f^{(1,0)} &= \frac{e\tau}{\hbar} {\bs E} \cdot \nabla_\kk  f_0 \\
    f^{(0,1)} &= -\tau \vv^{(0)} \cdot \left(\frac{\nabla_\rr T}{T}\right)\left(T\frac{\partial}{\partial T}\right) f_0 \\
    f^{(1,1)} &= -\tau \vv^{(0)} \cdot \left(\frac{\nabla_\rr T}{T}\right)\left(T\frac{\partial}{\partial T}\right) f^{(1,0)} 
    + \frac{e\tau}{\hbar} {\bs E} \cdot \nabla_\kk f^{(0,1)}
    + e\tau ({\bs E} \times \mathbf\Omega)^{(0)} \left(\frac{\nabla_\rr T}{T}\right)\left(T\frac{\partial}{\partial T}\right)f_0 \\
    f^{(2,0)} &= \frac{e\tau}{\hbar} {\bs E} \cdot \nabla_\kk  f^{(1,0)} \\
    f^{(0,2)} &=  -\tau \vv^{(0)} \cdot \left(\frac{\nabla_\rr T}{T}\right)\left(T\frac{\partial}{\partial T}\right) f^{(0,1)} 
\end{align}

\subsection{Quantum metric corrections to group velocity and Berry curvature}\label{sec:formalism:quantum_metric}
The group velocity and anomalous velocity due to Berry curvature are modified in the presence of electric field, the corrections of which are determined by quantum metric dipoles~\cite{PhysRevLett.127.277202,kaplan2024unification}.
These corrections affect the nonlinear transport.
In this subsection we first briefly sketch the derivation of this correction to the group velocity and Berry curvature and then extends this theory to temperature gradient driven responses.

Consider the Hamiltonian of a non-interacting electron system in an external electric field ${\bs E}$:
\begin{equation}
    H = \sum_{mn} \left( \varepsilon^{(0)}_n \delta_{nm} - e{{\bs E}}\cdot \langle n |\hat{\bs r} | m\rangle  \right) | n\rangle \langle m |,
\end{equation}
where $|n\rangle$ is the unperturbed wavefunction of the $n$-th band eigenstate at zero field. 
Apply the Schrieffer-Wolff transformation, which is the perturbation theory of operators. The operator $\cal O$ is transformed to
\begin{align}\label{eqn:O_correction_sm}
    {\cal O} \longrightarrow e^S {\cal O} e^{-S} = {\cal O} + [S,{\cal O}] + \frac12 [S,[S,{\cal O}]] + \frac16 [S,[S,[S,{\cal O}]]] + \dots
\end{align}
where $S_{nm} = {-eE^a A^a_{nm}}/{\varepsilon_{nm}}$.
Then the Hamiltonian in the new basis is
\begin{align}
    H\longrightarrow H' &=H_0+\frac12[S,H_1] + \frac13 [S,[S,H_1]] +\dots
\end{align}
which gives the following corrections to the eigen-energies:
\begin{equation}\label{eqn:E_correction_sm}
    \epsilon_{n\kk}' = \epsilon_{n\kk} + e^2 G^{ab}_n E_aE_b + O(E^3) \,,
\end{equation}
where the band-normalized quantum metric \begin{equation}
    G_n^{ab}=\sum_{m\neq n}\frac{A_{nm}^a A_{mn}^b+A_{mn}^a A_{nm}^b}{2\varepsilon_{nm}} \,,
\end{equation}
and $A_{nm}^a$ is the inter-band Berry connection. 
This term give corrections to the group velocity
\begin{align}\label{eqn:c_correction_sm}
    v_{n\kk}^a &= \frac{1}{\hbar} \frac{\partial\epsilon_{n\kk}}{ \partial k_a} + \frac{e^2}{\hbar}  \partial_{k_a} G_n^{bc} E_bE_c + O(E^3) \,,\\
\end{align}

The Berry curvature can be expanded in powers of $\mathbf{E}$ by applying the Schrieffer-Wolff transformation to Berry connection:
\begin{align}\label{eqn:A_correction_sm}
    \mathbf A \longrightarrow \mathbf A' &= \mathbf A+[S, \mathbf A]+\frac12[S,[S,\mathbf A]]\dots
\end{align}
which gives the following corrections to Berry curvature:
\begin{align}\label{eqn:omega_correction_sm}
    \Omega^{c\prime}_{n\kk} &= \Omega^{c}_{n\kk} - e E_d \epsilon^{abc} \partial_a G^{bd}_n + O(E^2) \,.
\end{align}

We now adopt Luttinger’s gravitational field approach to incorporate the effects of a temperature gradient~\cite{Luttinger1964}. To leading order in the gradient, the effective Hamiltonian can be written as
\begin{align}
    H &= \int_{\mathbf r} (1+\psi(\mathbf r))\left[ h(\mathbf{r}) - e \mathbf{E}\cdot \mathbf{r}  \right]\\
    &\approx \sum_{mn} \left( (\varepsilon^{(0)}_n-\mu) \left(\delta_{nm}  + {\frac{\Nabla T}{T}}\cdot \langle n |\hat{\bs r} | m\rangle  \right) - e{{\bs E}}\cdot \langle n |\hat{\bs r} | m\rangle  \right) | n\rangle \langle m | \\ 
    &= \sum_{mn} \left( (\varepsilon^{(0)}_n-\mu) \delta_{nm}  + {{\bs F}}\cdot \langle n |\hat{\bs r} | m\rangle  \right) | n\rangle \langle m |
\end{align}
where the gravitational field satisfies $\Nabla \psi(\mathbf r) = \frac{\Nabla T}{T}$.
Within this framework, the temperature gradient enters on the same footing as the electric field, allowing the preceding derivations to be straightforwardly generalized to this case. Physically, the temperature gradient acts as an effective statistical force and modifies the Berry curvature in a manner analogous to Eq.~(\ref{eqn:omega_correction_sm}), upon replacing the electric force $\mathbf{F}=-e\mathbf{E}$ with $\mathbf{F}=-e\mathbf{E} + \frac{\epsilon-\mu}{k_B T} k_B\Nabla T$. 
This framework will be justified by comparing to wave packet expansion approach in Sec.~\ref{sec:formalism:magnetization}.

\subsection{Anomalous current}\label{sec:formalism:anomalous_current}
When entering the nonlinear regime, the Berry curvature and quantum metric lead to additional contributions to the conductivity tensors.
In this subsection, we summarize the formalism of nonlinear transport for a generic system with external electric field and temperature gradient and include  the Berry curvature and quantum metric effects.

To start with, we briefly review the quantum geometric tensor and its relation to the Berry curvature and quantum metric.
The quantum geometric tensor of a band labeled by $n$ is defined as
\begin{equation}
    {\cal Q}^n_{ab}(\bs{k}) = \langle D_{k_a} u_{\bs{k}}^n|D_{k_b} u_{\bs{k}}^n\rangle = g_{ab}^n + \frac{i}{2} \Omega_{ab}^n \,,
\end{equation}
where $D_{k_a}u_{\bs{k}}^n=\partial_{k_a}u_{n\bs{k}} - iA^a_{n\bs{k}}u_{n\bs{k}}$ and the Berry connection is defined by
\begin{equation}
    A^a_{n\bs{k}} = -i\langle u_{n\bs{k}}|\partial_{k_a} u_{n\bs{k}}\rangle  \,.
\end{equation}
Its real part $g_{ab}^n$ is the quantum metric and the imaginary part $\Omega_{ab}^n=\partial_{k_a} A^b_{n\bs{k}} - \partial_{k_b} A^a_{n\bs{k}}$ is the Berry curvature.

Consider a wave packet of an electron in band $n$, which is described by the wave function $|\Psi(\rr,t)\rangle = \sum_n \int d\kk \phi_n(\rr,\kk,t) |u_{n\kk}\rangle$, where $\phi_n(\rr,\kk,t)$ is the wave packet envelope function.
Berry curvature gives rise to the anomalous velocity in its semiclassical equations of motion Eq.~(\ref{smeqn:vr}) and Eq.~(\ref{smeqn:vk}).
In transport problems, the current density is thus modified by the anomalous velocity. 
Here we will focus on the electric current density ${\bs j}$ and heat current density ${\bs j}_Q$.
It has been shown that the Drude terms which arise from the normal velocity and the anomalous terms which arise from the anomalous velocity should be treated separately.

Denote the charge current and heat current as $\bs j_1$ and $\bs j_2$.
The Drude current densities for a partially filled single band are:
\begin{align}
    {\bs j}^{(D)}_1 &= -e \int_\kk {\bs v} f \,, \\
    {\bs j}^{(D)}_2 &= \int_\kk (\epsilon_k - \mu) {\bs v} f \,,
\end{align}
where ${\bs v} = \frac{1}{\hbar} \frac{\partial \epsilon_n(\kk)}{\partial \kk}$ is the group velocity and $f$ is the distribution function.
Here we have suppressed the band index $n$ for simplicity, which could be added back by summing over $n$ for multiple bands.

The Berry curvature current densities have various formalisms.
Here we follow Matsumoto and Murakami's edge current formula which gives the current densities with generic external potential~\cite{Matsumoto2011Theoretical}:
\begin{align}
    {\bs j}^{(\Omega)}_1 &=e \Nabla \times \frac{1}{\hbar} \int_\kk \int^\infty_{\epsilon_\kk} \dd \epsilon f \mathbf{\Omega}_\kk  \,, \label{eqn:key1}\\
    {\bs j}^{(\Omega)}_2 &= -\Nabla \times \frac{1}{\hbar} \int_\kk \int^\infty_{\epsilon_\kk} \dd \epsilon (\epsilon-\mu) f \mathbf{\Omega}_\kk \,,  \label{eqn:key2}
\end{align}
where $\mathbf{\Omega}_\kk$ is the Berry curvature and $f$ is the distribution function. 
We consider the current density under varying external electric field ${\bs E}$ and temperature gradient $\nabla T$.
The electric current is ${\bs j}_1 = -e{\bs j}_N$ and the heat current is ${\bs j}_2 = {\bs j}_E - \mu {\bs j}_N$ where $\bs j_N$ and $\bs j_E$ are the particle current and energy current.
Using Eqs.~(\ref{eqn:key1}) and (\ref{eqn:key2}), we get the electric field ${\bs E}$ driven currents:
\begin{align}
    \bs j_1 &= \frac{e^2 \bs E}{\hbar} \times  \int_\kk \int \frac{\partial f}{\partial \epsilon} \bs \Omega \Theta(\epsilon-\epsilon_\kk) \\
    \bs j_2 &= -\frac{e \bs E}{\hbar} \times  \int_\kk \int (\epsilon -\mu) \frac{\partial f}{\partial \epsilon} \bs \Omega \Theta(\epsilon-\epsilon_\kk)
\end{align}
and the temperature gradient $\Nabla T$ driven currents:
\begin{align}
    \bs j_1 &= e\frac{\Nabla T}{T} \times \frac{1}{\hbar} \int_k \int \dd \epsilon \left(-\beta \frac{\partial f}{\partial \beta}\right) \bs \Omega \Theta(\epsilon-\epsilon_\kk) \\
    \bs j_2 &= -\frac{\Nabla T}{T} \times \frac{1}{\hbar} \int_k \int \dd \epsilon (\epsilon -\mu) \left(-\beta \frac{\partial f}{\partial \beta}\right) \bs \Omega \Theta(\epsilon-\epsilon_k) 
\end{align}
These equations serve as the starting point for the derivation of the transport coefficients.

We will perturbatively expand the current density in powers of the electric field and temperature gradient, and derive the corresponding nonlinear coefficients.
The $(n,m)$-th order conductivity responses are defined for current ${\bs j}^{(n,m)}$ (or ${\bs j}^{(n,m)}_Q$) generated by $n$ powers of ${\bs E}$ and $m$ powers of $\nabla T$:
\begin{equation}
\label{eqn:SM_sig_def}
j^c_l = L_{l, \underbrace{1 \ldots 1}_{n\ \text{times}},\,\underbrace{2 \ldots 2}_{m\ \text{times}}}^{c, a_1\dots a_n b_1\dots b_m} ~ E_{a_1}\dots E_{a_n} (\nabla_{b_1} T)\dots (\nabla_{b_m}T) 
\end{equation}
where $l=1,2$ labels the charge and heat current respectively.

\subsection{Second order current from wave packet expansion}\label{sec:formalism:magnetization}
In this section, we compare our formalism of nonlinear response with the wave packet approach of Ref.~\cite{Gao2018Orbital}. 
We show that our formalism reproduces the effects discussed there while also capturing additional contributions.

In semiclassical linear-response theory, the charge and heat transport currents are obtained from the wave-packet expansion~\cite{Xiao2006Berry},
\begin{align}
    \mathbf j_1 &= \int \left( -e  \dot \rr + \nabla \times \mathbf m \right)f - \nabla \times \mathbf M
    \\
    \mathbf{j}_2 &= \int \left((\epsilon-\mu)  \dot \rr - {\mathbf E} \times \mathbf m \right)f + {\mathbf E} \times {\mathbf M} 
\end{align}
where $\mathbf m$ denotes the local orbital magnetization of a wave packet and $\mathbf M$ is the macroscopic magnetization.  
The magnetization dipole induced current consists of a local circulating contribution associated with $\mathbf m$ and a net edge current.  
Only the latter contributes to transport, reproducing the edge-current formulas in Eqs.~(\ref{eqn:key1}) and (\ref{eqn:key2})~\cite{Matsumoto2011Theoretical}.

In Ref.~\cite{Gao2018Orbital}, the semiclassical expression for the current was systematically expanded to second order in the external driving fields, yielding
\begin{align}\label{eqn:key3}
\mathbf{j}_1
= - \boldsymbol{\nabla} \times \int \frac{d\mathbf{k}}{(2\pi)^3}\, \mathcal{G}\, \boldsymbol{\Omega}
+ \boldsymbol{\nabla} \times \left(
\hat{\mathbf{e}}_i \, \partial_j
\int \frac{d\mathbf{k}}{(2\pi)^3}\, \theta_{ij}\, \mathcal{G}
\right) \, ,
\end{align}
where 
$ \theta_{ij} = 2 \mathrm{Re} \frac{({A}_j)_{0n}(\mathbf{v}_0 \times {A}_{n0})_i} {\varepsilon_0 - \varepsilon_n}$
and $\mathcal{G} = - k_B T \ln \left[ 1 + \exp\left( \beta(\varepsilon - \mu)\right) \right]$.
The second term in Eq.~(\ref{eqn:key3}) can be interpreted as the difference between a macroscopic magnetization quadrupole current and a local orbital magnetization quadrupole contribution, such that only a pure edge current remains and contributes to transport.  
After straightforward algebras, Eq.~(\ref{eqn:key3}) can be rewritten into the compact form
\begin{align}\label{eqn:key3a}
\mathbf{j}_1
= - \boldsymbol{\nabla} \times \int \frac{d\mathbf{k}}{(2\pi)^3}\, \mathcal{G}\, ( \boldsymbol{\Omega} + \boldsymbol{\Omega}^{(1)} )
\end{align}
where the correction to Berry curvature is given by
\begin{align}
    (\boldsymbol{\Omega}^{(1)})^c = F_d~ \epsilon^{abc} \partial_a G^{bd}_n + O(F^2) \,.
\end{align}
and the generalized force $\mathbf{F}=-e\mathbf{E}+\frac{\epsilon-\mu}{k_B T}\Nabla T$ which applies to electric-field and temperature-gradient drives, respectively.

Equation~(\ref{eqn:key3a}) reproduces the result obtained in Sec.~\ref{sec:formalism:quantum_metric}, thereby justifying our framework from the perspective of the wave-packet expansion.
Note our approach not just provide a quantum theory for quantum metric induced thermal correction to Berry curvature, but also produce quantum metric induced thermal correction to group velocity which was not captured in the semiclassical wave packet expansion method.

\subsection{Useful formulas}
We find the transport coefficients can all be expressed in terms of the following integrals:
\begin{align}
    I_n(\Phi_k)&\equiv\int_{-\infty}^{\infty} \left[ \left(\frac{\partial}{\partial \mu}\right)^n f_0 \right] \Theta(\epsilon-\epsilon_\kk) \Phi_\kk \dd \epsilon  \,, \label{eqn:def_In} \\
    J_n(\Phi_k) &\equiv \int_{-\infty}^{\infty} (\epsilon - \mu) \left[\left(\frac{\partial}{\partial \mu}\right)^{n}f_0 \right] \Theta(\epsilon-\epsilon_\kk) \Phi_k \dd \epsilon \,. \label{eqn:def_Jn} \\ 
    K_n(\Phi_k) &\equiv \int_{-\infty}^{\infty} (\epsilon - \mu)^2 \left[\left(\frac{\partial}{\partial \mu}\right)^{n}f_0 \right] \Theta(\epsilon-\epsilon_\kk) \Phi_k \dd \epsilon \,. \label{eqn:def_Kn} \\ 
    T_n(\Phi_k) &\equiv \int_{-\infty}^{\infty} (\epsilon - \mu)^3 \left[\left(\frac{\partial}{\partial \mu}\right)^{n}f_0 \right] \Theta(\epsilon-\epsilon_\kk) \Phi_k \dd \epsilon \,. \label{eqn:def_Tn}
\end{align}
where $\Phi_\kk$ is a function of $\kk$, $\Theta(\epsilon-\epsilon_\kk)$ is the Heaviside step function and $f_0$ is the Fermi-Dirac distribution function.
This motivates us the summarize the results of the integrals in this subsection.

The order $n$ dependence of $I_n$, $J_n$, $K_n$ and $T_n$ are:
\begin{align}
    I_n(\Phi_k)&= \left(\frac{\partial}{\partial \mu}\right)^{n} I_0(\Phi_k) \,, \label{eqn:rec_In} \\
    J_n(\Phi_k)&= \frac{\partial J_{n-1}(\Phi_k)}{\partial \mu} + I_{n-1}(\Phi_k) 
    \label{eqn:rec_Jn} \\ 
    K_n(\Phi_k)&= \frac{\partial K_{n-1}(\Phi_k)}{\partial \mu} + 2J_{n-1}(\Phi_k) \label{eqn:rec_Kn}  \\ 
    T_n(\Phi_k)&= \frac{\partial T_{n-1}(\Phi_k)}{\partial \mu} + 3K_{n-1}(\Phi_k) \label{eqn:rec_Tn}  
\end{align}
for $n\geq 1$.
It is also helpful to note that the modified integrals $\tilde{I}_n$, $\tilde{J}_n$, $\tilde{K}_n$ and $\tilde{T}_n$ are derivatives of $I_n$ and $J_n$ with respect to $\mu$:
\begin{align}
    \tilde{I}_n(\Phi_k) &\equiv \int_{-\infty}^{\infty} \left[ \left(\frac{\partial}{\partial \mu}\right)^n f_0 \right] \frac{\partial\Theta(\epsilon-\epsilon_k)}{\partial \epsilon} \Phi_k \dd \epsilon = I_{n+1}(\Phi_k) \\
    \tilde{J}_{n}(\Phi_k) &\equiv  \int_{-\infty}^{\infty} (\epsilon - \mu) \left[\left(\frac{\partial}{\partial \mu}\right)^{n}f_0 \right] \frac{\partial\Theta(\epsilon-\epsilon_k)}{\partial\epsilon} \Phi_k \dd \epsilon = \frac{\partial}{\partial \mu} J_{n}(\Phi_k)  \\ 
    \tilde{K}_{n}(\Phi_k) &\equiv  \int_{-\infty}^{\infty} (\epsilon - \mu)^2 \left[\left(\frac{\partial}{\partial \mu}\right)^{n}f_0 \right] \frac{\partial\Theta(\epsilon-\epsilon_k)}{\partial\epsilon} \Phi_k \dd \epsilon = \frac{\partial}{\partial \mu} K_{n}(\Phi_k)  \\ 
    \tilde{T}_{n}(\Phi_k) &\equiv  \int_{-\infty}^{\infty} (\epsilon - \mu)^3 \left[\left(\frac{\partial}{\partial \mu}\right)^{n}f_0 \right] \frac{\partial\Theta(\epsilon-\epsilon_k)}{\partial\epsilon} \Phi_k \dd \epsilon = \frac{\partial}{\partial \mu} T_{n}(\Phi_k)  
\end{align}

The Fermi-Dirac distribution function $f_0=\frac{1}{e^{\beta(\epsilon-\mu)}+1}$ is a function of $\epsilon$, $\mu$ and $\beta=\frac{1}{k_B T}$. 
It has the following properties:
\begin{align}
    -\frac{1}{\beta}\frac{\partial f_0}{\partial \mu} = \frac{1}{\epsilon_\kk-\mu}\frac{\partial f_0}{\partial \beta} &= f_0(1-f_0) \approx \delta(\epsilon_\kk-\mu) \,, \\
    \epsilon_\kk-\mu &= \frac{1}{\beta} \ln (\frac{1}{f_0}-1)
\end{align}
The integrals $I_0$, $J_0$ and $J_1$ can be evaluated using these properties:
\begin{align}
    I_0(\Phi_\kk) &= -\frac{1}{\beta} \ln(1-f_0) \Phi_\kk \\
    I_1(\Phi_\kk) &= c_0 \Phi_\kk \\
    J_0(\Phi_\kk) &= \frac{1}{\beta^2} \left( \operatorname{Li}_2(f_0) + \ln f_0\ln(1-f_0) - \frac12 \ln^2(1-f_0) \right) \Phi_\kk \\
    J_1(\Phi_\kk) &= \frac{1}{\beta}c_1 \Phi_\kk + I_0(\Phi_\kk)
\end{align}
where $\operatorname{Li}_2(x)$ is the dilogarithm and we have used properties of dilogarithm. Here:
\begin{align}
    c_q \equiv \int_{-\infty}^{\infty} \dd \epsilon \frac{\partial f_0}{\partial \mu} \Theta(\epsilon-\epsilon_\kk) (\beta(\epsilon-\mu))^q 
\end{align}
which can be shown that 
\begin{align}
    c_0 &= f_0 \\
    c_1 &= f_0 \ln f_0 + (1-f_0) \ln(1-f_0) \\
    c_2 &= \frac{\pi^2}{3} + f_0 \ln^2\left( \frac{1}{f_0} - 1 \right) - \ln^2(1 - f_0) - 2\, \operatorname{Li}_2(1 - f_0) \\
    &= f_0 \ln^2\left( \frac{1}{f_0} - 1 \right) + 2\beta^2 J_0(1)
\end{align}
The higher order terms can be straightforwardly generated by Eqs.~(\ref{eqn:rec_In}) and (\ref{eqn:rec_Jn}). 
However, it is not transparent to extract the Wiedemann-Franz and Mott types of relations near zero temperature from these exact results. 
Expansion of these results with respect to temperature is needed.

Then we use Sommerfeld expansion to re-evaluate and analyze the integrals up to leading two orders in $T$.
The first several terms of $I_n$ and $J_n$ are:
\begin{align}
    I_0(\Phi_k) &\approx ((\mu-\epsilon_k) \Phi_k \Theta) + \frac{\pi^2}{6\beta^2}\left(\frac{\partial}{\partial \mu}\right) \left(\Phi_k \Theta\right)\bigg|_{\mu} + O(T^4) \\
    I_1(\Phi_k) &\approx (\Phi_k \Theta) + \frac{\pi^2}{6\beta^2}\left(\frac{\partial}{\partial \mu}\right)^2 \left(\Phi_k \Theta\right)\bigg|_{\mu} + O(T^4) \\
    J_0(\Phi_k) &\approx -\frac12(\mu-\epsilon_k)^2\Phi_k\Theta + \frac{\pi^2}{6\beta^2}(\Phi_k\Theta) + O(T^4) \\
    J_1(\Phi_k) &\approx \frac{\pi^2}{3\beta^2} \frac{\partial}{\partial \mu}\left(\Phi_k \Theta\right)\bigg|_{\mu} + O(T^4) \\
    J_2(\Phi_k) &\approx \left(\Phi_k \Theta\right) + \frac{\pi^2}{2\beta^2} \left(\frac{\partial}{\partial \mu}\right)^2 \left(\Phi_k \Theta\right)\bigg|_{\mu} + O(T^4) 
\end{align}
and for $K_n$ and $T_n$ are:
\begin{align}
    K_0(\Phi_k) &\approx  \frac13(\mu-\epsilon_k)^3\Phi_k\Theta 
    + \frac{7\pi^4}{180\beta^4} \frac{\partial}{\partial \mu} \left(\Phi_k \Theta\right)\bigg|_{\mu} + O(T^6)  \\ 
    K_1(\Phi_k) &\approx \frac{\pi^2}{3\beta^2}\left(\Phi_k \Theta\right)\bigg|_{\mu}  + O(T^4) \\ 
    T_0(\Phi_k) &\approx  -\frac14(\mu-\epsilon_k)^4\Phi_k\Theta - \frac{7\pi^4}{60\beta^4}  \left(\Phi_k \Theta\right)\bigg|_{\mu}   + O(T^6)\\
    T_1(\Phi_k) &= 0
\end{align}
Expansion of the higher order terms can be straightforwardly generated by Eqs.~(\ref{eqn:rec_In}) (\ref{eqn:rec_Jn})  (\ref{eqn:rec_Kn})  (\ref{eqn:rec_Tn}). 
These equations will be directly used in the next subsection.

Next we introduce the $\kk$-dependent terms $\Phi_\kk$ that enter the expression of our transport coefficients.
For convenience, we will separate the transport current according to their origins labeling by ``Drude'', ``anomalous'' and ``quantum metric''.
We introduce following notation to simplify our expressions:
\begin{align}
    C_1^{ab} &= \int_\kk v^a v^b \Theta(\mu-\epsilon_k) \\
    C_2^{abc} &= \int_\kk v^a v^b v^c \Theta(\mu-\epsilon_k) \\
    \widetilde{C}_2^{abc} & = \int_\kk v^a\frac{\partial v^c}{\hbar\partial k_b} \Theta(\mu - \epsilon_k) \\ 
    D_1^{ab} &= \int_\kk \epsilon^{abd}\Omega^d \Theta(\mu-\epsilon_k) \\
    D_2^{abc} &= \int_\kk \epsilon^{abd}\Omega^d v^c \Theta(\mu-\epsilon_k) \\
    F^{abc}_{\Omega} &= \int_\kk (\partial_a G^{bc}-\partial_b G^{ac})\Theta(\mu-\epsilon_k) \\
    F^{abc}_D &= \int_\kk \partial_a G^{bc} \Theta(\mu-\epsilon_k)
\end{align}
where $\int_\kk \equiv \int_{\kk \in \text{BZ}} \frac{d^D k}{(2\pi)^D}$ and $D$ is the spatial dimension of the lattice.

\section{Additional details of response coefficients}
In this section, provide additional details of the first order and second order response coefficients for the electric current and heat response functions in the presence of electric field ${\bs E}$ and temperature gradient $\Nabla T$.
The coefficients are separated into three parts: the Drude response, the Berry curvature response and the quantum metric response.
Then we discuss the symmetry properties of these functions.

\subsection{First order responses}
Drude part of the response:
\begin{align}
    L_{1,1}^{(D)a,b} &=e^2\tau \tilde{I}_1(C_1^{ab}) = e^2\tau (C_1^{ab})' \\
    L_{2,2}^{(D)a,b} &= \frac{\tau}{T} D_\beta \tilde{J}_0(C_1^{ab}) = \tau \frac{\pi^2 k_B^2 T}{3} (C_1^{ab})' \\
    L_{1,2}^{(D)a,b} &= \frac{e\tau}{T}  D_\beta \tilde{I}_0(C_1^{ab}) = e\tau \frac{\pi^2 k_B^2 T}{3} (C_1^{ab})''\\
    L_{2,1}^{(D)a,b} &=-e\tau \tilde{J}_1(C_1^{ab}) = -e\tau \frac{\pi^2 k_B^2 T^2}{3} (C_1^{ab})''
\end{align}
where $D_\beta \equiv -\beta \frac{\partial}{\partial \beta}$ and prime indicates derivative with respect to $\mu$.

Berry curvature part of the response:
\begin{align}
    L_{1,1}^{(\Omega)a,b} &=-\frac{e^2}{\hbar} I_1(D_1^{ab}) = -\frac{e^2}{\hbar} D_1^{ab} \\
    L_{2,2}^{(\Omega)a,b} &=-\frac{1}{\hbar T} D_\beta J_0(D_1^{ab})= -\frac{1}{\hbar} \frac{\pi^2 k_B^2 T}{3} D_1^{ab}  \\
    L_{1,2}^{(\Omega)a,b} &=-\frac{e}{\hbar T} D_\beta I_0(D_1^{ab}) = -\frac{e}{\hbar} \frac{\pi^2 k_B^2 T}{3} (D_1^{ab})'  \\
    L_{2,1}^{(\Omega)a,b} &=\frac{e}{\hbar} J_1(D_1^{ab}) = \frac{e}{\hbar} \frac{\pi^2 k_B^2 T^2}{3} (D_1^{ab})' 
\end{align}

\subsection{Second order responses}
\begin{figure}[ht]
    \centering
    \includegraphics[width=0.4\linewidth]{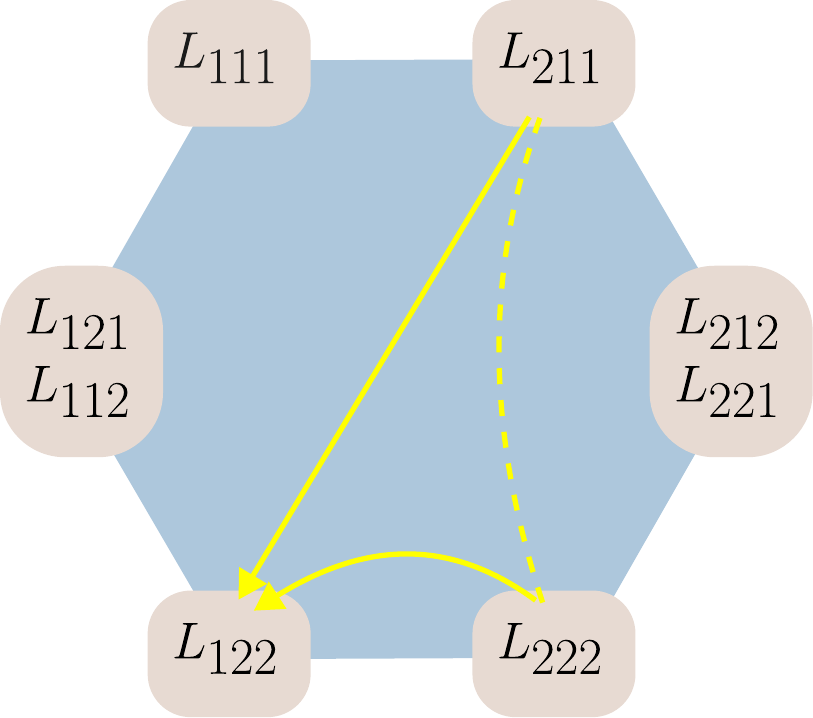}
    \caption{Relations for Drude terms of second order responses.}
    \label{fig:relation_Drude}
\end{figure}

Drude part of the response:
\begin{align}
    L_{1,11}^{(D)a,bc} &= -e^3\tau^2 \left(\tilde{I}_2({C}_2^{abc}) - \tilde{I}_1(\widetilde{C}_2^{abc})\right) =-e^3\tau^2 (({C}_2^{abc})''-(\widetilde{C}_2^{abc})') \\
    L_{2,22}^{(D)a,bc} &= \frac{\tau^2}{T^2 \beta} D_\beta \beta D_\beta \tilde{J}_0(C_2^{abc}) = \tau^2 \frac{\pi^2 k_B^2}{3} (C_2^{abc})' \\
    L_{1,22}^{(D)a,bc} &= -\frac{e\tau^2}{T^2 \beta} D_\beta \beta D_\beta \tilde{I}_0(C_2^{abc})= -e\tau^2 \frac{\pi^2 k_B^2}{3} (C_2^{abc})'' \\
    L_{2,11}^{(D)a,bc} &= e^2\tau^2 \left(\tilde{J}_2({C}_2^{abc}) - \tilde{J}_1(\widetilde{C}_2^{abc})\right) = e^2\tau^2 ({C}_2^{abc})'\\
    L_{1,12}^{(D)a,bc} &= -\frac{e^2\tau^2}{T} D_\beta \left(\tilde{I}_1(2{C}_2^{abc})-\tilde{I}_0(\widetilde{C}_2^{abc}) \right) = -e^2\tau^2\frac{\pi^2 k_B^2 T}{3} \left(2(C_2^{abc})''' -(\widetilde{C}_2^{abc})''\right)\\
    L_{2,12}^{(D)a,bc} &= 2 \frac{e\tau^2}{T} D_\beta \left(\tilde{J}_1(C_2^{abc}) - \tilde{J}_0(\widetilde{C}_2^{abc}) \right)= e\tau^2\frac{\pi^2 k_B^2 T}{3} \left(4(C_2^{abc})'' -(\widetilde{C}_2^{abc})'\right)
\end{align}
We summarize their generalized relations in Fig.~\ref{fig:relation_Drude}.

Berry curvature part of the response:
\begin{align}
    L_{1,11}^{(\Omega)a,bc} &=\frac{e^3 \tau}{\hbar}I_2(D_2^{abc}) = \frac{e^3 \tau}{\hbar} \left( (D_2^{abc})'+\frac{\pi^2}{6\beta^2}(D_2^{abc})'''\right) \\
    L_{2,22}^{(\Omega)a,bc} &=-\frac{\tau}{\hbar T^2 \beta} D_\beta \beta D_\beta J_0(D_2^{abc})= -\frac{\tau}{\hbar}\frac{\pi^2 k_B^2}{3}(D_2^{abc}) \\
    L_{1,22}^{(\Omega)a,bc} &=\frac{e \tau}{\hbar T^2\beta} D_\beta \beta D_\beta  I_0(D_2^{abc}) = \frac{e\tau}{\hbar} \frac{\pi^2 k_B^2}{3}(D_2^{abc})'  \\
    L_{2,11}^{(\Omega)a,bc} &=-\frac{e^2 \tau}{\hbar} J_2(D_2^{abc}) = -\frac{e^2 \tau}{\hbar} \left( D_2^{abc} +  \frac{\pi^2}{2\beta^2} (D_2^{abc})'' \right) \\
    L_{1,12}^{(\Omega)a,bc} &=\frac{e^2 \tau}{\hbar T}D_\beta \left(2I_1(D_2^{abc})+\tilde{I}_0(D_2^{bca}) \right)=\frac{e^2\tau}{\hbar}\frac{\pi^2 k_B^2 T}{3}\left(2(D_2^{abc})''+(D_2^{bca})'' \right) \\
    L_{2,12}^{(\Omega)a,bc} &= -\frac{e \tau}{\hbar T}D_\beta \left( 2J_1(D_2^{abc}) + \tilde{J}_0(D_2^{bca})\right) =-\frac{e\tau}{\hbar}\frac{\pi^2 k_B^2 T}{3}\left(4(D_2^{abc})'+(D_2^{bca})' \right)
\end{align}
Notice the $D^{bca}$ term corresponds to nonparallel sources ${{\bs E}}\times \Nabla T $ which and dose not contribute to the pure transverse responses $L^{xyy}$ and $L^{yxx}$.

Quantum metric part of the response:
the leading order contribution is from Berry curvature correction by correcting $\Omega^c$ with $\Omega^{(1)c}=F_d \epsilon^{abc}\partial_a G^{bd}$. 
\begin{align}
    L^{(G)a,bc}_{1,11} &= \frac{e^3}{\hbar} (F^{abc}_\Omega + {F}_D^{abc}) \\
    L^{(G)a,bc}_{2,11} &= - \frac{e^2}{\hbar} \frac{\pi^2 k_B^2 T^2}{3} (F^{abc}_\Omega + {F}_D^{abc})'  \\
    L^{(G)a,bc}_{2,21} &= \frac{e}{\hbar} \frac{\pi^2  k_B^2 T}{3} (F^{abc}_\Omega + {F}_D^{abc})  = \mathfrak{L}T L^{(G)a,bc}_{1,11}\\
    L^{(G)a,bc}_{1,21} &=  \frac{e^2}{\hbar} \frac{\pi^2 k_B^2 T}{3}  (F^{abc}_\Omega + {F}_D^{abc})' = e(L^{(G)a,bc}_{221})' \\ 
    L^{(G)a,bc}_{1,22} &= -\frac{e}{\hbar} \frac{\pi^2  k_B^2}{3}  (F^{abc}_\Omega + {F}_D^{abc})' = -L^{(G)a,bc}_{121}/eT \\ 
    L^{(G)a,bc}_{2,22} &= 0 
\end{align}

In addition to the symmetry resolved relations mentioned in the maintext, at the second order, a Mott-type relation exists for the full response coefficients in the $\cal T$ broken systems which receive contributions from both the Berry curvature and the second order Drude contribution,
\begin{align} \label{eqn:Mott_full}
    L_{1,22}^{a,b c} &= -e \partial_\mu L_{2,22}^{a,b c} \equiv -e (L_{2,22}^{a,b c})' \,,
\end{align}
where $L_{1,22}^{a, bc}$ and $L_{2,22}^{a, bc}$ are the nonlinear Nernst and nonlinear thermal Hall coefficients.

\subsection{Separation of Drude, Berry curvature and quantum metric parts}
\begin{table}[t]
    \centering
    \begin{tabular}{|c|c|c|c|c|c|}
        \hline
           & Expression &Permutation &$\cal P$ &$\cal T$ &$\cal PT$  \\
        \hline 
        1st order Drude &$C_1^{ab}$ &$[a,b]$ &$+1$ &$+1$ &$+1$ \\
        \hline 
        1st order Berry curvature &$D_1^{ab}$ &$\{a,b\}$ &$+1$ &$-1$ &$-1$ \\
        \hline 
        2nd order Drude &$C_2^{abc}$, $\widetilde{C}_2^{abc}$ &$[a,b]$, $[b,c]$ &$-1$ &$-1$ &$+1$ \\
        \hline 
        2nd order Berry curvature &$D_2^{abc}$ &$\{a,b\}$ &$-1$ &$+1$ &$-1$ \\
        \hline 
        2nd order Quantum metric &$F_\Omega^{abc}$ & $\{a,b\}$, $[b,c]$ &$-1$ &$-1$ &$+1$ \\
        \hline 
        2nd order Quantum metric &$F_D^{abc}+\frac12 F_\Omega^{abc}$ & $[a,b]$, $[b,c]$ &$-1$ &$-1$ &$+1$ \\
        \hline 
    \end{tabular}
    \caption{Symmetry of Drude, Berry curvature and quantum metric terms in 1st and 2nd order responses.
    A term vanishes if it is odd under the symmetry. $[a,b]$ indicates the two indices are interchangeable and $\{a,b\}$ indicates exchanging the two indices generates a minus sign for the quantity.}
    \label{tab:symmetry}
\end{table}
The second order response coefficients can be separated into three parts: the Drude part, the Berry curvature part and the quantum metric part.
Each part has its own generalized Wiedemann-Franz and Mott relations.
Therefore, it is important to separate the three parts of the response coefficients in experiments.
In this subsection, we discuss how to separate the three parts from different perspectives of symmetries and relaxation time.

First, there are situations where some of the three parts vanish due to symmetries.
The well-known example is the vanishing of anomalous Hall effect in systems with time-reversal symmetry $\cal T$.
In general, the (magnetic) space group symmetry constraint on the Drude or the quantum geometric tensor $\cal Q$ with the following condition~\cite{Fang2024Quantum}:
\begin{equation}
\label{eqn:symmetry_constraint}
    {\cal Q} = \frac{1}{|G|} \sum_{g\in G} g {\cal Q} 
\end{equation}
where $|G|$ is the order of $G$. This condition is equivalent to that ${\cal Q}=g {\cal Q},~\forall g\in G$.
If there is a set of symmetry elements $g_1,\dots,g_M$ that satisfies $g_1{\cal Q}+\dots+g_M{\cal Q}=0$, then Eq.~(\ref{eqn:symmetry_constraint}) equals zero. 
Here we focus on three symmetries, time-reversal $\cal T$, inversion $\cal P$ and the combination of both $\cal PT$.
The transformation of the Drude terms $C_2$, $\widetilde{C}_2$, the Berry curvature terms $D_2$, and the quantum metric dipole term $F_\Omega$ and $F_D$ under these symmetries are summarized in Table~\ref{tab:symmetry}. 
All the terms transform as $3$-tensors under space group symmetries. 
Notice that Berry curvature/quantum metric is odd/even under time-reversal symmetry $\cal T$.
Now focus on the second order response coefficients.
For $\cal P$-symmetric systems, all three parts vanish and it is less interesting.
For $\cal T$-symmetric systems, only the second order Berry curvature term $D_2$ is nonzero. 
For $\cal PT$-symmetric systems, the Drude part and the quantum metric part are nonzero, while the Berry curvature part vanishes. 
Therefore, one needs to separate them for this case from other perspectives.

Second, the spatial indices of the terms for the three parts have different symmetry properties. 
It is well known that for linear order response, the Drude term $C_1$ and the Berry curvature term $D_1$ can be separated by (anti-)symmetrizing conductivities with reversed indices.
Consider $\cal T$-broken systems, the linear order response coefficients $L_{1,1}^{a,b} = L_{1,1}^{(D)a,b} + L_{1,1}^{(\Omega)a,b}$ can be separated as:
\begin{align}
    L_{1,1}^{(D)a,b} &= \frac{1}{2}\left(L_{1,1}^{ab}+L_{1,1}^{ba}\right)  \\
    L_{1,1}^{(\Omega)a,b} &= \frac{1}{2}\left(L_{1,1}^{ab}-L_{1,1}^{ba}\right)      
\end{align}
Then we consider the second order response coefficients and use $L_{1,11}$ as an example.
Under permutation of the spatial indices, all three indices of the coefficients are symmetric for $C_2$, $\widetilde{C}_2$; the first two indices are antisymmetric for $D$, $F_\Omega$; the last two indices are symmetric for $F_\Omega$, $F_D$.
From symmetry perspective they belong to different irreducible representations of permutation group over spatial indices.
We also summarize these properties in table~\ref{tab:symmetry} where we used brackets $[a,b]$ to denote the permutation symmetry and $\{a,b\}$ to denote the anti-symmetry under exchange of the two indices.
In the $\cal PT$-symmetric systems, the the quantum metric part $F_\Omega$ can be separated from other terms by antisymmetrizing the first two indices:
\begin{align}
    L_{1,11}^{(G)a,bc}(F_\Omega) &= \frac{1}{2}\left(L_{1,11}^{a,bc} - L_{1,11}^{b,ac}\right)     
\end{align}
though $F_D$ is still not separable from the Drude part $C_2$ and $\widetilde{C}_2$.

Third, in the second order response, the Drude part, Berry curvature part and quantum metric part can be separated by the scaling with relaxation time $\tau$.
The Drude part is proportional to $\tau^2$, the Berry curvature part is proportional to $\tau$ and the quantum metric part is independent of $\tau$.
In experiment, one can compare the temperature behavior of the nonlinear coefficients with the normal longitudinal conductivity $\sigma_{xx}$, which is linear in $\tau$~\cite{QMDnature,QMDscience}.
For a generic system, one can fit the coefficients with the following form:
\begin{align}
    L_{1,11} &= \tau^2 A + \tau B + C \nonumber \\
    L_{1,11}^{(D)} &= \tau^2 A, \quad L_{1,11}^{(\Omega)} = \tau B, \quad L_{1,11}^{(G)} = C
\end{align}

Finally, there are some cases where additional properties of the system can be used to separate the three parts.
For example, in the case of bilayer graphene, the Berry curvature part and the quantum metric part can be separated by (anti-)symmetrizing under flipping displacement field $u_D$ since the Berry curvature part is odd under $u_D$ while the quantum metric part is even under $u_D$ as shown in Fig.~2 (c) and (d) in the maintext.

In conclusion, the three parts of the response coefficients can be separated by symmetry properties, scaling with relaxation time and additional properties of specific systems.

\section{Details of the calculations on Bernal bilayer graphene}
Bernal-stacked graphene possesses an emergent fourfold band degeneracy due to spin and valley degeneracies. 
However, electron-electron interactions can lift these degeneracies at different fillings, leading to various symmetry-broken phases in the presence of external perturbations such as displacement fields and strains.
Notably, in addition to the unpolarized fully symmetric phases, experiments have identified partially polarized phases and isospin ferromagnet (fully polarized) phases~\cite{zhou2022isospin}.
Due to the negligible spin–orbit coupling, it is effectively spin-agnostic.
Therefore, we will treat bilayer graphene as a spinless model and discuss both fully symmetric and valley-polarized phases.

\subsection{Symmetry of bilayer graphene}
Graphene is a two dimensional material with layer group $p6/mmm$ and time-reversal symmetry. 
The Bernal-stacked bilayer graphene has layer group $p\bar{3}m1$ and time-reversal symmetry, since the mirror $M_z$ is broken.
More precisely, the physical properties of bilayer graphene are constrained by its magnetic layer group, which provides a complete description of the system's symmetries including both unitary and anti-unitary operations.
We also care about symmetry of single valley at $K$ (or $K'$). Here we summarize the magnetic little co-group at $K$ and the symmetry allowed valley Hall response.
We list its magnetic layer group and corresponding symmetry properties for the Bernal-stacked bilayer graphene with and without inversion symmetry in Table~\ref{tab:bilayer_graphene_sym}.

Since both inversion $\cal P$ and time-reversal $\cal T$ are broken while either $\cal PT$ or $C_{2z}\cal T$ is preserved for a single valley, quantum metric dipole induced second order valley-Hall effect exists for all the cases. 

When $\cal P$ is broken, the leading order response for a single valley is the Berry curvature induced first order valley-Hall effect while the leading order nonlinear Hall response is quantum metric quadrupole induced 3rd order Hall effect.

When both $\cal P$ and $C_{3z}$ are broken, the leading order response for a single valley is the Berry curvature induced first order valley-Hall effect while the leading order nonlinear Hall response is Berry curvature dipole induced 2nd order Hall effect.

\begin{table}[t]
\centering
\begin{tabular}{|c|c|c|c|c|c|}
\hline
 & \textbf{MLG} & \textbf{MPG} & \makecell{\textbf{magnetic little}\\\textbf{co-group at $K$}} & \textbf{valley-Hall} & \textbf{Hall} \\
\hline
Fully symmetric & $p\bar{3}m11'$ & $\bar{3}m1'$ & $\bar{3}'m'$  & 2nd & 3rd \\
\hline
$\cal P$ breaking & $p3m11'$ & $3m1'$ & $3m'$ & 1st & 3rd \\
\hline
$\cal P$ and $C_{3z}$ breaking & $pm11'$ & $m1'$ & $m'$ & 1st & 2nd \\
\hline
\end{tabular}
\caption{Symmetry for bilayer graphene with and without inversion symmetry. MLG: Magnetic Layer Group, MPG: Magnetic Point Group.}
\label{tab:bilayer_graphene_sym}
\end{table}


\subsection{Hamiltonian, band structure and quantum geometry}

\begin{figure}[t]
    \centering
    \includegraphics[width=0.9\linewidth]{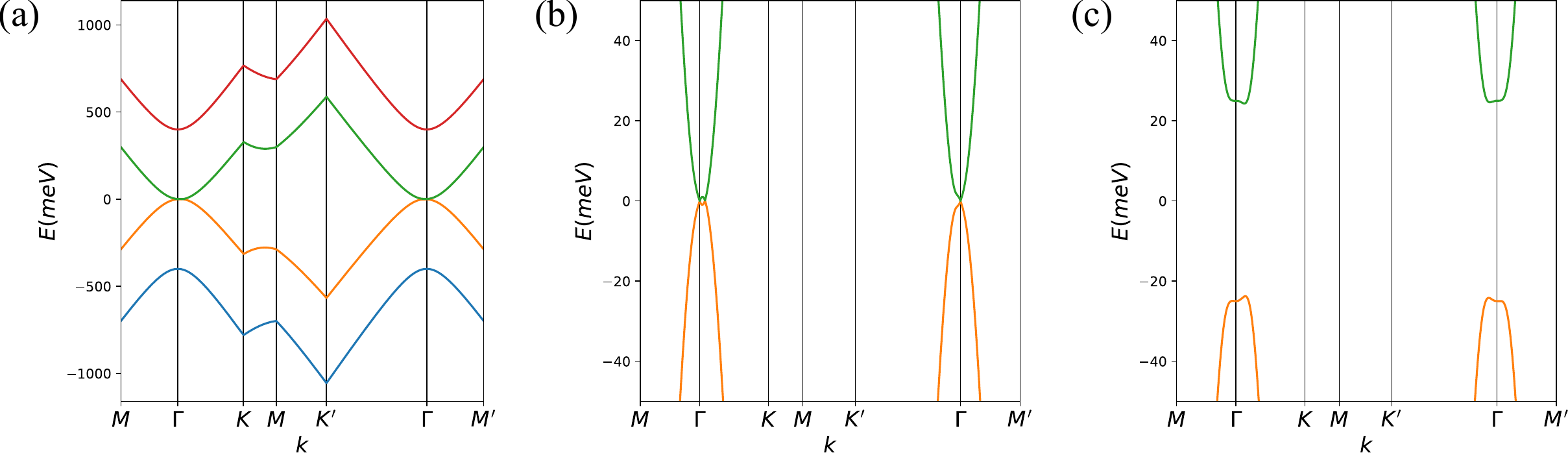}
    \caption{(a) Band structure of bilayer graphene with Hamiltonian Eq.~(\ref{eqn:H0}) and cutoff $k_0=1/a$ at $u_D=0$.
    (b) Zoom-in plot of (a). There is a Weyl point at $\Gamma$ and three Weyl points slightly away from $\Gamma$.
    (c) Band structure at $u_D=50\, {\rm meV}$}
    \label{fig:bands}
\end{figure}

\begin{figure}[t]
    \centering
    \includegraphics[width=0.8\linewidth]{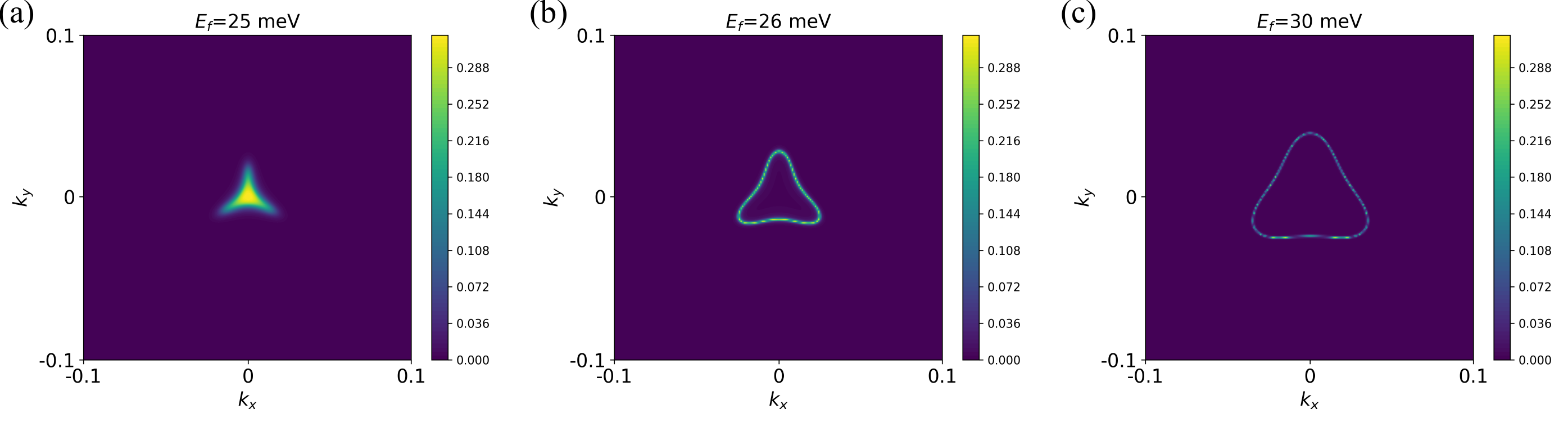}
    \caption{Fermi surface for $u_D=20\, {\rm meV}$ and $\delta=0$ at (a) $E_f=25\, {\rm meV}$, (b) $E_f=26\, {\rm meV}$ and (c) $E_f=30\, {\rm meV}$.}
    \label{fig:Fermi_surface}
\end{figure}

The effective Hamiltonian of Bernal-stacked multi-layer graphene have been derived for generic $N$-layers in, e.g. Ref.~\cite{Koshino2011Landau}.
Here we adapt the Hamiltonian and parameters for bilayer graphene. 
The basis is $(A_1,B_1,A_2,B_2)$ where $A/B$ labels sublattice and $1/2$ labels the layer index. The $\kk \cdot \pp$ expansion of the Hamiltonian around a valley for a single spin is:
\begin{align}\label{eqn:H0}
    H_0=\begin{pmatrix}
        0 & v\pi^\dagger &- v_4 \pi^\dagger & v_3 \pi \\
        v\pi & \Delta &\gamma_1 & - v_4 \pi^\dagger \\
        - v_4 \pi & \gamma_1 &\Delta &v\pi^\dagger\\
        v_3 \pi^\dagger & - v_4 \pi &v\pi &0
    \end{pmatrix} \,,
\end{align}
where $\pi = \zeta k_x + i k_y$, $\zeta = \pm1$ labels valley $K$, $K'$, and $v = \frac{\sqrt{3} a \gamma_0}{2 \hbar}$, $v_3 = \frac{\sqrt{3} a \gamma_3}{2 \hbar}$, $v_4 = \frac{\sqrt{3} a \gamma_4}{2 \hbar}$ are the hopping parameters.
The $v_3$ term is the trigonal warping term that breaks the continuous symmetry to $C_{3z}$ symmetry.
We choose $\gamma_0=3$~meV, $\gamma_1=0.4$~meV, $\gamma_3=0.3$~meV, $\gamma_4=0.04$~meV and $\Delta=0$ for our calculations~\cite{Koshino2011Landau}.

Without inter-layer hoppings, this four-band model describes two Weyl semimetals of a single valley.
The inter-layer hoppings gap our one Weyl point and push the band to high energy. 
The low energy two bands can be described by the following effective Hamiltonian:
\begin{align}
    H_{\text{eff}} =- \frac{v^2}{\gamma_1} \left[ (k_x^2 - k_y^2)\, \sigma_x + 2k_x k_y\, \sigma_y \right] + v_3 \left( k_x\, \sigma_x + k_y\, \sigma_y \right) \nonumber + \frac{2 v v_4}{\gamma_1} (k_x^2 + k_y^2)\, \sigma_0
\end{align}
In order to break the $\cal PT$ symmetry and $C_{3z}$ symmetry, we add
\begin{equation}\label{eqn:H1}
    H_1 = \frac{u_D}{2} \sigma_z + \delta v_3 \left( k_x\, \sigma_x - k_y\, \sigma_y \right)
\end{equation}
where $u_D$ is the displacement field that breaks $\cal PT$ and $\delta$ is a parameter to break $C_{3z}$ symmetry.

\begin{figure}[t]
    \centering
    \includegraphics[width=0.6\linewidth]{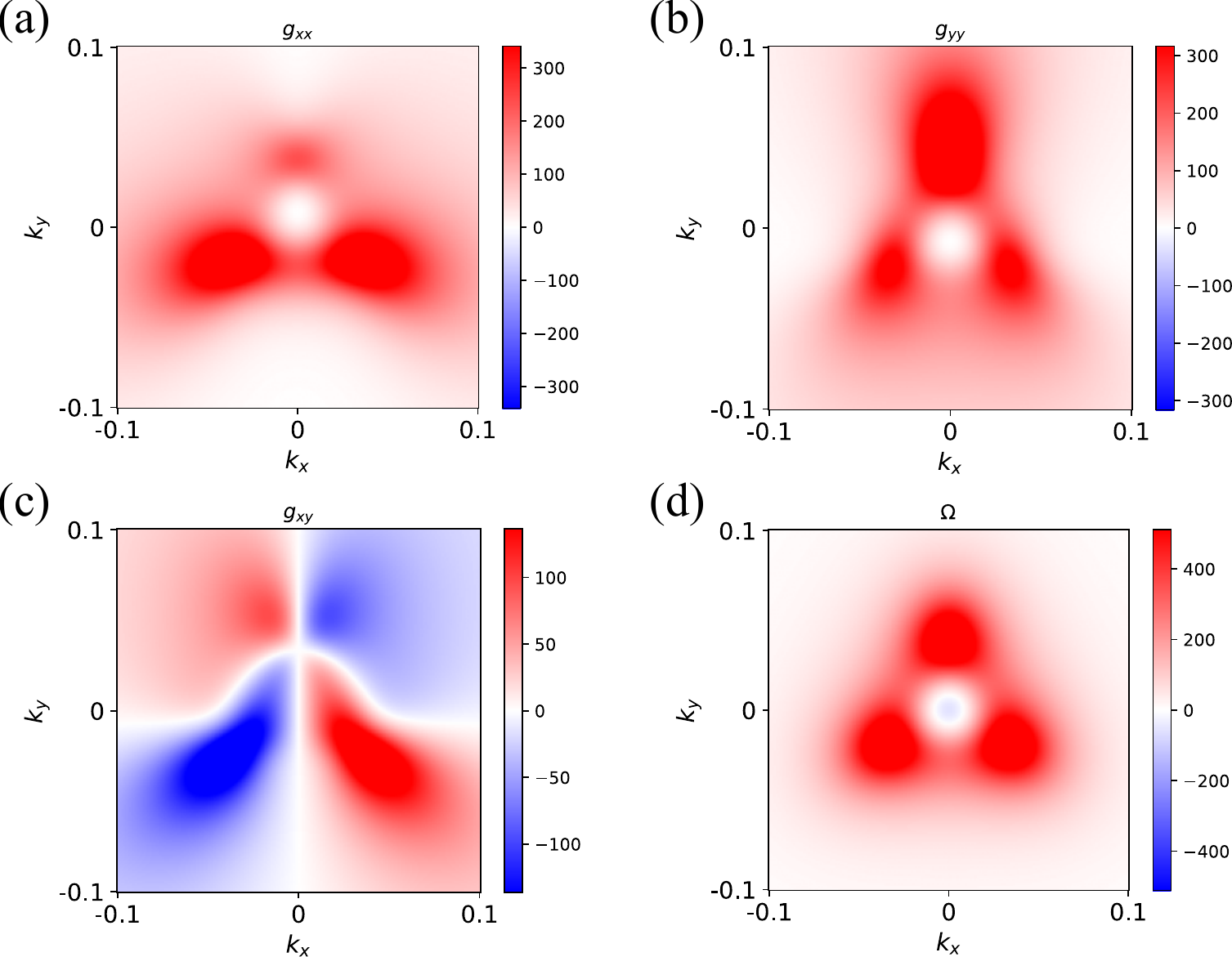}
    \caption{Quantum geometric tensor of the upper band at $u_D=50\,{\rm meV}$ and $\delta=0$.}
    \label{fig:QGT}
\end{figure}

\begin{figure}
    \centering
    \includegraphics[width=0.8\linewidth]{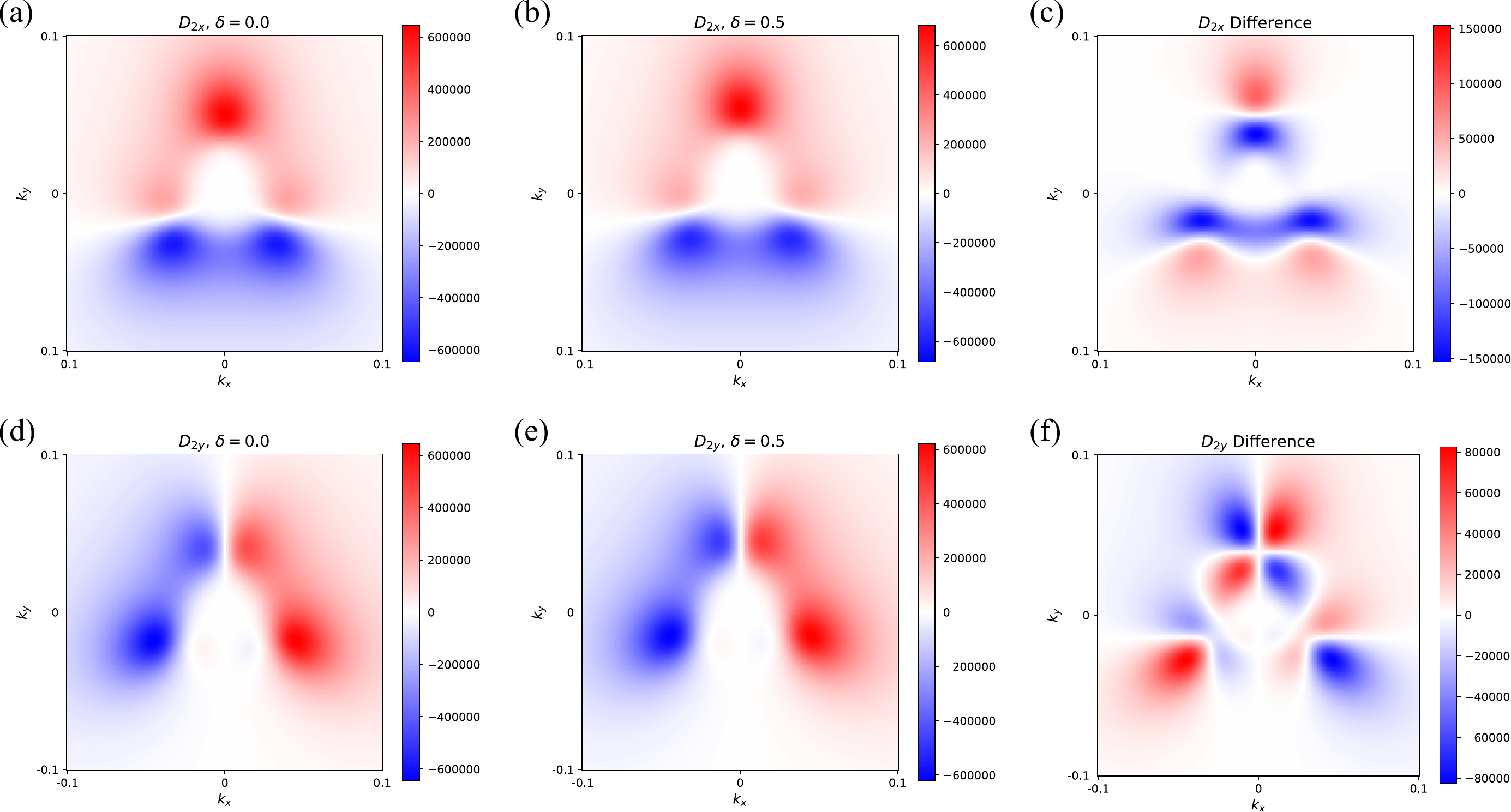}
    \caption{Berry curvature dipole term $D_{2x}$ for (a) $\delta=0$ and (b) $\delta=0.5$ and $c$ their difference. 
    Berry curvature dipole term $D_{2y}$ for (a) $\delta=0$ and (b) $\delta=0.5$ and $c$ their difference.}
    \label{fig:BCD}
\end{figure}

\begin{figure}[t]
    \centering
    \includegraphics[width=0.6\linewidth]{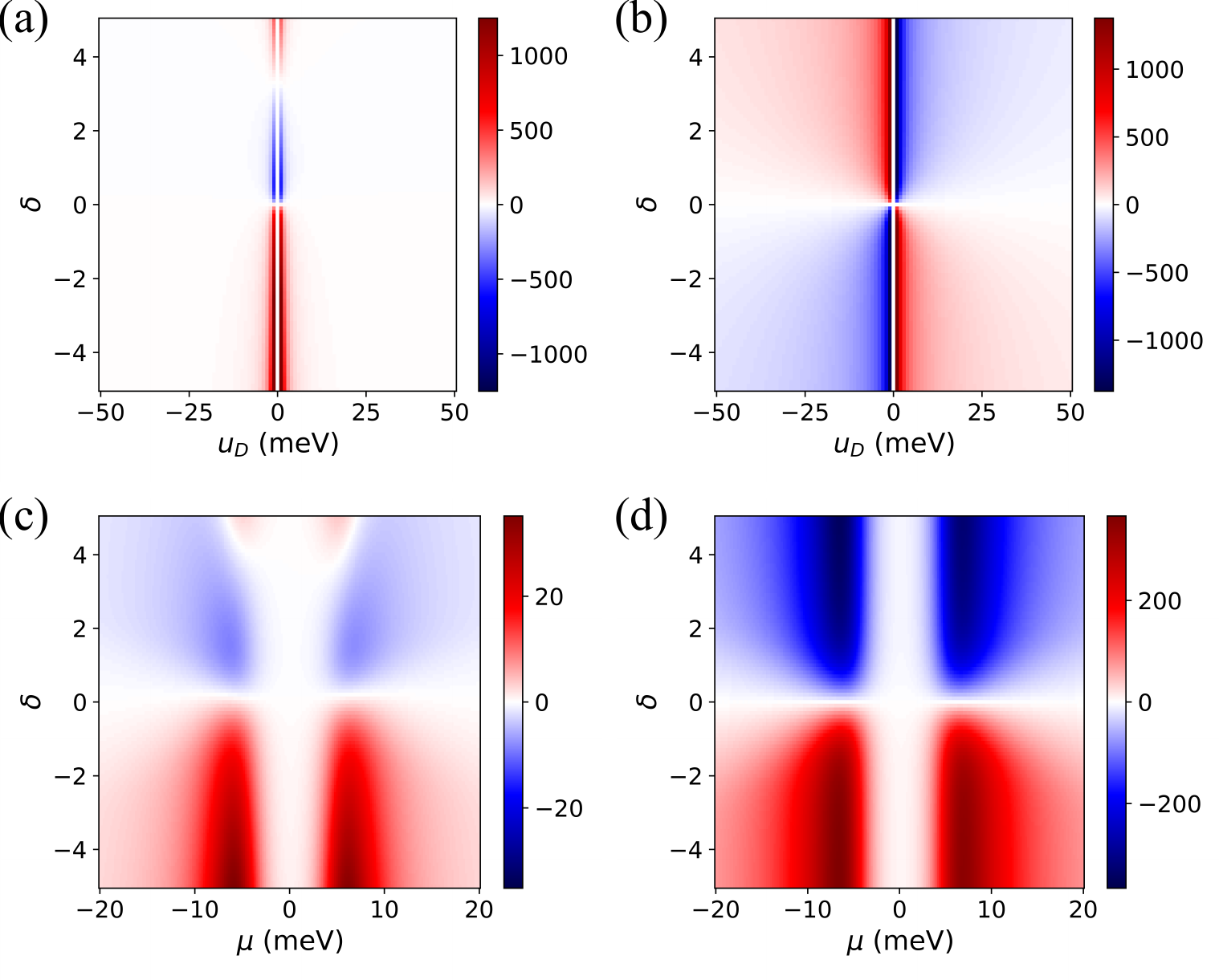}
    \caption{(a) Quantum metric dipole $F_D^{xyy}$ and (b) Berry curvature dipole $(D_2^{xxy})'$ in the $(u_D,\delta)$ phase diagram with $\mu=u_D/2$.
    (c) Quantum metric dipole and (d) Berry curvature dipole in the $(\mu,\delta)$ phase diagram with $u_D=10$meV.
    }
    \label{fig:pd2}
\end{figure}

Fig.~\ref{fig:bands}(a) and (b) show the band structure of the bands with $u_D=0\,{\rm meV}$ and $\delta=0$. 
Fig.~\ref{fig:bands}(c) shows the band structure of the bands with $u_D=50\,{\rm meV}$ and $\delta=0$. 
Fig.~\ref{fig:Fermi_surface} shows the Fermi surface of this model at different energies.
We compute the spectral function with $\rho(E_f)=-\frac{1}{\pi} {\rm Im\,Tr}  \frac{1}{(E_f-H)+i\Gamma}$.
The spectral function plotted is normalized as $\rho \Gamma$.
The grid size is $200\times 200$ on region $[-k_0,k_0]\times[-k_0,k_0]$ and $k_0=0.1/a$.

Fig.~\ref{fig:QGT} shows the quantum metric $g^{ab}$ and Berry curvature $\Omega$ for the upper band.
Fig.~\ref{fig:BCD} shows the Berry curvature dipole term $D_2^{xxy}=\int_\kk D_{2x} \Theta(\mu-\epsilon_\kk)$ and $D_2^{yxy}=\int_\kk D_{2y} \Theta(\mu-\epsilon_\kk)$.
Due to mirror $M_x$ symmetry, the integral $D_2^{yxy}=0$ for both cases.
When $\delta=0$, $C_{3z}$ symmetry forces $D_2^{xxy}=0$. 
Fig.~\ref{fig:BCD}(c) shows the difference between $\delta=0.5$ and $\delta=0$, which is clearly nonzero after ingratiation with $\Theta$.

Fig.~\ref{fig:pd2} shows the quantum metric dipole and Berry curvature dipole in the phase diagram of $(u_D,\delta)$ with $\mu=u_D/2$ and $(\mu,\delta)$ with $u_D=10$meV.
It shows that within a large parameter space of $\delta$, the magnitude of quantum metric dipole and Berry curvature dipole remain large.

\begin{figure}[t]
    \centering
    \includegraphics[width=0.6\linewidth]{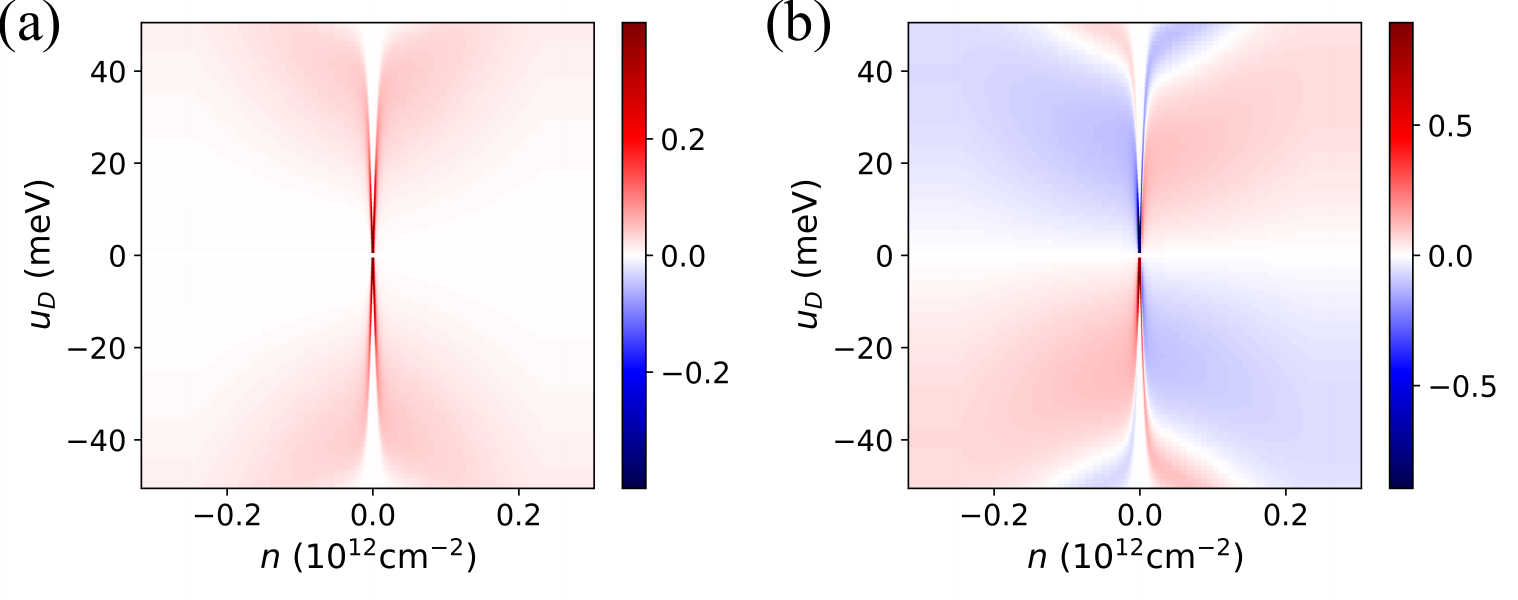}
    \caption{The $(n,u_D)$ phase diagram of (a) quantum metric dipole $F_{D}^{xyy}$ for $\delta=0$ and (b) the Berry curvature dipole $(D_2^{xxy})'$ for $\delta=0.5$.}
    \label{fig:pd3}
\end{figure}

In the maintext Fig.~2 (c) and (d) we have shown the $(u_D, \mu)$ phase diagram of quantum metric dipole $F_{D}^{xyy}$ for $\delta=0$ and the Berry curvature dipole $(D_2^{xxy})'$ for $\delta=0.5$.
Here we convert $\mu$ to electron filling $n$ and show the phase diagram in Fig.~\ref{fig:pd3}.
In this calculation we used lattice constant $a=2.46${\AA} and grid size $1000\times 1000$ on region $[-k_0,k_0]\times[-k_0,k_0]$ and $k_0=0.1/a$.
As shown in the figures, the quantum metric dipole concentrates at small $n$ and the region smear when $u_D$ is enlarged.
The Berry curvature dipole is nonzero for a relative large region in the phase diagram.

\section{Potential experimental protocols}
Recent experiments have reported giant nonlinear Hall responses in gate-tunable two-dimensional materials. These observations provide a natural platform for testing the generalized Wiedemann--Franz and Mott relations derived in the maintext. In particular, our generalized relation,
\begin{equation}\label{eqn:generalized_relation}
    L_{1,22}^{a,bc} = \mathfrak{L} L_{1,11}^{a,bc} \,,
\end{equation}
suggests that a large second-order electrical response should be accompanied by a correspondingly large nonlinear thermoelectric response, including nonlinear Nernst and Ettingshausen effects.
Here $\mathfrak{L}$ is the Lorenz number.

To estimate the required experimental conditions, we compare the electric field and temperature gradient needed to generate nonlinear currents of comparable magnitude. Suppose that an electric field \(E\) produces a charge current through the second-order electrical response, 
$j^a \sim L_{1,11}^{a,bc} E_b E_c$.
The corresponding nonlinear thermoelectric response induced by a temperature gradient is controlled by $L_{1,22}^{a,bc}$. Using Eq.~\eqref{eqn:generalized_relation}, the temperature gradient required to produce a current of the same order is estimated as
\begin{equation}
    \nabla T \sim \frac{E}{\sqrt{\mathfrak{L}}} \,. 
\end{equation}
As a concrete estimate, Ref.~\cite{chichinadze2024observation} reported a giant nonlinear electrical response of order $L_{1,11} \sim 1\,{\rm \frac{\mu m}{\Omega \cdot V}}$.
Since currents below $10^{-10}\,{\rm A}$ can be accurately resolved in such devices, this corresponds to an electric field scale of approximately
\begin{equation}
    E \sim 10^{-2}\,{\rm V/m} \,.
\end{equation}
Then the equivalent temperature gradient is then estimated to be
\begin{equation}
    \nabla T \sim 10^2\,{\rm K/m}.
\end{equation}
This scale is well within current experimental capabilities. For example, recent thermopower measurements in Bernal bilayer graphene were performed with temperature gradients as large as
$\nabla T \sim 6.0 \times 10^3\,{\rm K/m}$. 
Therefore, the temperature gradients required to observe the nonlinear thermoelectric counterparts of the giant nonlinear Hall response should be experimentally accessible.

A straightforward experimental test of Eq.~\eqref{eqn:generalized_relation} is to measure nonlinear electrical and thermoelectric transport in the same gate-tunable two-dimensional device. Gate-dependent measurements have already enabled tests of the conventional Mott relation in the linear-response regime by comparing the measured Seebeck coefficient $S(n)$ with the value inferred from $d\ln R/dn$, where $R$ is the measured electrical resistivity~\cite{Ghosh2025Thermopower}. The generalized Mott relation can be tested in an analogous manner by comparing the gate dependence of the nonlinear thermoelectric coefficient with the chemical-potential derivative of the nonlinear electrical coefficient.

As in linear thermoelectric measurements, the application of a magnetic field can be important for separating longitudinal and transverse responses and for reducing the mixing between different transport channels. In addition, temperature, electric field, and gate potentials provide useful tuning knobs for driving phase transitions and testing the generalized relations across a broad range of quantum phases, including strongly interacting regimes.

\end{document}